\documentclass{aa}
\usepackage[varg]{txfonts}
\usepackage{graphicx}
\usepackage{natbib}
\bibpunct{(}{)}{;}{a}{}{,}
\usepackage{lscape}
\usepackage[utf8]{inputenc}

\begin{document}

\title{The Fornax Deep Survey (FDS) with the VST:} 

\subtitle{III. Low Surface Brightness (LSB) dwarfs and Ultra Diffuse Galaxies (UDGs) in the center of the Fornax cluster} 

\author{Aku Venhola\inst{1,2}
  \and Reynier Peletier\inst{2} 
  \and Eija Laurikainen\inst{1}
   \and Heikki Salo\inst{1}
    	 \and Thorsten Lisker\inst{3}
    	 \and Enrichetta Iodice\inst{5}
    	  \and Massimo Capaccioli\inst{4}
    	   \and Gijs Verdoes Kleijn\inst{2}
    	    \and Edwin Valentijn\inst{2}
    	    	 \and Steffen Mieske \inst{6}
    	    	  \and Michael Hilker \inst{7}
    	    	   \and Carolin Wittmann \inst{3}
    	    	    \and Glenn van de Ven \inst{8}
    	    	    	 \and Aniello Grado \inst{5}
    	    	    	  \and Marilena Spavone \inst{5}
    	    	    	   \and Michele Cantiello \inst{9}
    	    	    	   	\and Nicola Napolitano \inst{5}
    	    	    	   	 \and Maurizio Paolillo \inst{4,5}
    	    	    	   	 \and Jes\'us Falc\'on-Barroso \inst{10,11}} 

\offprints{A. Venhola, \email{aku.venhola@oulu.fi}}

\institute{Astronomy Research Unit, University of Oulu, Oulu, FI-90014  Finland 
  \and Kapteyn Institute, University of Groningen, Postbus 800, 9700 AV Groningen, the Netherlands 
  \and Zentrum für Astronomie der Universität Heidelberg, Mönchhofstrasse 12-14, D-69120 Heidelberg, Germany
  \and University of Naples Federico II, C.U. Monte Sant'Angelo, Via Cinthia, 80126 Naples, Italy
  \and INAF - Astronomical Observatory of Capodimonte, Salita Moiariello 16, I80131, Naples, Italy 
  \and European Southern Observatory, Alonso de Cordova 3107, Vitacura,  Santiago, Chile
  \and European Southern Observatory, Karl-Schwarzschild-Strasse 2, D-85748 Garching bei München, Germany
  \and Max-Planck-Institut für Astronomie, Königstuhl 17, D-69117 Heidelberg, Germany
  \and INAF Osservatorio Astronomico di Teramo,  Via Maggini, 64100, Teramo, Italy
  \and Instituto de Astrofisica de Canarias, C/ Via L'actea s/n, 38200 La Laguna, Spain 
  \and Depto. Astrofisica, Universidad de La Laguna,  C/ Via L'actea s/n, 38200 La Laguna, Spain
}

\date{Received \today / Accepted ---}

\abstract { Studies of low surface brightness (LSB) galaxies in nearby
  clusters have revealed a sub-population of extremely diffuse
  galaxies with central surface brightness of $\mu_{0,g'}$ > 24 mag
  arcsec$^{-2}$, total luminosity M$_g'$ fainter than -16 mag and
  effective radius between 1.5 kpc < R$_e$ < 10 kpc. The origin of
  these Ultra Diffuse Galaxies (UDGs) is still unclear, although
  several theories have been suggested. As the UDGs overlap with the
  dwarf-sized galaxies in their luminosities, it is
  important to compare their properties in the
  same environment. If a continuum is found between the properties of 
  UDGs and the rest of the LSB population, it would be consistent with 
  the idea that they have a common origin. }
  { Our aim is to exploit the deep g', r' and i'-band images of
  the Fornax Deep Survey (FDS), in order to identify LSB galaxies
  in an area of 4 deg$^2$ in the center of the Fornax cluster. The 
  identified galaxies are divided into UDGs and dwarf-sized LSB 
   galaxies, and their properties are compared.
  }  
  {We identify visually all extended structures having r'-band central
  surface brightness of $\mu_{0,r'}$ > 23 mag arcsec$^{-2}$. We classify the 
  objects based on their appearance into galaxies and tidal structures, 
  and perform 2D S\'ersic model fitting with GALFIT to measure the properties 
  of those classified as galaxies. We analyze their radial distribution 
  and orientations  with respect of the cluster center, and with respect 
  to the other galaxies in our sample. We also study their colors and 
  compare the LSB galaxies in Fornax with those in other environments.
  }  
  {Our final sample complete in the parameter space of the previously
  known UDGs, consists of 205 galaxies of which 196 are LSB dwarfs 
  (with R$_e$ < 1.5kpc) and 9 are UDGs (R$_e$ > 1.5 kpc). We show that 
  the UDGs have (1) g'-r' colors similar to those of LSB dwarfs of 
  the same 
  luminosity. (2) The largest UDGs (R$_e$>3kpc) in our sample 
  appear different from the other LSB galaxies, in that they are significantly 
  more elongated and extended, whereas (3) the smaller UDGs differ from
  the LSB 
  dwarfs only by having slightly larger effective radii. (4) We 
  do not find clear differences between the structural parameters of 
  the UDGs in our sample and those of UDGs in other galaxy 
  environments.  (5)
  We find that the dwarf LSB galaxies in our sample are less concentrated 
  in the cluster center than the galaxies with higher surface brightness, 
  and that their number density drops within 180 kpc from 
  the cluster center. We also compare the LSB dwarfs in Fornax with the 
  LSB dwarfs in the Centaurus group, where data of similar 
  quality to ours is available. (6)  We find the smallest LSB
  dwarfs to have similar colors, sizes and S\'ersic profiles regardless of their 
  environment. However, in the Centaurus group the colors  become bluer with increasing galaxy magnitudes, an effect 
  which is probably due to smaller mass and hence weaker environmental
  influence of the Centaurus group.
  } 
  {Our findings are consistent with the small UDGs forming the tail 
  of a continuous distribution of less extended LSB galaxies. However, 
  the elongated and distorted shapes of the large UDGs could imply that 
  they are tidally disturbed galaxies. Due to limitations of the automatic
  detection methods and uncertainty in the classification the objects, 
  it is yet unclear what is the
  total contribution of the tidally disrupted galaxies in the UDG 
  population. 
  } {}

\maketitle

\section{Introduction}

For several decades it has been known that some galaxies have
much lower surface brightnesses than
  others. \citet{Davies1994} demonstrated, using the ESO-Uppsala
galaxy catalog \citep{Lauberts1989}, how galaxy samples with fixed
magnitude and size limits are biased by missing the
low surface brightness galaxies due to limits in depth and sensitivity, resulting in a lower data quality. Historically, all galaxies with B-band
central surface brightness of $\mu_{0,B}$ > 23 mag arcsec$^{-2}$ are
considered as Low Surface Brightness (LSB) galaxies
\citep{Impey1997}. However, this definition is relatively broad and
includes galaxies ranging from almost Milky Way sized galaxies to the
smallest Milky Way satellite dwarf spheroidals (dSph). Early
studies of LSB galaxies with detailed photometric measurements
(\citealp{Romanishin1983}, \citealp{Bothun1985},
\citealp{Sprayberry1995}, \citealp{Impey1989} and \citealp{Impey1990})
concentrated mostly on relatively massive galaxies. These giant LSB
galaxies, like Malin 1 \citep{Impey1989}, for which they measured a
disk scale length of 55 kpc and a V-band central surface
brightness of $\mu_{0,V}$ = 25.5 mag arcsec$^{-2}$, form an
interesting class of objects in terms of galaxy evolution processes,
as it is not well understood how they have managed to grow so massive without increasing their surface brightness. The appearance of such giant LSB galaxies has been proposed to be a result of their high angular momentum, and low star formation rate \citep{Jimenez1998}. These galaxies are predicted to be relatively rare and only appear in low density environments. In dense environments like the Fornax cluster, galaxies experience frequent tidal interactions with other galaxies and are affected by the cluster potential. This process, called \textit{harassment} \citep{Moore1998}, tends to rip off material from the galaxies that enter deep into the cluster core, and eventually makes them denser \citep{Smith2015}. 

\indent In the end of 1980's, large galaxy catalogs of the Virgo \citep{Binggeli1985} and Fornax clusters \citep{Ferguson1989} included also many dwarf sized (B-band absolute magnitude M$_B$ > -18 mag\footnote{M$_B$ > -18 corresponds to M$_{r'}\sim$-19 mag. See Appendix C for details.}) LSB galaxies, which greatly increased the number of known LSB galaxies. The automated photometry of \citet{Davies1988} allowed the detailed analysis of dwarf galaxies in these samples.  Later studies, like the ones of \citet{Davies1990} or \citet{Sabatini2005}, showed that dwarf LSB galaxies actually form the bulk of the population at the low luminosity end of the luminosity function. Most of these galaxies were found in clusters. In fact, galaxies with central V-band surface brightness of  $\mu_{0,V}$ = 22 - 25 mag arcsec$^{-2}$ have been estimated to contribute significant amount (even 50\%) of the light emitted by galaxies  \citep{Impey1997}. 

\indent It is only recently that new instruments, such as OmegaCAM \citep{Kuijken2002}, Suprime-Cam \citep{Miyazaki2002}, DECam \citep{Flaugher2015}, and MegaCAM \citep{Boulade1998}, have made it possible to perform multi-band surveys with limiting V-band surface brightness down to $\mu_V$ $\sim$28 mag arcsec$^{-2}$, over large areas in the sky. These new observations allow to study also the distribution and properties of the dwarf LSB galaxies in statistically significant samples. Indeed, new imaging surveys like the Next Generation Virgo Survey \citep{Ferrarese2012}, have revealed a large number of low surface brightness galaxies. Although  the wide-field instruments attached to large telescopes have shown their efficiency performing deep surveys, other approaches have proven effective as well. One of them used the Dragonfly instrument \citep{Abraham2014}, which consists of a set of small 143 mm cameras, and was used when \citet{VanDokkum2015}  discovered a large number of extended LSB galaxies in the Coma cluster, which they named Ultra Diffuse Galaxies (UDG).

\indent The UDGs discovered by \citet{VanDokkum2015} are defined to be extended (effective radius $\mathrm{R}_{e}$ > 1.5 kpc) and faint (V-band absolute magnitude -16 mag < $\mathrm{M}_{\mathrm{V}}$ < -13 mag), and have similar central surface brightnesses ($\mu_{0,g}$ > 24 mag arcsec$^{-2}$) as the faintest dwarf galaxies. However, their effective radius can be even 10 times larger. What makes these galaxies particularly interesting is that UDGs reside in the cluster environment where they appear in large numbers (\citealp{Yagi2016}, \citealp{VanDerBurg2016},\citealp{Wittmann2017}, \citealp{VanDerBurg2017}). UDGs have now been found in all clusters where they have been searched for. For example, \citet{VanDerBurg2016} used an automated algorithm to find UDGs in clusters in the redshift range 0.044 < z < 0.063. They found that their abundance increases with increasing cluster halo mass, reaching $\sim$ 200 UDGs in typical halo masses of M$_{200}$ $\approx$ 10$^{15}$ M$_{\odot}$. Recently UDGs have been reported also in some nearby galaxy groups (\citealp{Merrit2016}, \citealp{Toloba2016}, \citealp{Crnojevic2016}, \citealp{Roman2017})  and low density environments (\citealp{MartinezDelgado2016} and \citealp{Leisman2017}) showing that these galaxies appear in all kind of galaxy environments. 

\indent In our study, we hereafter define all the galaxies with $\mu_{0,r'}$ > 23 mag arcsec$^{-2}$ as LSB galaxies, and the ones additionally having absolute r'-band magnitude M$_{r'}$ > -19 mag as dwarf LSB galaxies. LSB galaxies that have an effective radius R$_e$ > 1.5 kpc are called UDGs\footnote{For typical LSB galaxies g'-r' $\sim$ 0.6, so our limits correspond to $\mu_{g',0}$ = 23.6 mag arcsec$^{-2}$ and M$_{g'}$ = -18.4 mag.}.

\indent The formation mechanism of UDGs is still unclear. They have been suggested to form from medium mass (halo mass of 10$^{10-11}$M$_{\odot}$) galaxies as a result of strong gas outflows due to star formation feedback \citep{DiCintio2016}, whereas \citet{Baushev2016} suggested  that UDGs can form via head-on collisions of gas-rich systems in the centers of galaxy clusters. UDGs have also been suggested to be the high spin tail of the typical dwarf elliptical (dE) galaxy population \citep{Amorisco2016I}. Indeed, it is important to study the photometric properties of UDGs and LSB dwarf galaxies in different environments to see if there is a continuum between their properties.

\indent The Fornax cluster, with a virial mass of 7$\times$10$^{13}$ M$_{\odot}$ \citep{Drinkwater2001}, is less massive than the Coma (1.4$\times$10$^{15}$ M$_{\odot}$, \citealp{Lokas2003}) and Virgo clusters  (1-3$\times$10$^{14}$ M$_{\odot}$ , \citealp{McLaughlin1999}). However, in spite of its fairly low mass the Fornax cluster has a high fraction of early-type galaxies (E+S0+dE+dS0)/all = 0.87 (\citealp{Ferguson1989b}), and a central galaxy density similar to that of the more massive Virgo cluster. The core of the Fornax is also filled with hot X-ray emitting gas (\citealp{Paolillo2002} and references therein), which makes the galaxies vulnerable to ram pressure stripping, which removes the cold gas from the galaxies. This implies that any galaxy which has spent a long time in the core of the Fornax cluster, should consist only of fairly old stellar populations.

\indent So far only a few studies have mapped the LSB galaxy population in the Fornax cluster.  \citet{Bothun1991} studied the properties of the LSB galaxies with central B-band surface brightnesseses of $\mu_{0,B}$ > 23 mag arcsec$^{-2}$, and showed that there are tens of relatively metal-poor galaxies, which make a significant contribution to the faint end of the luminosity function. Due to their low surface brightnesses it is problematic to spectroscopically confirm their distances, and therefore often the cluster membership has been deduced from the clustering of these galaxies. However, the cluster membership of several LSB galaxies were spectroscopically confirmed in the sample of \citet{Drinkwater1999}, which included galaxies with B-band total apparent magnitudes brighter than m$_B$ < 19.7 mag (M$_B$ < -11.7) and central surface brightnesses between 20 and 24 mag arcsec$^{-2}$. Also the study of \citet{Mieske2007}, which used surface brightness fluctuation analysis to define galaxy distances, confirmed the cluster membership of several Fornax LSB galaxies. These studies support the idea that at least 2/3 of the LSB galaxies in the area of Fornax are real cluster members. A recent study of \citet{Munoz2015}, performed with the DECam instrument, searched for new faint galaxies in the central parts of the Fornax cluster. Their observations reaching g'-band point sources down to 26.6 mag with S/N>5, reveal more than hundred previously non-detected dwarf sized LSB galaxies. The faintest LSB dwarf galaxies in their sample have R$_{e}$ $\approx$ 100 pc, which means that they have similar sizes as the Local Group dSph's (\citealp{McConnachie2012} and the references therein).

\indent In this paper, we perform a systematic search for  LSB galaxies in the images of the Fornax Deep Survey (FDS), which is an ongoing survey using the VLT Survey Telescope (VST) at ESO / Cerro Paranal. It covers a larger field-of-view than any of the previous Fornax surveys with deep multi-band observations, and has similar depth as the Next Generation Virgo Survey \citep{Ferrarese2012}. 
The survey has already obtained several results, such as the discovery of an extended globular cluster population \citep{DAbrusco2016}, the characterization of the extended stellar halo of NGC 1399 \citep{Iodice2016}, and the analysis of the merger system around NGC 1316 \citep{Iodice2017}. Since we have deep data in g', r' and i'-bands, we can determine also the colors of the galaxies.

\indent We present a sample of LSB galaxies in the Fornax cluster, based on our analysis, and combined with those obtained in the previous studies. In sections 2 and 3 we describe the data used in this work, and briefly describe the reduction steps. In sections 4 and 5 we present the sample selection and the photometric measurements performed to obtain the structural parameters. We show the cluster-centric radial distributions (section 6.1), orientations (section 6.2) and colors (section 7) of the galaxies. In section 8 we discuss the results in the context of different formation theories of UDGs, and in section 9 give the conclusions of this paper. For the Fornax cluster we use the distance of 19.95 Mpc \citep{Tonry2001} corresponding to the distance modulus of 31.43 and scale of 0.0967 kpc arcsec$^{-2}$.

\section{Data}

We use the ongoing Fornax Deep Survey (FDS), which consists of the combined data of the Guaranteed Time Observation Surveys FOCUS (P.I. R. Peletier) and VEGAS (P.I. E. Iodice), dedicated to the Fornax cluster. Both surveys are performed with the ESO VLT Survey Telescope (VST), which is a 2.6-meter diameter optical telescope located at Cerro Paranal, Chile \citep{Schipani2012}. The imaging is done with the OmegaCAM instrument \citep{Kuijken2002}, using the u', g', r' and i'-bands, and $1^{\circ} \, \times \, 1^{\circ}$ field of view. OmegaCAM consists of an array of 8 $\times$ 4 CCDs, each with 2144 $\times$ 4200 pixels. The pixel size is 0.21 arcsec and the average FWHM of the observations is $\approx$ 1 arcsec, so that the PSF is well sampled. For further information about the data see \citet{Iodice2016} and Peletier et al. (\textit{in prep.}).
 
\indent The observations used in this work were gathered in visitor mode runs during November 2013, 2014 and 2015 (ESO P92, P94 and P96, respectively). All the observations were performed in clear (photometric variations $<$ 10 \%) or photometric conditions. The observations in u' and g'-bands were obtained in dark time, and those of the other bands in grey or dark time. 

\begin{figure}[!ht]
    \centering
        \resizebox{\hsize}{!}{\includegraphics{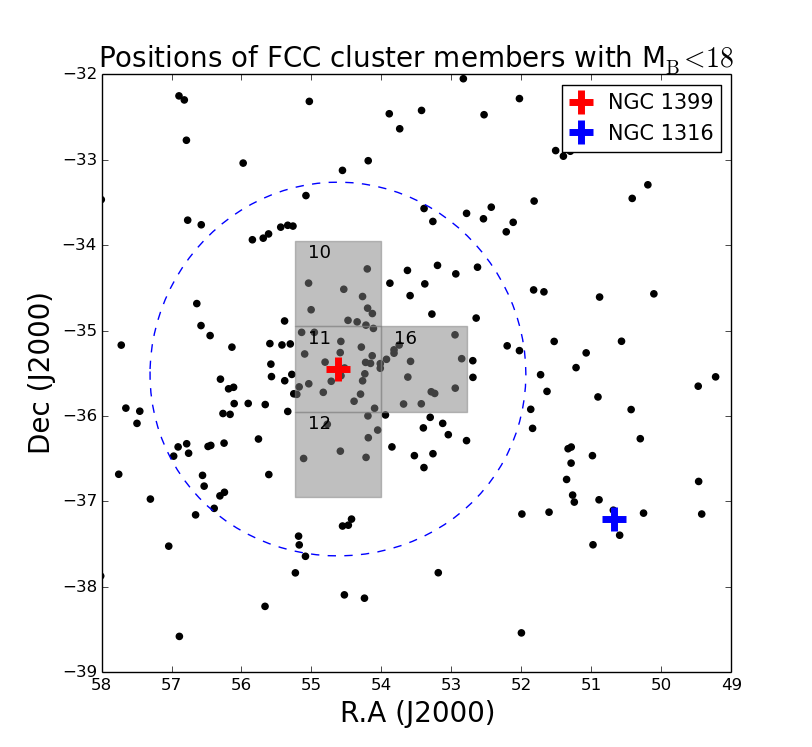}}
    \caption{Locations of the analyzed  1$^{\circ}$ x 1$^{\circ}$ Fornax fields (highlighted with \textit{grey}), plotted over the FCC galaxies \citep{Ferguson1989} with m$_{\mathrm{B}} \, < \, 18$ mag and the member classification 1 or 2 (confirmed or likely member, respectively, in \citealp{Ferguson1989}).  The \textit{blue dashed circle} is the virial radius 2.2$^{\circ}$ (or 0.7 Mpc at the distance of Fornax cluster) adopted from  \citet{Drinkwater2001}. The \textit{red} cross shows the cD galaxy NGC 1399 (see \citealp{Iodice2016}) in the core of the Fornax cluster, and the \textit{blue cross} shows the elliptical shell galaxy NGC 1316(see \citealp{Iodice2017}) , which is the central galaxy of the Fornax-SW subcluster.}
    \label{fig:fieldlocaitons}
\end{figure}

\indent The observation area of FDS is divided into $1^{\circ} \, \times \, 1^{\circ}$ fields (see Fig. \ref{fig:fieldlocaitons}) with some overlaps between the fields. The observations of each field were performed using offsets larger than 1 deg between consecutive exposures, always covering an adjacent field. During the sequence of multiple exposures, additional dithers of $\lesssim$10 arcmin width have been added to the 1 deg offsets with a fixed pattern for all fields. This ensures that the whole area has a uniform depths, i.e. whenever an exposure is dithered from the field center, and thus only partially covers the original field, there will be exposures in the adjacent fields with identical dithers providing a full coverage over all fields. This offset and dither strategy makes it possible to perform accurate sky subtraction without spending time for separate sky exposures (see section 3.2  for details). The total exposure times in all fields are 11000, 8000, 8000, and 5000 sec in u', g', r' and i'-bands, respectively. The total exposure times are divided into 150 s exposures meaning that each field is covered with a minimum of 75, 55, 55 and 35 exposures in u', g', r' and i', respectively. 

\begin{figure*}
    \centering
    \resizebox{\hsize}{!}{\includegraphics[width=17cm,height=22cm]{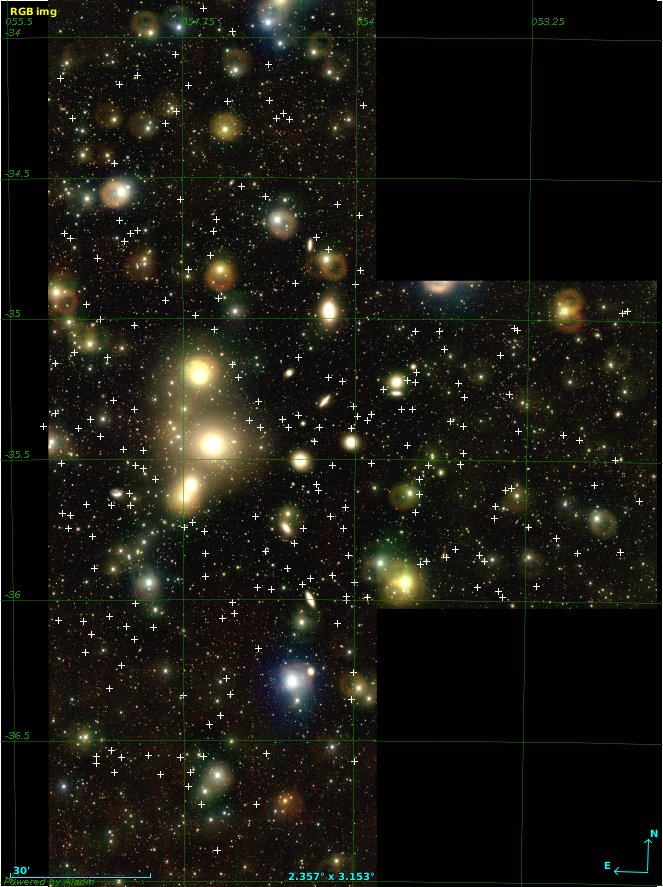}}
    \caption{Locations of the galaxies identified by us (marked with crosses). The background image shows the fields 10, 11, 12 and 16 using the combined i', r' and g' bands. \textit{Aladin} \citep{Boch2011} is used to generate the graphics.}
    \label{fig:locations}
\end{figure*}

\indent The data used in this work cover an 4 deg$^{2}$ area  centered on NGC 1399 (see  Fig. \ref{fig:locations}). All the fields are imaged  using the four bands. The limiting  $5\sigma$ magnitudes (in AB system) for 1 arcsec$^2$ area are 27.6, 28.5, 28.5 and 27.1 mag in u', g', r' and i', respectively. In this work only g', r' and i' bands are used, due to their deeper surface brightness limit and higher sensitivity for the relatively red galaxies in the Fornax cluster. The sources studied in this work have a central r'-band surface brightness of $\mu_{0,r'}$ $\geq$ 23 mag arcsec$^{-2}$, so that most of them are not bright enough in the u'-band to detect them. In this work we analyse the fields 10, 11, 12 and 16 (the \textit{grey areas} in Fig. \ref{fig:fieldlocaitons}).

\section{Data reduction}

We have developed a pipeline to reduce FDS OmegaCAM data adapting the general OmegaCAM pipeline within the AstroWISE environment \citep{McFarland2013}. The pipeline used in this work is different from the one described in \citet{Iodice2016}. However, the results of these two pipelines are consistent with each other (details will be given in Peletier et al. {\it in prep.}). The main reduction steps are provided in the next paragraphs.

\subsection{Instrumental corrections}

The instrumental corrections include removal of bias and correcting the sensitivity variations over the field of view (flat-field and illumination correction).

\indent Bias is first defined from the overscan regions of the images, and the row-wise median values are subtracted from each science image. For not to add too much noise in the bias subtraction, we used the overscan corrected master bias image, obtained by median combining 10 bias images taken each night.

\indent After removing bias from the  science images, they are corrected for the sensitivity variations over the focal plane by dividing the images by a master flat-field image. We adopted the flat-fielding method that was used by the Kilo Degree Survey (KiDS, \citealp{deJong2015}): the master flat-field is achieved by first median-combining and normalizing 8 dome flat-fields and 8 twilight flat-fields, and then multiplying the averaged flat-fields with each other. High-frequency spatial Fourier modes are corrected using the dome flats and low frequencies using the twilight flats. This is based on the pre-assumption that the large-scale illumination of the twilight flat-field matches better the observational situation than that obtained from the dome. On the other hand, in the dome flats the {\it S/N} is high, which can be used to capture the pixel-by-pixel variations in the pixel sensitivity.

\indent However, even after applying the flat-field correction small systematic flux variations remain across the instrument. These variations can be corrected by applying an illumination correction. We use the correction models made for the KiDS (see \citealp{Verdoes-Kleijn2013} for details). The models are made by mapping the photometric residuals across the CCD array using a set of dithered Landolt Selected Area (SA) field \citep{Landolt1992} observations, and fitting a linear model to the residuals. The images are multiplied with this illumination correction.  The correction is applied after sky subtraction (see the next sub-section), to avoid the sky residuals being amplified by the illumination correction procedure.

\subsection{Background subtraction and de-fringing}

The images contain sky background flux composed of direct and scattered atmospheric emission, and scattered light of bright celestial sources. A careful removal of the atmospheric background light is essential when studying LSB objects such as UDGs. When considering a single image at low luminosity levels, we cannot directly tell which part of the light is diffuse light coming from the sources and which is background light. To bypass the problem, we can make an assumption that the pattern of background light stays constant if the telescope pointing direction is not changed by more than a few degrees. Due to the large dithers between the consecutive integrations the objects are not likely to appear twice in the same pixel, which allows us to produce background models by averaging a stack of images. The intensity of the sky changes thorough the night (1--10\% between exposures), especially at the beginning and at the end of the night, which forces to scale the images used for the background model before combining them. The pattern of the scattered light changes also as a function of telescope pointing direction and the positions of the Sun and the Moon, so that any accurate model of those variations is not available.

\indent A unique background model is made for each CCD of each exposure by scaling and stacking 12 consecutive dithered exposures (six before and six after the frame, whose background is modelled). First, SExtractor \citep{Bertin1996} is used to mask all stars and galaxies from all individual images (12 consecutive pointings). Specifically, we masked the objects with 5 pixels above the 5$\sigma$ threshold, using a 50$\times$50 pixel grid to estimate the background in the masking process. We ensured that we are not masking systematically any shapes (like vignetted edges) of the background by comparing a set of masks of background models. After the masking, the images are scaled with each other assuming that the shape of the background scales linearly with the total level of the background. The scaling factor $s$ can be found by:
\begin{equation}
s = \mathrm{Median}\left( \frac{m_{1,i}}{m_{2,i}}\right), \, i=1,...,96,
\end{equation}
where $m_{1,i}$ is a set of medians measured within 96 90 $\times$ 90-pixel areas in image 1, used as a reference image (i.e. the image for which we are making the background model). Correspondingly, $m_{2,i}$ is a set medians measured at the same locations in image 2, which we are scaling. Before combining the frames we exclude the images that are not suitable for obtaining the background model. We exclude frames, where more than 1/3 of the area is masked. Excluded are also frames which have a large scatter in $\frac{m_{1,i}}{m_{2,i}}$, since they have either large unmasked objects or have otherwise peculiar background. If only 6 or less frames are found to be useful for the background model (e.g. if an extended source fills the whole CCD being modelled, it is excluded), the scaling is done by using the median values of all 32 CCDs (instead of just one) and then using the equation (1). After masking and multiplying the images with $s$, the selected frames are combined by taking the median of the stacks. This background model is then subtracted from the final image.

\indent OmegaCAM has interference patterns (\textit{fringes}) in i'-band images due to the internal reflections in the CCDs. Intensity of the fringe patterns is proportional to the total light coming to the CCD. As long as the background and the filter does not vary, the fringes have always the same shape since they are related to the properties of the CCD.  Luckily, the fringes appear also  in the background model and are subtracted with it. No other fringe correction is applied. The intensity of the fringe pattern in our image is lower than the masking threshold, which leaves them unmasked.

\subsection{Weight maps}

The pipeline also generates a weight frame for each exposure. The weight frame carries the information about the noise level, bad pixels, cosmic  rays, satellite tracks, and saturated pixels in the image. The weight of a given pixel $W_{ij}$ can be written as:
\begin{equation}
W_{i,j} = \frac{1}{\sigma^2}\frac{F_{ij}}{I_{ij}} P_{hot} P_{cold} P_{saturated} P_{cosmic} P_{satellite},
\end{equation}

\indent where $\sigma^2$ is the variance of the image measured from the raw frame,  $F$ is the flat-field pixel value, $I$ is the  illumination  correction, and $P_{hot},P_{cold}, P_{saturated}, P_{cosmic}$, and $P_{satellite}$, are the hot-, cold-, saturated-, cosmic ray-, and satellite- pixel maps, respectively (good $\rightarrow$ 1, bad $\rightarrow$ 0). The weight images are used for masking the unwanted pixels from the science images before they are stacked as a final mosaic.

\indent Pixels are masked by giving a value of 0 to their fluxes, while for the other pixels a value of 1 is given. In hot-pixel maps the pixels which have high values in the bias images compared to the other pixels, are masked. In cold-pixel maps the pixels which deviate clearly (either high or low values) from the other pixels in the flat-field images are masked. Cosmic rays are detected from the images using the SExtractor cosmic ray detection algorithm \citep{Bertin1996}. Satellite tracks are detected by first applying a Hough transform \citep{Vandame2001} to increase the linear patterns in the images: the lines consisting of more than 1000 pixels with intensity above the 5-$\sigma$ level relative to the background are then masked.

\subsection{Astrometric and photometric calibrations}

The first-order astrometric calibration is done by first matching the pixel coordinates to RA and DEC using the World Coordinate System (WCS) information from the fits header. Point source coordinates are then extracted using SExtractor and associated with the 2 Micron All Sky Survey Point Source Catalog (2MASS PSC, \citealp{Skrutskie2006}). The transformation is then extended by a second-order two-dimensional polynomial across the focal plane. SCAMP \citep{Bertin2006} is used for this purpose. The polynomial is fitted iteratively five times, each time clipping the 2$\sigma$-outliers. The astrometric solution gives typically RMS errors of 0.3 arcsec (compared to 2MASS PSC) for a single exposure, and 0.1 arcsec for the stacked final mosaic.

\indent The absolute photometric calibration is performed by observing standard star fields each night and comparing their OmegaCAM magnitudes with the Sloan Digital Sky Survey Data Release 11  (SDSS DR11, \citealp{Alam2015}) catalog values. The OmegaCAM point source magnitudes are first corrected for the atmospheric extinction by subtracting a term $kX$, where $X$ is airmass and $k$ is the atmospheric extinction coefficient with the values of 0.182, 0.102 and 0.046 for g', r' and i', respectively. The zero-point for a given CCD is the difference between the object's corrected magnitude measured from a standard star field exposure and the catalog value. The zero-point for each CCD is kept constant for the whole night, only correcting for the varying airmass.

\indent Fornax is poorly covered with stellar catalogs (in the optical) which could be used to check the accuracy of the photometric calibration. The American Association of Variable Star Observers Photometric All-Sky Survey catalog (APASS, \citealp{Henden2012}) is the only catalog with a large coverage over our observed Fornax fields. However, as the photometric errors of stars in this catalog with M$_{r'}$ > 16 mag are as high as 0.05 mag, and because the photometric accuracy of our data is expected to be better than that, we do not use APASS for comparison. We made principal color analysis in a similar manner as was done for the Sloan Digital Sky Survey (SDSS) data by \citet{Ivezic2004}. We did this test for the stacked 1$^{\circ}\,\times$ 1$^{\circ}$ mosaics. We measured standard deviations of  0.035, 0.029, and 0.046 mag for the widths of the stellar locus principal colors s, w and x. The corresponding values for SDSS are 0.031, 0.025, and 0.042 mag. The offsets of stellar locii are  -0.009, 0.003, and 0.009 mag for s, w and x, respectively. The typical rms scatter of the same offsets for SDSS are 0.007, 0.005, and 0.009 mag for s, w and x showing that our photometric accuracy is comparable to that of the SDSS. Zeropoint errors for SDSS are 0.01, 0.01, and 0.02 mag for g', r' and i' \citep{Ivezic2004}. As the scatter in the stellar locii measured from our data is $\approx$ 10 \% larger than in SDSS,  we estimate that our photometric errors are 0.02, 0.02, and 0.03 mag in g', r' and i'-bands, respectively.

\subsection{Creating mosaics} 

After the astrometric and photometric calibrations, the images are sampled to 0.20 arcsec pixel size and combined using the SWarp software \citep{Bertin2010}. Before combining the images cosmic rays and bad pixels are removed using the weight maps. Regardless of this removal of contaminated pixels, the resulting pixel distribution in the final mosaic is often non-Gaussian. To obtain better stability against outlier pixels, we decided to use median instead of mean when combining the mosaics. In order to achieve maximal depths in the images, the mosaics are combined from all the overlapping exposures. As a result the pipeline produces  $1^{\circ} \times 1^{\circ}$ mosaics with a $0.2$ arcsec pixel resolution, and the corresponding weight images. The pixel values $W_{x,y}$ for the final weight mosaic are obtained as \citep{Kendall1977}:
\begin{equation}
W_{x,y} = \left\{\begin{matrix}
\frac{2}{\pi} \left( \frac{\sum_k \sqrt{w_k}}{n_{\neq 0}} \right )^2 \left ( n_{\neq0} +\frac{\pi}{2}-1 \right), \, \textup{if } n_{\neq 0}\textup{  is even} \\ 
\frac{2}{\pi} \left( \frac{\sum_k \sqrt{w_k}}{n_{\neq 0}} \right )^2 \left ( n_{\neq0} +\pi - 2 \right), \, \textup{otherwise.}
\end{matrix}\right.
\end{equation}
where $W_{x,y}$ is the weight of the median in a re-sampled mosaic pixel,$\sum_k$ is a sum over all the images that overlap with the pixel, $w_k$ is the weight value in the image $k$, and $n_{\neq0}$ is the number of non-zero pixels.

\section{Catalog of Low Surface Brightness objects}

\subsection{Quantitative selection criteria}

We aim to identify and  classify the LSB galaxies in the selected Fornax fields, in particular the extended UDGs. According to our experience (see also \citealp{Munoz2015} and \citealp{Muller2015}), automatic detection using e.g. SExtractor at low surface brightness levels is at present not as reliable as the eye, so that the catalogue is created by visually inspecting the images (done by AV). In future this catalog can be used as a control sample to test the completeness of any automatic detection method. Our sample includes diffuse sources fulfilling the criteria listed below, i.e., we are deliberately excluding compact galaxies with the r'-band central surface brightness of $\mu_{0,r'} < 23$ mag arcsec$^{-2}$. Most of the galaxies with surface brightnesses brighter than that are already identified in the Fornax Cluster Catalog (FCC; \citealp{Ferguson1989}) or can be easily found with automatic detection algorithms.  

\begin{figure}
    \centering
        \resizebox{\hsize}{!}{\includegraphics{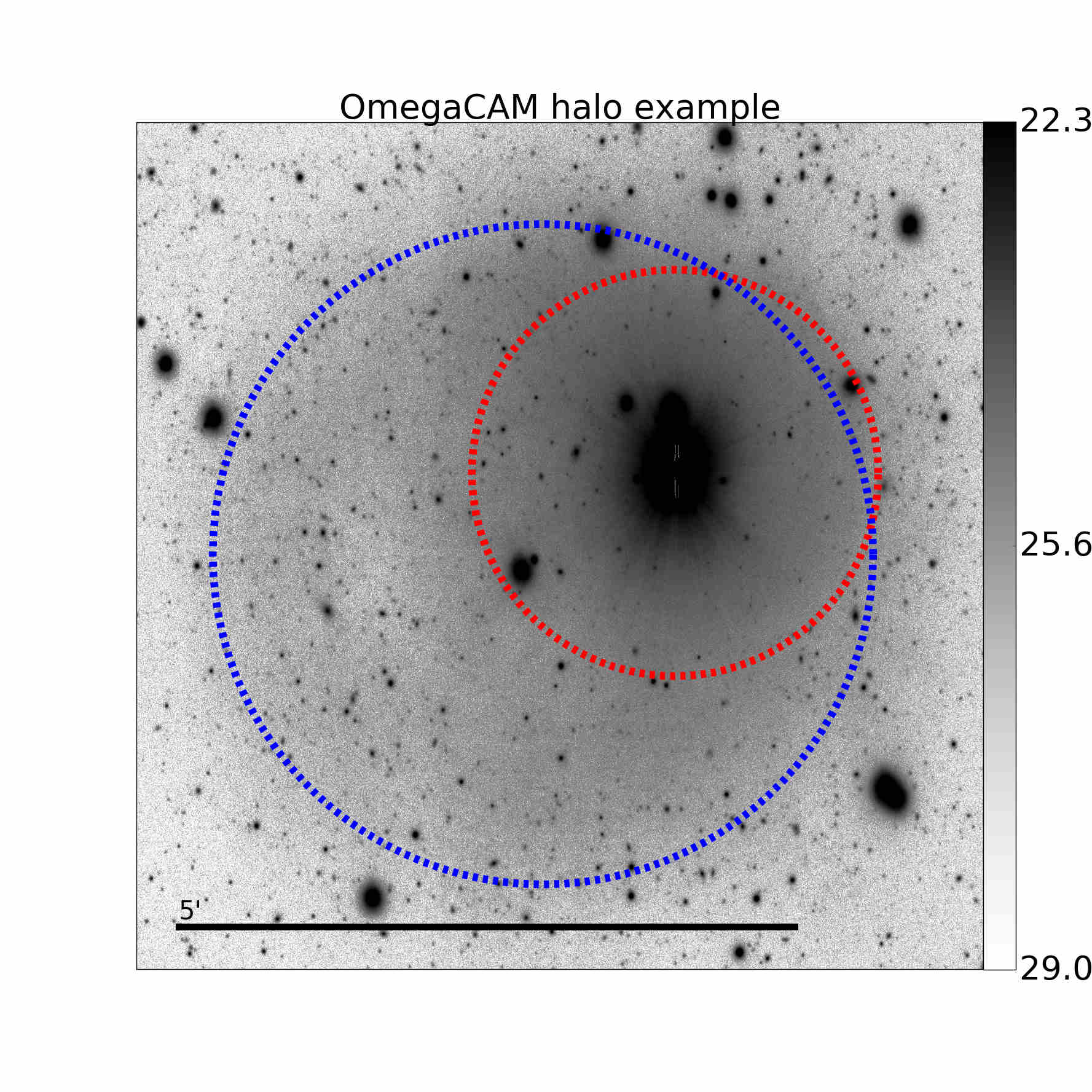}}
    \caption{Example of the reflection halo of OmegaCAM in an r'-band mosaic image. The position of the reflection halo (marked with the \textit{blue dashed line}) depends on the location of its parent star (\textit{red dashed line}) on the instrument's focal plane. Objects residing within the \textit{red} and \textit{blue dashed circles} are excluded from the sample due to the halo contamination. The surface brightness of the area highlighted with the blue circle is $\sim$15 magnitudes fainter than the peak surface brightness of the source star \citep{Capaccioli2015}. In typical seeing conditions, the stars brighter than $\mathrm{m}_{r'} =$ 8.3 mag have central surface brightness of $\mu_{0,r'}$ = 9 mag  arcsec$^{-2}$, which will cause a halo with the surface brightness of 24 mag arcsec$^{-2}$. In the studied 4 deg$^2$ area, there were $\sim$40 stars bright enough to cause halos which had to be excluded. }
    \label{fig:halo_example}
\end{figure}

\indent The selection criteria for the sources are:
\begin{enumerate}

\item \textbf{Low surface brightness}: the object has a central r'-band surface brightness of $\mu_{0,r'}\,\gtrsim \, 23$ mag arcsec$^{-2}$.

\item \textbf{Extended:} the object has a diameter of d$_{27}$ $\gtrsim$ 10 arcsec (corresponding to 0.9 kpc) in the r'-band, at the surface brightness level of 27 mag arcsec$^{-2}$. So, we are excluding small dwarf galaxies, and  sources too faint for any reliable fitting of the surface brightness profiles. The Point Spread Function (PSF) for a source with a central surface brightness of $\mu_{0,r'}$ = 23 mag arcsec$^{-2}$ has d$_{27}$ $<$ 4 arcsec, so that there is no danger to mix faint stars with objects.

\item \textbf{Multi-band detection:} the object can be recognized visually, and it has similar shapes in g', r' and i'-bands. 

\item \textbf{No contamination from bright sources:}  we have excluded all the areas which have severe contamination from stellar halos (see Fig. \ref{fig:halo_example}), i.e. the overlapping halo is brighter than 24 mag arcsec$^{-2}$. This is to make sure that possible halo patterns are not confused with sources or cause biases to the photometry. In the vicinity of bright and extended galaxies like NGC 1399, we include only objects, which are located outside the 24 mag arcsec$^{-2}$ isophotes of the bright galaxy.
\end{enumerate}

The identification criterion 1 is set to guarantee that also faint
sources possibly omitted in the previous studies, are systematically identified. The objects with the lowest surface brightness in FCC have $\mu_{0,B} \approx$ 24 mag arcsec$^{-2}$, which corresponds to $\mu_{0,r'} \approx$ 23 mag arcsec$^{-2}$. The galaxies fainter than this limit have been only rarely mapped before (\citealp{Munoz2015}, \citealp{Hilker2003}, \citealp{Bothun1991}, \citealp{Mieske2007} and \citealp{Kambas2000}). 

\subsection{Accounting for imaging artefacts}

Due to their low surface brightness true objects can easily be confused with imaging artifacts, such as residuals from the background subtraction, or faint reflections from the instrument's optics. The criterion 3 is set to filter out such false detections.

\indent OmegaCAM, like many other wide-field instruments such as MegaCAM, is known to have strong halos (see Fig. \ref{fig:halo_example}) around bright stars, caused by reflections from the secondary mirror. These halos appear on the extension of the line connecting the bright source and the focal point of the CCD array. They are easy to identify as they are always associated with a bright star, and their brightness scales with that of the source. The criterion 4 ensures that these reflections will not bias our photometric analysis.

\indent OmegaCAM is also known to have cross-talk between the CCDs 93-96 (see OmegaCAM user manual\footnote{https://www.eso.org/sci/facilitiess/parnal/instruments/omegacam/\\doc.html} provided by ESO), which can lead a bright source (a star or a galactic nucleus) to appear as a faint ghost image in an adjacent CCD. The cross-talk can manifest either as positive or negative patterns, which have the same shape as the object causing that pattern. The negative crosstalk cannot be confused with the sources, but the positive crosstalk, ending up to 4 \% of the surface brightness of the source causing it, is more problematic. A crosstalk pattern may therefore have the same appearance as a faint diffuse source. However, as the crosstalk appears always in the same pixels in both CCDs, it is possible to identify that pattern by looking for a bright source with a similar shape within a distance of 7.5 arcmin. Even if a faint source appears in all three bands with a similar shape, it does not automatically exclude the possibility of being a crosstalk ghost. Indeed, in the vicinity of bright point sources the shapes and locations of the LSB sources always need to be compared with possible crosstalk sources.

\indent Earlier works, such as that by \citet{Duc2015} have shown that Milky Way's dust (\textit{cirrus}) can be easily confused with LSB galaxies or streams. However, \citet{Iodice2016} showed that neither cirrus nor zodiacal light can cause contamination to the images in the Fornax cluster.

\subsection{Producing the object catalog}

\begin{figure*}[t!]
	\centering
	\includegraphics[width=17cm]{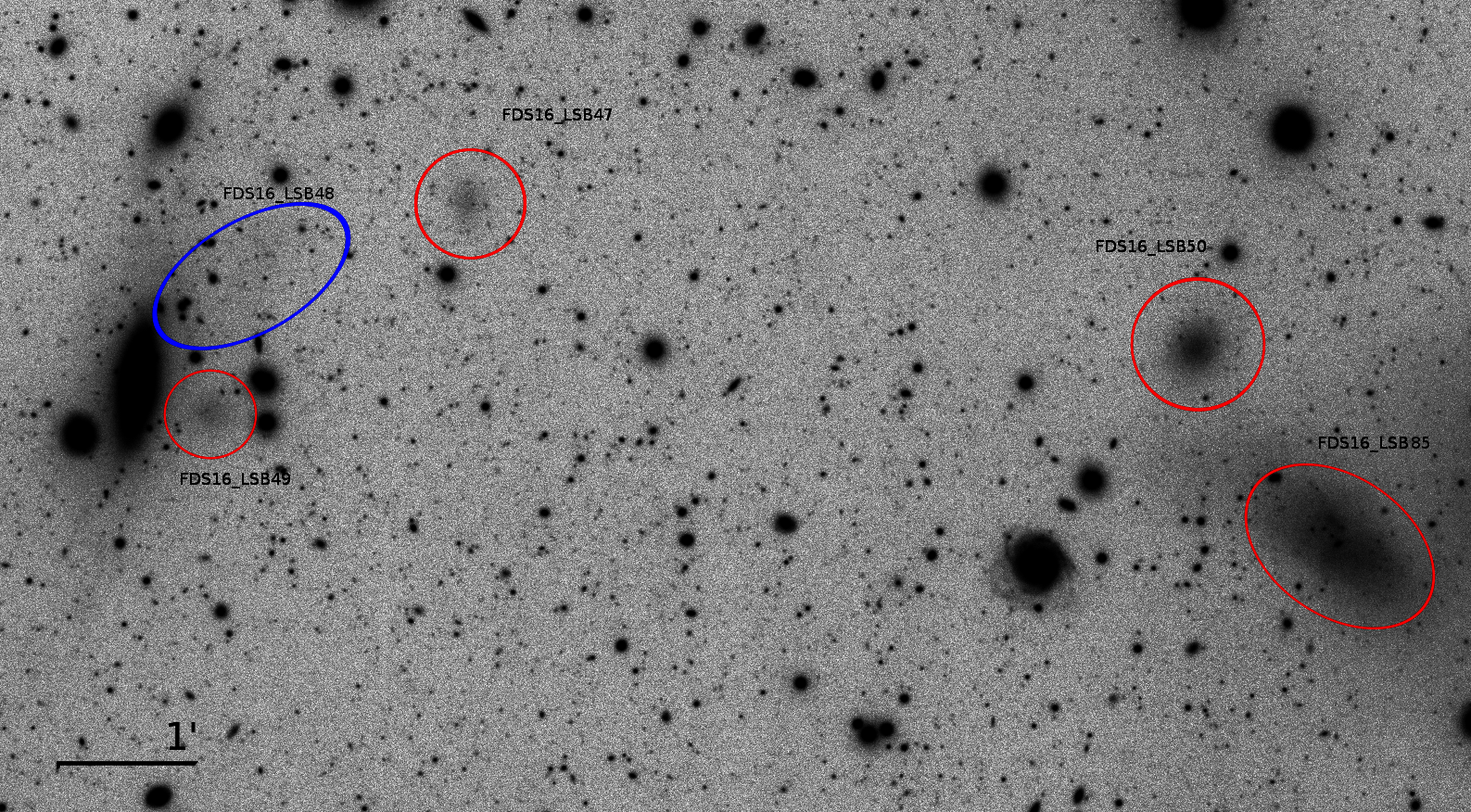}
	\caption{An r'-band cut from the field 16. In the image North is up and East is left. The red ellipses mark the objects which were classified as galaxies. The blue ellipse shows an object which was identified as a tidal feature. In the case  of FDS16\_LSB48, the object is not classified as a galaxy as it is asymmetric, has no clear center, and is connected to the nearby galaxy at its South-East side. The galaxies FDS16\_LSB50 and FDS16\_LSB85 appearing on the right are already identified in \citet{Bothun1991} as F2L7 and F2L8. Also several brighter galaxies appear in the image, but they do not match the surface brightness criteria of our sample. The image is shown in the same logarithmic scale used for the identification i.e. with lower and upper limits of $-1 \times 10^{-12}$ ADU pix$^{-1}$ (or "-26.5" mag arcsec$^{-2}$) and $4 \times 10^{-11}$ ADU pix$^{-1}$ (or 22.5 mag arcsec$^{-2}$), respectively.}
	\label{fig:lsbidentification}
\end{figure*}

The four fields used in this study are inspected visually to detect the sources (see Fig. \ref{fig:lsbidentification}). They are first identified from the 4 $\times$ 4 re-binned images and then from the unbinned r'-band images. Using both rebinned and unbinned images helps detecting LSB galaxies and structures in different intensity scales. As the objects are identified and selected visually, some of the smallest objects in our sample do not fulfil the criterion 2 after taking into account the effects of the {\it PSF}. However, as they are yet LSB galaxies, they are included in the analysis. We used the field definitions and running numbers to name the objects: for example, the sources found within the field 11 are called as FDS11\_LSBn, where n is replaced with a running number.

\begin{figure*}[!ht]
	\centering
	\resizebox{\hsize}{!}{\includegraphics[width=17cm]{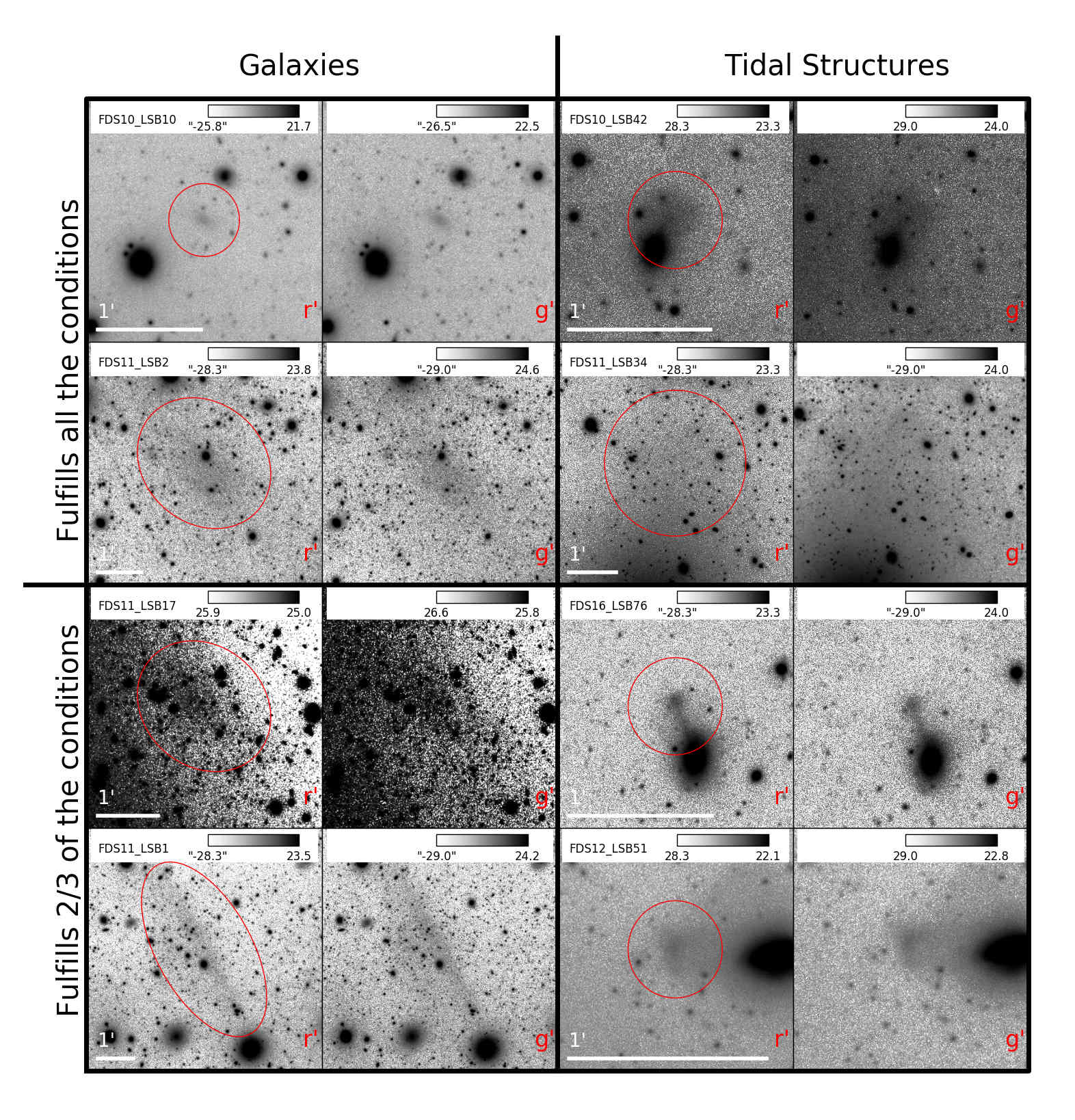}}
	\caption{The two left columns show g'- and r'-band cut-outs of objects classified as \textit{galaxies}. Correspondingly the two rightmost panels show examples of \textit{tidal structures}. The two top rows show typical examples of both classes, which fulfill all three classification criteria of their classes. The two lower rows show examples of objects which only fulfill two of the three criteria of their classes. Specifically: FDS11\_LSB17 has a center with excess light and symmetrical shape, but overlaps with the outskirts of NGC1399. FDS11\_LSB1 has a very elongated shape, but still has radially fading light profile away from its center, and it is not connected to any nearby galaxies.   FDS12\_LSB51 has an area with excess light which could be called a center, but its irregular shape and strong connection to the larger galaxy next to it makes it a \textit{tidal structure}. The criteria for classifying FDS16\_LSB76 as a \emph{tidal structure} are the same as for FDS12\_LSB51.}
	\label{fig:classification_examples}		
\end{figure*}

\indent After detection, the coordinates of the object are stored and postage stamp images (see Fig. \ref{fig:classification_examples} for an example) are cut in all three bands. The size of the image is adjusted manually to be at least 8 times the isophotal (d$_{27}$) diameter of the object so that a sufficient number of background pixels are included, but as little as possible light is coming from the other sources. The maximum size was limited to 500 $\times$ 500 pixels (corresponding to 1.7 arcmin $\times$ 1.7 arcmin) to make the data easily editable. The postage stamps which were larger than the maximum size were rebinned to fit the limits. 

\indent After checking the original mosaic for possible crosstalk or false detections, masks are generated using SExtractor \citep{Bertin2010}. All the sources are masked, and a 2D-plane with 3 degrees of freedom (intensity level and gradients along the x- and y-axes) is least square fit to the masked image and subtracted. This gives the first order approximation for the background level, and helps in judging whether a faint structure is real or not. {We emphasize here that this model is not the final background model used for photometry, but only for editing the masks and identifying the object.} If necessary, the masks are then modified, as in some cases the SExtractor masks were partly masking also the  galaxy, or were not masking the PSF wings properly.

\indent A total of 251 diffuse sources are initially identified from the four inspected fields. From this sample 17 objects are considered as imaging artifacts or they are too faint to be analyzed properly, leaving a sample of 234 objects. As discussed in the next section, 205 of these appeared to be galaxies.

\subsection{Distinguishing galaxies from tidal structures}

In this work we are particularly interested in the effects of the environment on the properties of LSB galaxies, and therefore want to exclude structures which most probably are tidal debris of an ongoing interaction. Therefore, the objects are classified either as \textit{galaxies} or \textit{tidal structures} based on their appearance. At this stage this distinction is qualitative and has an intrinsic uncertainty that is impossible to remove without spectroscopic data. We are aware of the possibility that some of the galaxies might have been born as a result of stripped tidal debris \citep{Bournaud2010}, a possibility which is not ruled out in our approach.

\indent  In our classification \textit{galaxies} can be identified as separate objects from their surroundings. They are structures that have distinguishable centers with excess light, and/or have apparent symmetry. Nevertheless, an irregular structure which is not connected to any other nearby sources will still be classified as a galaxy. \textit{Tidal structures} have elongated or irregular shapes. They are typically connected to a pair or a group of galaxies that have disturbed appearance. They do not have a center with excess light. A structure that has a connecting bridge with another object is a galaxy if it has a clear center, but is otherwise classified as a tidal structure.

\indent The classification criteria for both classes are listed in Table \ref{tab:classification}. As the objects appear in a variety of shapes it is not always immediately clear in which group an object should be classified. Therefore, a peculiar object is classified to the class where it fulfills at least 2 of the criteria. Examples of objects belonging to the two classes are shown in Fig. \ref{fig:classification_examples}. From the total sample of 234 objects 205 are classified as galaxies and 29 as tidal structures. We compared these galaxies with the catalogs of \citet{Munoz2015}, \citet{Mieske2007}, \citet{Bothun1991}, and \citet{Ferguson1989}. 59 of our galaxies are not included in any of those catalogs and are therefore new identifications.

\begin{table}
\caption{Definitions of the two classification classes used. The objects that fulfill at least two of the criteria in a class, are classified as a members of that class.}
\centering
\begin{tabular}{ccc}
\hline\hline
& "Galaxy" & "Tidal structure" \\
\hline\hline
Center with excess light & yes & no \\
Connected to other objects & no & yes \\
Symmetric & yes & no \\
\hline\hline
\end{tabular}
\label{tab:classification}
\end{table}

\section{Structural parameters and photometry}
We use the calibrated r'-band postage stamp images to measure the effective radius ($\mathrm{R}_{e}$), the apparent magnitude ($\mathrm{m}_{\mathrm{r'}}$), and the minor-to-major axis ratio ($b/a$), for all identified galaxies. Two different methods  are used. In the first method we produce the radial light profile for each galaxy using azimuthally averaged bins, and then fit the 1D-profile with a single S\'ersic function. In the second method, we use GALFIT 3.0 \citep{Peng2010,Peng2002} for fitting the 2D flux distribution of the galaxy, again using a single S\'ersic function, and if needed an additional \textit{PSF} component is fitted to the nucleus.

\subsection{Azimuthally averaged surface brightness profiles}
Radial profiles can be defined by fitting the isophotes with a series of ellipses leaving the center, axis ratio, and position angle as free parameters. However, since the $S/N$ of the galaxies in our sample is low, the shape of the ellipses in this approach becomes unreliable in the galaxy outskirts. To prevent this from affecting the measured profiles, we decided to use fixed ellipticity for the elliptical annuli based on pixel value distribution moments. Also, as the galaxies we study typically do not have well defined central peaks, a special approach is needed for defining the centers.  

\indent The center of the galaxy and its shape is defined using pixel distribution moments. A detailed description of the method can be found in Appendix A. In the measurement of the profiles, we use a radial bin width of 4 pixels, or 0.8 arcsec (for comparison the \textit{FWHM} of the \textit{PSF} is $\approx$ 1.1 arcsec, see Fig. \ref{fig:ocampsf} below). The adopted bin width ensures that the bins are small enough to be able to capture the changes in the radial shape of the profile.  The radial bins extend to the distance where the intensity level within the bin drops to 1/3 of the pixel-to-pixel background {\it RMS}, which typically corresponds to $\mu_{r'} \sim$ 28 mag arcsec$^{-2}$. All the masked pixels are rejected, and three times $\sigma$-clipped averages are used as the bin values. For the error of the bin we adopt the standard deviation of the non masked pixels within the bin, divided by the square root of number of the pixels.

\indent At this point the images may still include some positive or negative residual sky, since the initial sky fit was done using SExtractor masks which often fail to cover the faint outskirts of the sources. For the final sky level we use the value measured at the radius of 4 R$_e$. The value of R$_e$ is obtained from the cumulative light profile, and the sky level is measured from an 8 pixel wide galactocentric annulus, placed at r = 4 R$_e$. The annulus is divided into 20 azimuthal sectors, and a median of each sector is taken. Finally we use 4 times $\sigma$-clipped average of the medians as the residual sky value, which is then subtracted from each bin.

\indent The sky corrected azimuthally averaged radial profiles are fit with a single S\'ersic function using intensity units:  

\begin{equation}
\mathrm{I} ( \mathrm{r} ) = \mathrm{I}_{e} \exp \left[ -\mathrm{b}_{n} \left ( \frac{ \mathrm{r} }{ \mathrm{R}_{e} } \right )^{\frac{1}{ \mathrm{n} } } -1 \right ],
\end{equation}
where $\mathrm{R}_{e}$ is the effective radius, $\mathrm{I}_{e}$ is the surface brightness at $\mathrm{R}_{e}$, and $n$ defines how peaked the S\'ersic profile is. The parameter $\mathrm{b}_{n}$ depends on $n$ as $\Gamma (2\mathrm{n}) = 2\gamma (2\mathrm{n},\mathrm{b}_{n})$, where $\Gamma$ and $\gamma$ are the complete and incomplete gamma functions, respectively (\citealp{Sersic1968}, \citealp{Ciotti1991}).  While least-square
fitting the S\'ersic function, the radial bins are weighted with their inverse variances. The values for $\mathrm{R}_{e}$, $n$ and the total apparent magnitude $\mathrm{m}_{r'}$ are obtained from the 1D-S\'ersic fit. Those values are also used as the initial values of the S\'ersic profile in the GALFIT models.

\subsection{GALFIT models}

 GALFIT has been successfully used in several works to model the 2D light-distributions of bright galaxies (see e.g. \citealp{Peng2010}, \citealp{Salo2015} and \citealp{Hoyos2011}) and also those of faint galaxies \citep{Janz2012,Janz2014}. However, \citealp{Munoz2015} claim that for the low surface brightness galaxies not all of their fits using GALFIT converged. Nevertheless, in this study we have successfully fitted all our sample galaxies with GALFIT, to obtain $\mathrm{R}_e$, $\mathrm{m}_r'$, $n$ and $b/a$. Most likely our success stems from using good initial parameters in GALFIT obtained from the 1D fits.

\indent GALFIT is capable of fitting several components simultaneously, taking into account the effects of the \textit{PSF} and proper weighting of the data. The weights created during the data reduction were used for obtaining the $\sigma$-images needed for the GALFIT fits. The pixel value $\sigma_{i,j}$ for the $\sigma$- image is calculated as follows:
\begin{equation}
\sigma_{i,j}^{2} = \frac{1}{W_{i,j}},
\end{equation}
where $W_{i,j}$ corresponds to the value of the weight image in pixel coordinate $i,j$.

\begin{figure}[!ht]
    \centering
        \resizebox{\hsize}{!}{\includegraphics{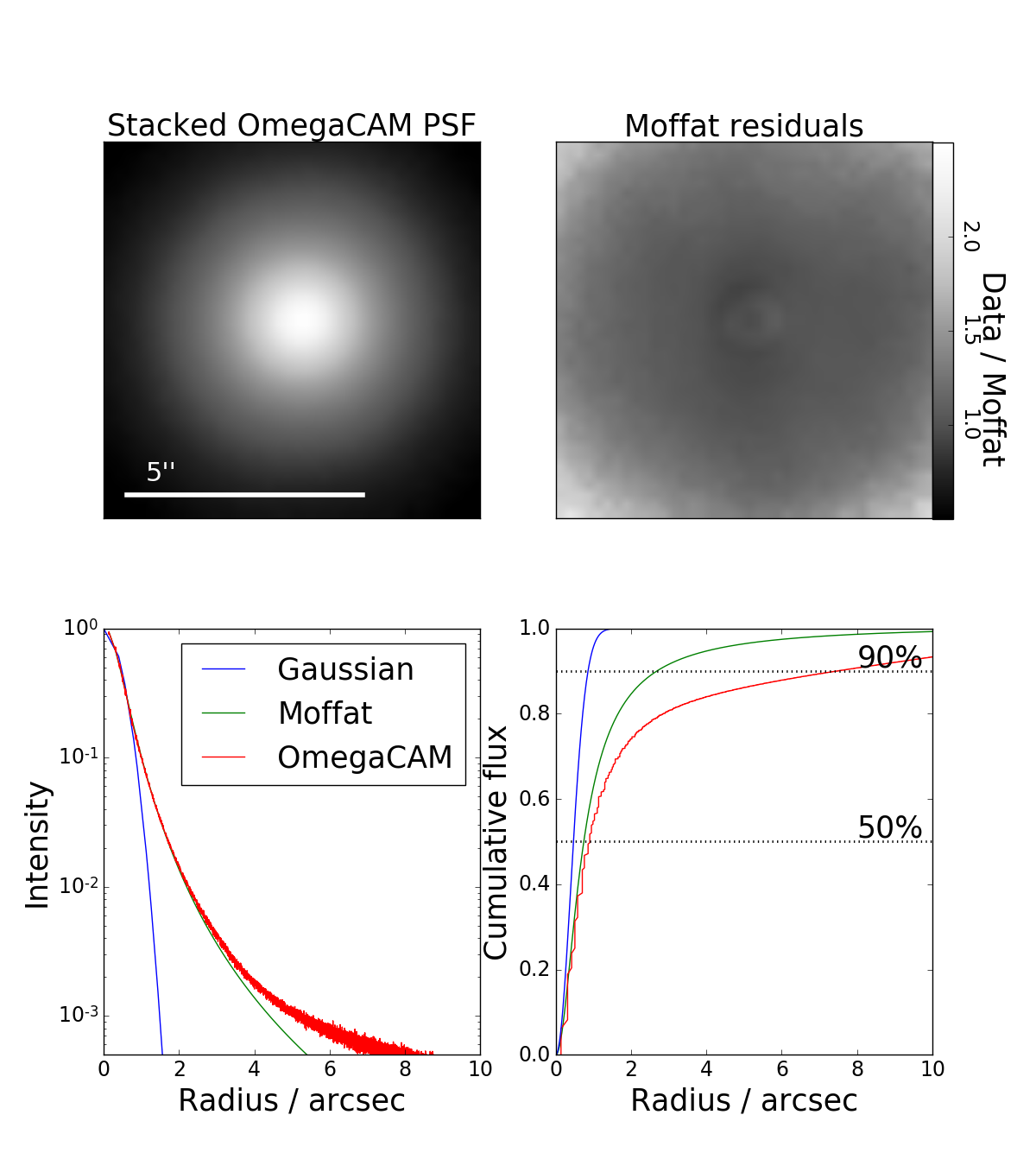}}
    \caption{The upper panels show the PSF images, averaged from the field 11 in the r'-band observations (\textit{upper left}) and the stack model divided with the best fit Moffat model (\textit{upper right}). \textit{Lower left:} the stacked radial intensity profile of the averaged PSF (\textit{OmegaCAM}) is compared with Gaussian and Moffat fits to that profile. All the models are normalized by their central intensity. \textit{Lower right:} cumulative profiles of the average observed PSF, and the Gaussian and Moffat fits. The horizontal dotted lines show the levels where 90\%  and 50\%  of the light are included.}
    \label{fig:ocampsf}
\end{figure}

\indent While fitting with GALFIT it is possible to convolve the models with a given \textit{PSF-}image. In order to take into account the variations of the \textit{PSF}, we made a separate model for each field. We used SExtractor to detect and select the stars to be used for the model. SExtractor has a parameter \emph{CLASS\_STAR}, which indicates the probability of an object to be a star (see \citealp{Bertin1996} for details).  We selected the objects with \emph{CLASS\_STAR} $>$ 0.8. We also excluded saturated and faint stars from the stack by using SExtractor's automatic aperture photometry \emph{MAG\_AUTO}. As the stars brighter than 15 mag are typically saturated, we included only the stars with 16 mag $<$ \emph{MAG\_AUTO} $<$ 19 mag. A pixel area 101 $\times$ 101 pixels was cut around each star, and normalized by the total flux within it. All the cut images were then median combined to get an average \textit{PSF} model. As the background has been subtracted in the images already in the reduction, we do not apply further background subtraction at this point. Any possible contamination from the nearby objects in the {\it PSF} - stack is averaged out when the stack is median combined. Fig. \ref{fig:ocampsf} shows an example of the stacked model (for field 11) as well the best fit Gaussian and Moffat \citep{Moffat1969} models. We decided to use the stacked model, as particularly the Gaussian function fail to fit the faint outermost parts of the \textit{PSF}. We did not take into account the PSF {\it FWHM} variations within the 1$^{\circ}\,\times\,1^{\circ}$ fields, as we found them to be less than 10\%  (consistent with \citealp{deJong2015}).

\indent Initially, we fitted all the images using a single S\'ersic function as a model for the galaxy, and a plane with three degrees of freedom (intensity level and gradients along the x- and y- axes) as the sky component. The parameters for $\mathrm{R}_e$, $\mathrm{m}_{r'}$, $n$, $b/a$ and the center coordinates obtained with the 1D fitting method (explained in Section 5.1), were used as input parameters in GALFIT. All the parameters except for the galaxy center, were fit as free parameters.

\indent In some of the fits the center of the galaxy was initially defined incorrectly. This can be seen as asymmetric residuals, so that part of the galaxy has positive residual values and the other part has negative values. In such cases we fixed all the other fit parameters except for the galaxy center, and let GALFIT to refit the center of the S\'ersic profile. After this the center was fixed again and the other S\'ersic parameters were fit again.  This procedure reduced the fit residuals significantly. 

\begin{figure*}[!ht]
    \centering
        \resizebox{\hsize}{!}{\includegraphics[width=17cm]{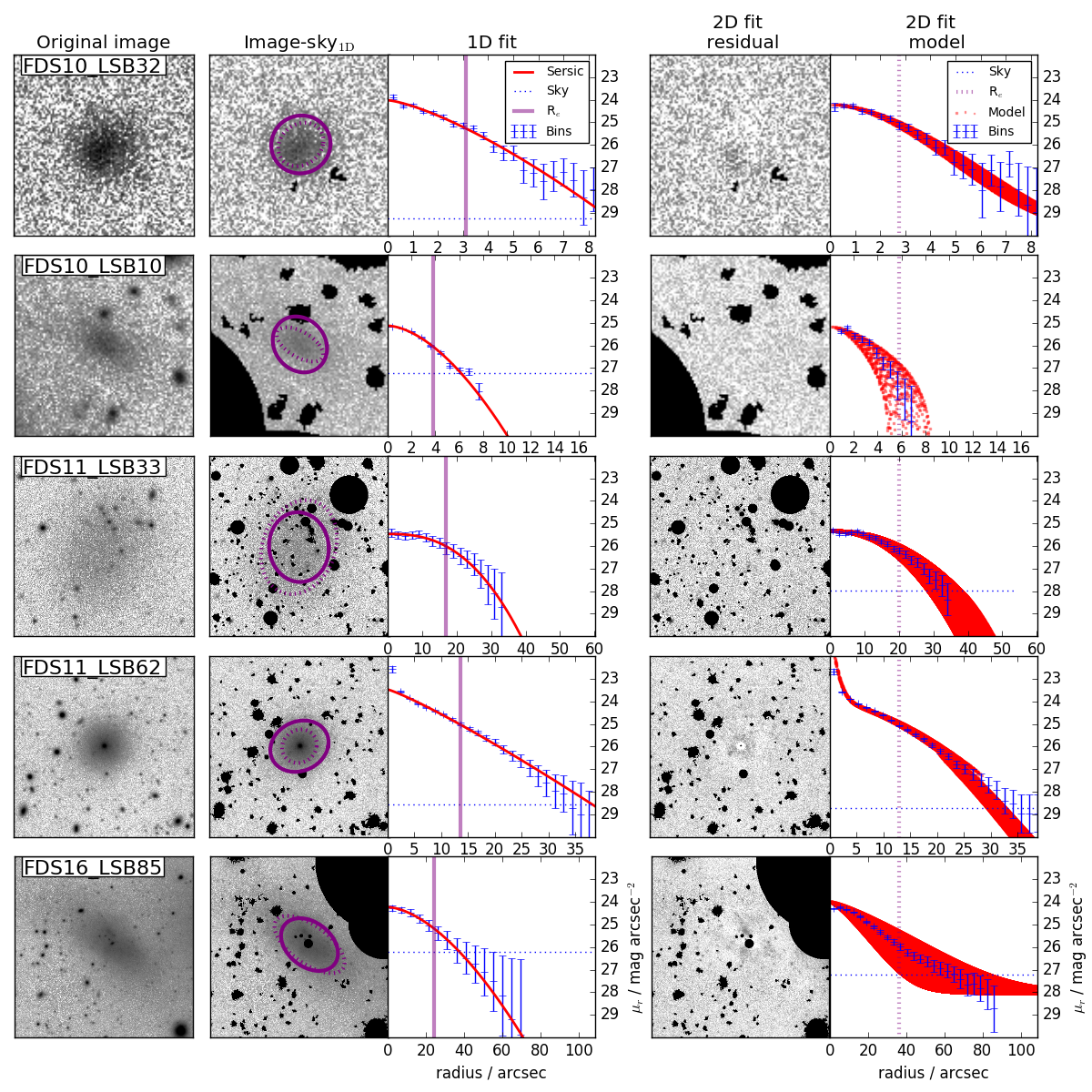}}
    \caption{A comparison of the two photometric methods for individual galaxies. Explanations for the columns from left to right: The {\it first frame} shows the r'-band postage image of the galaxy. The {\it second frame} shows the sky subtracted and masked image, with the elliptical bin shapes at $\mathrm{R}_e$ over-plotted using purple color ({\it solid line:} shape using pixel distribution moments, {\it dotted line:} shape using GALFIT). The {\it third frame } shows the sky-subtracted 1D radial profile derived using the azimuthally averaged bins, with the S\'ersic fit, {\it effective radius (R$_e$) from 1D fit}, and the measured sky level over-plotted. The blue dashed line marks the level of the residual sky measured using the sectors at 4 R$_{e}$. The {\it fourth frame} shows the residual image after subtracting the 2D GALFIT model. The {\it last frame } shows the azimuthally averaged radial profile measured using the elliptical shape from GALFIT. The red dots correspond to each pixel of the (2D) GALFIT model (pixel value as a function of the distance from the center), the purple dotted line shows the corresponding effective radius, and the blue dotted line shows the sky level from the GALFIT decomposition. The distribution moments fail to measure the actual isophotal shape of the second galaxy from the top. This can be seen as a misalignment between the purple solid ellipse and the actual galaxy shape in the second column image. The fourth galaxy from the top possesses a nucleus, which can be seen as a peak in the radial bins (the 3rd and 5th panels from the left). Only the GALFIT model contrives to fit the nucleus.}
    \label{fit_profiles}
\end{figure*}

\indent Some of the galaxies show also a peaked nucleus (see the 4th row in Fig. \ref{fit_profiles}) which is clearly a separate component in the galaxy center. In such cases, we manually place an additional \textit{PSF}-component to fit the nucleus. The {\it PSF}-component is added to improve the overall fit of the galaxy rather than aiming for the detailed analysis of the nuclear star clusters. We fit the {\it PSF}-component using a fixed center position, and leave the total magnitude of the nucleus as a free parameter to fit. 

\indent Using the steps described before, our fits converged for all the galaxies in our sample. In the end, we examined the fits for systematics in the residuals, and corrected the ones where the center of the galaxy  was wrong or the nucleus was missed. To ensure that the sky components are fit correctly, the radial surface brightness profiles and the cumulative radial profiles were inspected. Specifically we checked that GALFIT does not mix the sky component with the S\'ersic profile, which would give unrealistically large effective radii and total magnitudes to the galaxies. The opposite (absorbing galaxy light to sky) is unlikely, since all our images had large areas of background sky not covered by the galaxy.

\subsection{Comparison of the 1D and 2D methods}

\begin{figure}
    \centering
    \resizebox{\hsize}{!}{\includegraphics{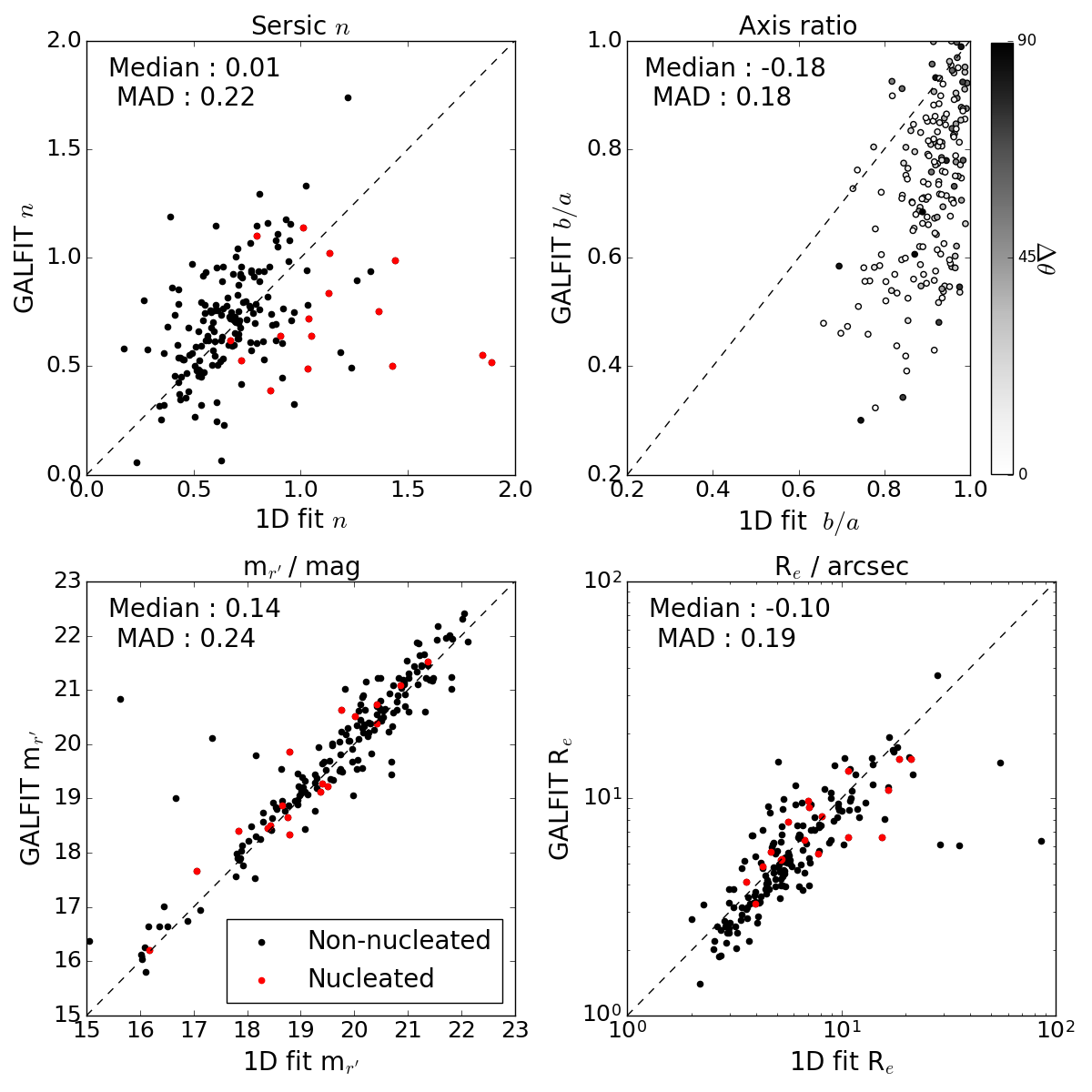}}
    \caption{The different panels show a comparison of the parameters obtained using the different photometric methods explained in Section 5: the x-axes correspond to the 1D fit and y-axes to the 2D GALFIT model. The upper panels compare the S\'ersic $n$-values ({\it upper left}) and axis ratios $b/a$ ({\it upper right}), whereas the lower panels compare the r'-band magnitudes m$_{r'}$ ({\it lower left}) and the effective radii R$_e$({\it lower right}). The differences in the position angles obtained by the two methods are indicated with grey scale colors in the upper right panel; dark colors correspond to a large difference, and light colors to a small difference.  The most obvious difference appears between the b/a-values, which are systematically closer to unity when calculated using the 1D method. The galaxies which have a nucleus are marked in red. The median differences ($n_{2d}-n_{1d}$ , $b/a_{2d}-b/a_{1d}$ , $m_{r',2d}-m_{r',1d}$, 2$\times$(R$_{e,2d}$-R$_{e,1d}$)/(R$_{e,2d}$+R$_{e,1d}$)),  and the corresponding median absolute differences (MAD=$\mathrm{Median} \left( \left| \mathrm{Median}(x_i)-x_i \right|\right)$), where $x_i$ is the difference between 2D- and 1D- measurements) are reported in the upper left corner of the plots.} 
    \label{fig:fitcomparison}
\end{figure}

Fig. \ref{fig:fitcomparison} shows a comparison of the S\'ersic $n$-values, axis ratios ($b/a$), total r'-band magnitudes (m$_{r'}$), and effective radii (R$_{e}$) obtained with the two methods. Both methods give similar distributions for the S\'ersic $n$ values, although the scatter is significant with RMS = 0.47. The effective radii and magnitudes are well in agreement in the two measurements  (the lower panels of \ref{fig:fitcomparison}), except for the four outliers. The outliers are explained due to the different fitting of their background levels in the 1D and 2D methods. Only for the axis ratios $b/a$ a systematic shift appears between the two methods (see upper right panel), so that the values defined using the distribution moments are systematically closer to unity than the ones obtained using GALFIT. By inspecting the residual images, we concluded that the axis ratios obtained by GALFIT resemble more the actual shapes of the galaxies (see e.g. the second row of Fig. \ref{fit_profiles}).

\begin{figure}
    \centering
    \resizebox{\hsize}{!}{\includegraphics{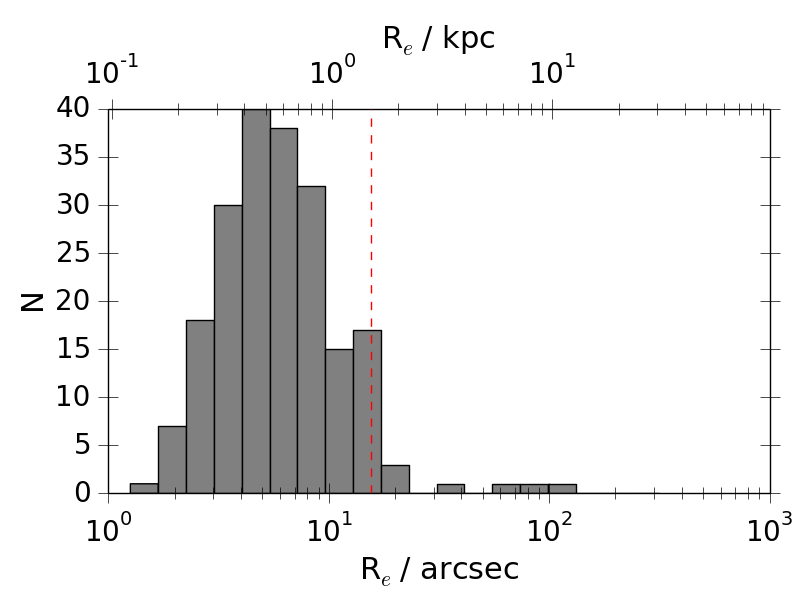}}
    \caption{The distribution of effective radii (R$_e$) obtained from the GALFIT fits. The red dashed line shows the 1.5 kpc limit, which divides the galaxies into UDGs and LSB dwarfs.} 
    \label{fig:reff_hist}
\end{figure}

\begin{figure}
    \centering
    \resizebox{\hsize}{!}{\includegraphics{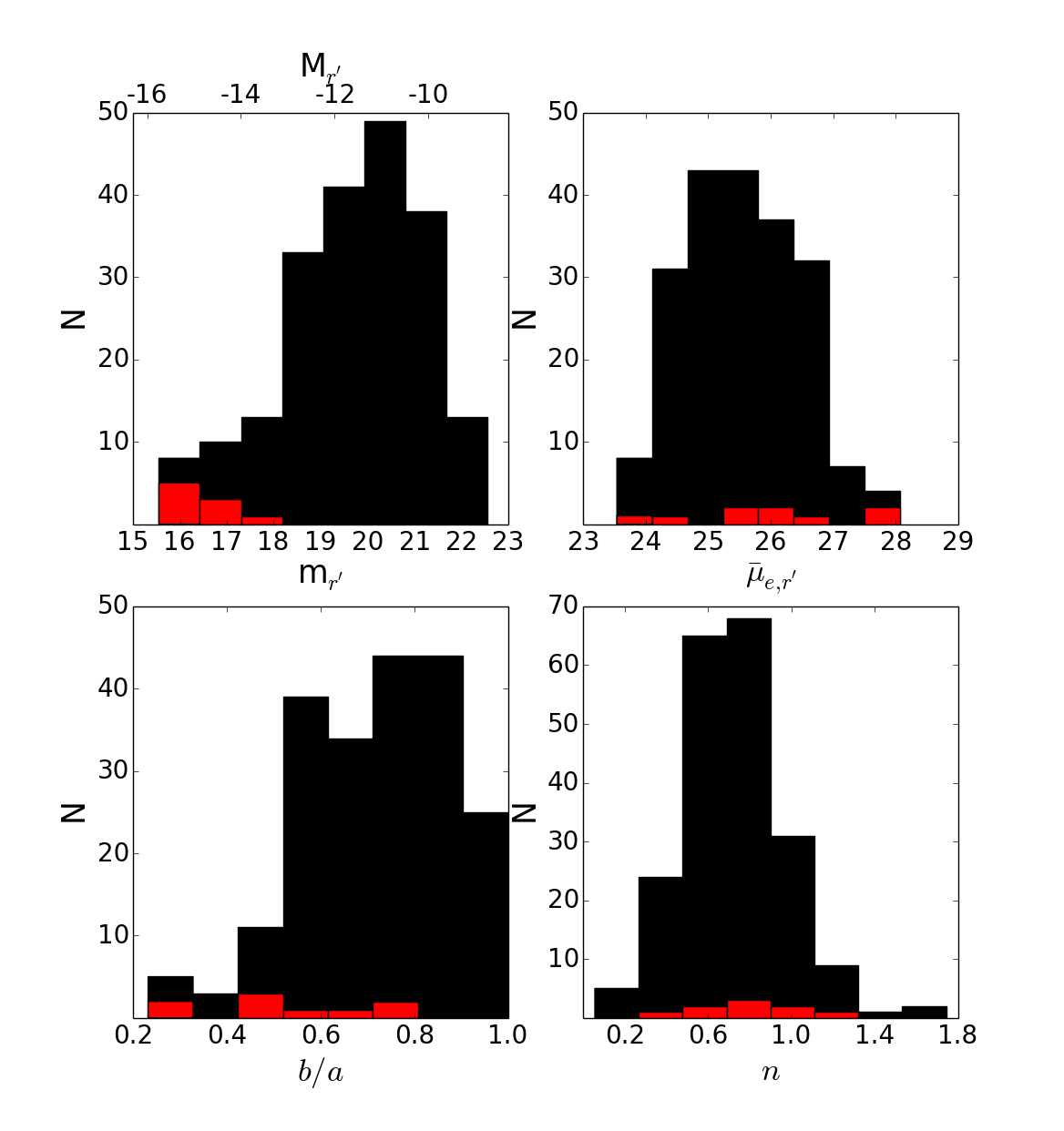}}
    \caption{The histograms of the  apparent magnitudes m$_{r'}$ (\textit{upper left panel}), mean effective surface brightnesses $\bar{\mu}_{e,r'}$ (\textit{upper right panel}), axis ratios $b/a$ (\textit{lower left panel}), and the S\'ersic $n$-values (\textit{lower right panel}) obtained with GALFIT. The black histograms show the distributions for the total sample, and the red histograms those for the UDGs.} 
    \label{fig:hist_dist}
\end{figure}

\indent As expected, the position angles are similar when both methods show small $b/a$ values. The galaxies which show a large difference between the elliptical shape measured with GALFIT and the distribution moments, are often located near to other sources. GALFIT seems to be more stable in the presence of such disturbances. Therefore, for the analysis we decided to use the values measured with GALFIT. The histogram for R$_e$ values used in the analysis is shown in Fig. \ref{fig:reff_hist}, and the histograms for magnitudes (m$_{r'}$), mean effective surface brightness\footnote{Mean effective surface brightness ($\bar{\mu}_{e}$) depends on central surface brightness ($\mu_0$) and S\'ersic $n$ as $\mu_0 = \bar{\mu}_{e}+2.5\times\log_{10}\left(n / b^{2n}\times\Gamma (2\mathrm{n})\right)$, where $b$ is defined as in equation 4. For n=0.5 / 1 / 1.5, central surface brightness is $\mu_0 - \bar{\mu}_{e}$ = 0.3, -1.1, -2.0 mag arcsec$^{-2}$ respectively.} ($\bar{\mu}_{e,r'}$), axis ratios ($b/a$), and S\'ersic $n$-values are shown in Fig. \ref{fig:hist_dist}. 

\indent We conclude that both of these methods can be used for studying LSB galaxies. However, GALFIT is more accurate in obtaining the axis ratios. Also,  as it is capable of correcting for the effect of the PSF, it should be used when the intrinsic shapes of these galaxies are analyzed. When comparing the background estimation of these methods, GALFIT has slight advantage as it uses all the non-masked pixels and allows the background to have gradients. However, our observation that the background level of GALFIT can be somewhat degenerated with the outer parts of the S\'ersic-profile is clearly an issue that should be acknowledged when this method is used.

\subsection{Accuracy of the photometric measurements and completeness of the UDG detections}

\subsubsection{Comparison to \citet{Munoz2015}}

We checked how the completeness of the galaxy sample obtained \\ \\ by us compares with that of \citet{Munoz2015}. Since the two samples have a different spatial coverage, we limited the comparison to field 11 (see Fig. \ref{fig:fieldlocaitons}) which is covered by both studies. There are 96 and 62 galaxies located in that field in the samples of Munoz et al. and this study, respectively. 52 of those galaxies are common to both studies. The sample of Munoz et al. has 44 objects that are not included in our sample, which is {explained by our } selection criteria given in Section 4.1. When these criteria are taken into account, 43 of the 44 galaxies appearing only in the sample of Munoz et al. are excluded\footnote{21 objects are excluded by the criterion 1., 25 objects are excluded by the criterion 2., and 9 by the criterion 4.} from our sample. The remaining object was classified as a tidal structure by us. In that same field, there are 10 objects that are not in the sample of Munoz et al., but appear in our sample.

\indent We also checked how the parameters obtained by us compare with those given in \citet{Munoz2015}, for the 126 sources identified in both studies. In both studies the parameters are obtained using GALFIT. Fig. \ref{fig:photomaccuracy} (\textit{right panels}) shows the differences for the galaxies as a function of the surface brightness, and Fig. \ref{fig:munozcomp} shows a comparison of S\'ersic $n$, R$_e$ and m$_{i'}$ between the two measurements. In order to convert the i'-band magnitudes of \citet{Munoz2015} to r'-band, we used the median r'-i' aperture color\footnote{r'-i' = 0.3 is consistent with the Virgo red sequence between -13 mag > M$_{g'}$ > -16 mag \citep{Roediger2017}, where the values are between 0.2 mag and 0.3 mag.} of $\langle$r'-i'$\rangle$ = 0.3 mag from the FDS data, measured within R$_{e}$. The measured offsets and their standard deviations are $\Delta$m$_{i'}$ = 0.3 with $\sigma$ = 0.3 mag, $\Delta\mathrm{R}_e / \mathrm{R}_e$ = 0.0 with $\sigma$ = 0.2, and $\Delta n$ = 0.0 with $\sigma$ = 0.2. We find that our R$_e$ and $n$ values are well in agreement with those of \citet{Munoz2015}. However, there is a small offset in the total magnitudes, so that the values measured by us are 0.3 magnitudes brighter.

\subsubsection{Tests with mock galaxies}
In order to test the completeness of our faint galaxy detections, and the reliability of the parameters obtained for them, we added mock-galaxies to our images. The detection efficiency is tested for UDGs by embedding mock UDGs with their parameters adopted from \citet{Mihos2015} and \citet{VanDokkum2015} to the science images (see Appendix B for details). Additionally, we tested the photometric accuracy of our measurements for the UDGs and LSB dwarfs using additional $\sim$150 mock galaxies with parameters typical for the galaxies in the sample of \citet{Munoz2015}. The parameter ranges of the used mock galaxies are shown in Appendix B.

\indent The detection efficiency for UDGs is tested by embedding mock UDGs with parameters adopted from 16 representative UDGs from \citet{Mihos2015} and \citet{VanDokkum2015}, to the r'-band science images of all the studied fields one by one. They were distributed randomly across the image and their coordinates were stored. 
After embedding the mock galaxies, the image was inspected and the sources were identified (see Section 4). In this test, we failed to find only one UDG, which should have been detected according to our selection criteria. This UDG was not detected as it was overshadowed by the diffuse light of a nearby star and the outskirts of a galaxy. The results of this test for different fields are listed in Appendix B. The test shows that our data and the visual detection is efficient ($\approx $92\% efficiency) in detecting UDGs such as those presented in \citet{VanDokkum2015} and \citet{Mihos2015}.

\begin{figure}[!ht]
	\centering
		\resizebox{\hsize}{!}
		{\includegraphics{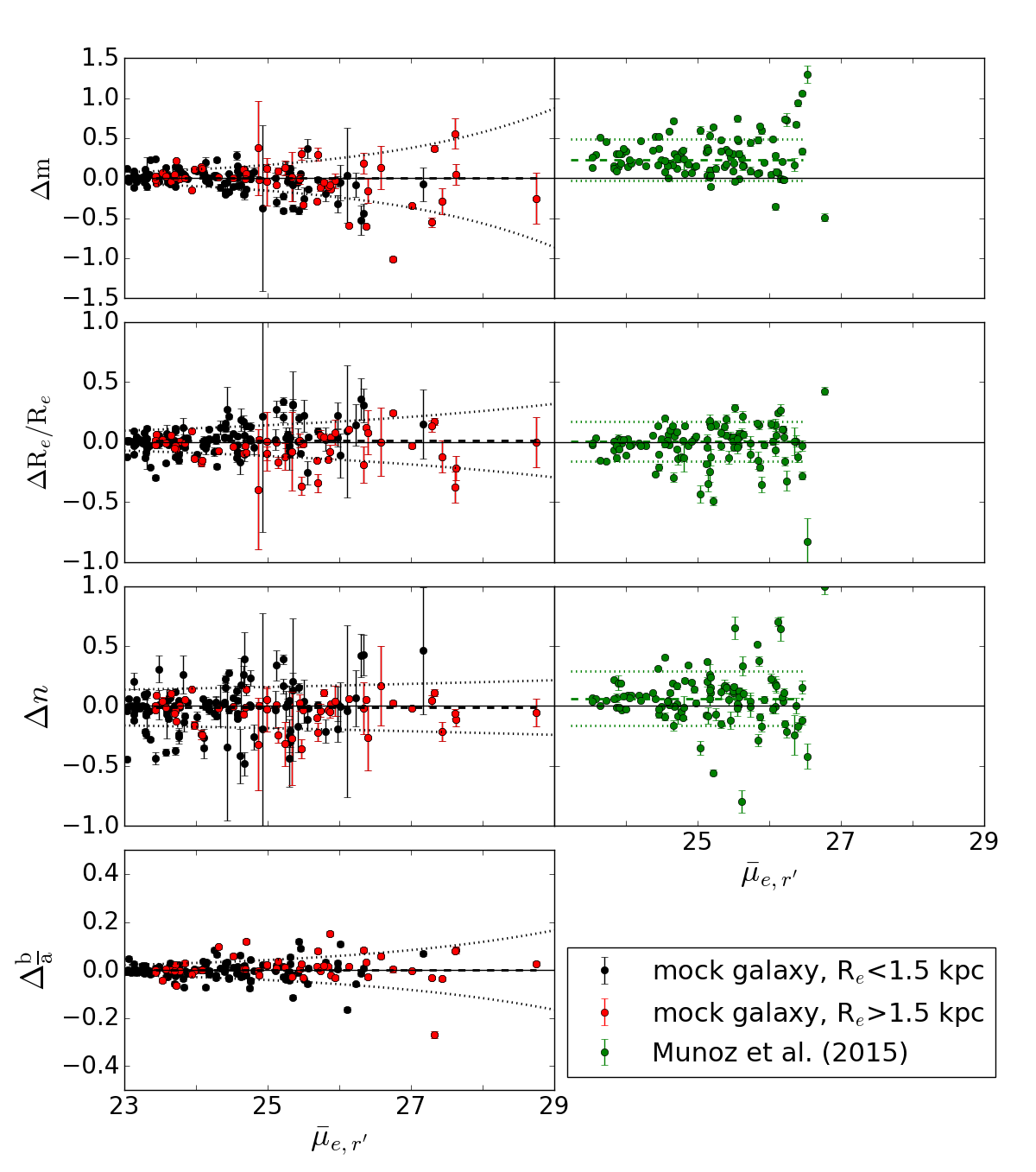}}
\caption{ {\it The left panels} show the difference between the input and output parameters for the mock galaxies. The mock galaxies with effective radii R$_e$>1.5 kpc, and those smaller than that are plotted with red and black dots, respectively. From top to bottom they show the differences (input - output) of apparent magnitudes ($\Delta m$), effective radii ($\Delta\mathrm{R}_{e}$/R$_{e}$), and values of the S\'ersic $n$ ($\Delta n$), as a function of the mean effective surface brightness $\bar{\mu}_e$. {\it The right panels} compare our measurements with the values measured by \citet{Munoz2015} for the galaxies common in the two studies.  The differences in axis ratios ($\Delta \frac{b}{a}$) are only shown for the mock galaxies, since these values are not available for the galaxies of Munoz et al.. Since the magnitudes in \citet{Munoz2015} are in i'-band, we have added the median r'-i'  color of 0.3 mag to their values before the comparison. The errorbars represent the formal errors from our GALFIT fits. The dotted lines in the left panels show the 1$\sigma$-deviations given by the equation 6. In the right panels, the dotted lines show the standard deviations of the differences of the compared  measurements. The dashed lines show the mean differences between the compared values.}  
	\label{fig:photomaccuracy}
\end{figure}

\begin{figure}[!h]
    \centering
        \resizebox{\hsize}{!}{\includegraphics{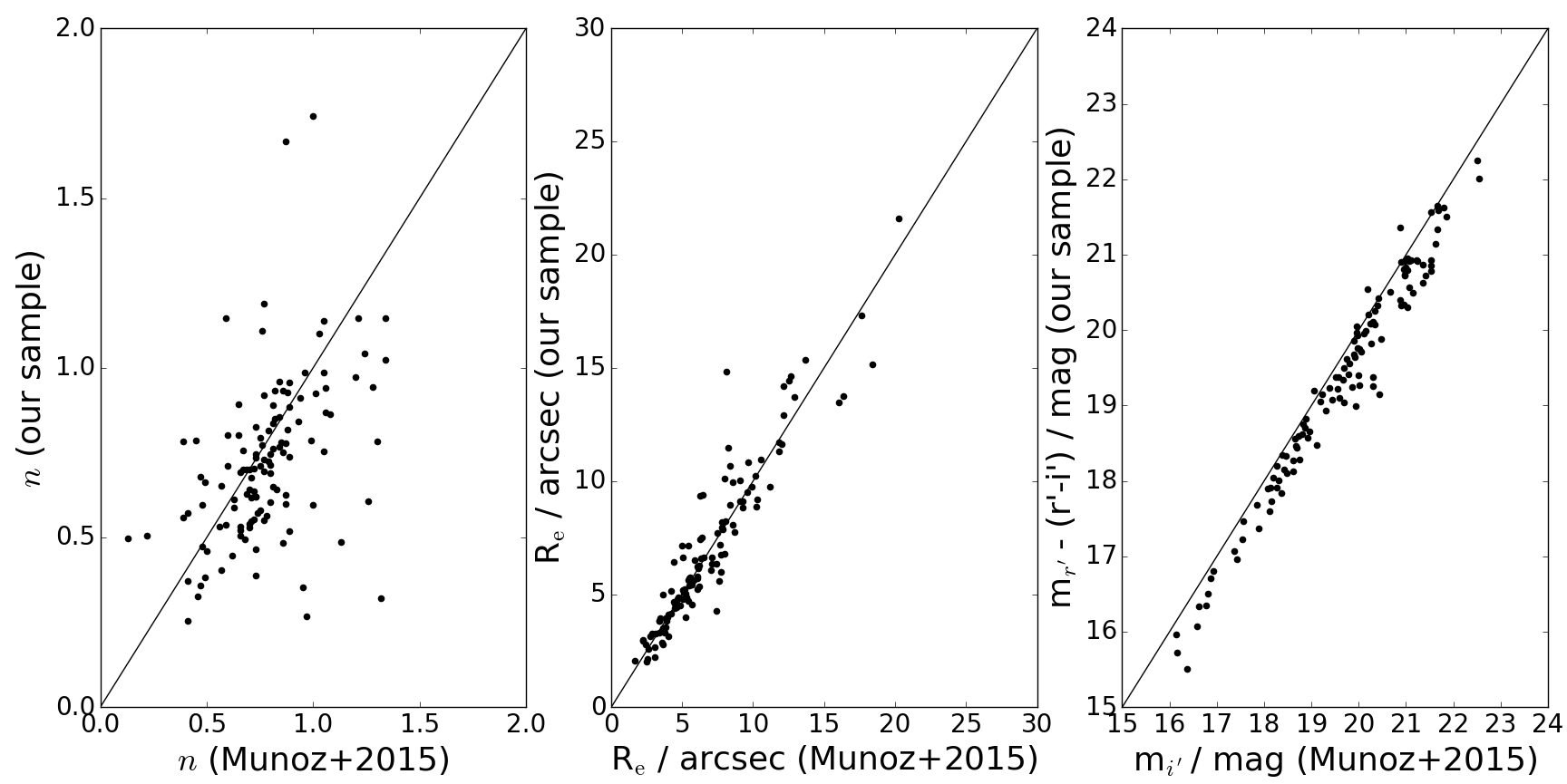}}
\caption{A comparison of the effective radii R$_e$, apparent i'-band magnitudes m$_i'$, and the values of S\'ersic $n$ of the LSB galaxies, as obtained by \citet{Munoz2015} and by us for the same galaxies. The median r'-i' color of the objects was used to transform our r'-band magnitudes to i'-band magnitudes. The black diagonal line shows the 1/1 ratio. The plots show that even though the scatter in $n$ (standard deviation: 0.24) is fairly large, R$_e$ and m$_{i'}$ still match very well with each other. In both works, GALFIT fits were used.}
	\label{fig:munozcomp}
\end{figure}

\indent We made GALFIT fits for all the mock galaxies  described in Appendix B. The photometry was perforemd as described in Section 5.2 Fig. \ref{fig:photomaccuracy} collects the differences in $\mathrm{m}_{r'}$, $\mathrm{R}_e$, $b/a$, and S\'ersic $n$, between the original and measured values (red and black dots), for all identified mock galaxies. Any possible systematic shifts are negligible in all studied parameters. The measured offsets are $\Delta$m$_{r'}$ = 0.00 mag, $\Delta\mathrm{R}_e / \mathrm{R}_e$ = 0.01, $\Delta$ $(b/a)$ = 0.00, and $\Delta$ $n$ = -0.01. 

\indent To characterize the typical uncertainties of the fit parameters, we tabulate the 1$\sigma$-deviations in Fig. \ref{fig:photomaccuracy} as a function of the mean effective surface brightness. Similarly as \citet{Hoyos2011}, we fit these intrinsic standard deviations ($\sigma$) , which we assume to follow a Gaussian distribution, with a simple linear function:
\begin{equation}
\log_{10}(\sigma) = \alpha\times\bar{\mu}_{e,r'} + \beta,
\end{equation}
where $\alpha$ and $\beta$ are free parameters. The fit results are listed in Table \ref{tab:fiteq6}, and the error estimates for the individual galaxies are given with their other photometric parameters in the Appendix D.

\begin{table}
\caption{The fit parameters of equation 6, which is used to estimate the 1$\sigma$-deviations of the photometric parameters as a function of the mean effective surface brightness. }
\label{tab:fiteq6}
\centering
\begin{tabular}{lcc}
\hline\hline
Fit parameters:  &  $\alpha$ & $\beta$ \\
\hline\hline
$\sigma_{\mathrm{m}}$ & 0.1829 & -5.3650 \\
$\log_{10}(\sigma_{\mathrm{R}_e})$ & 0.0945 & -3.2528 \\
$\log_{10}(\sigma_{b/a})$ & 0.1494 & -5.1128 \\
$\log_{10}(\sigma_n)$ & 0.0307 & -1.5330 \\ 
\end{tabular}
\end{table}

\section{Locations and orientations of LSBs within the Fornax cluster}

\subsection{Radial number density profile}

\begin{figure}[!h]
    \centering
        \resizebox{\hsize}{!}{\includegraphics{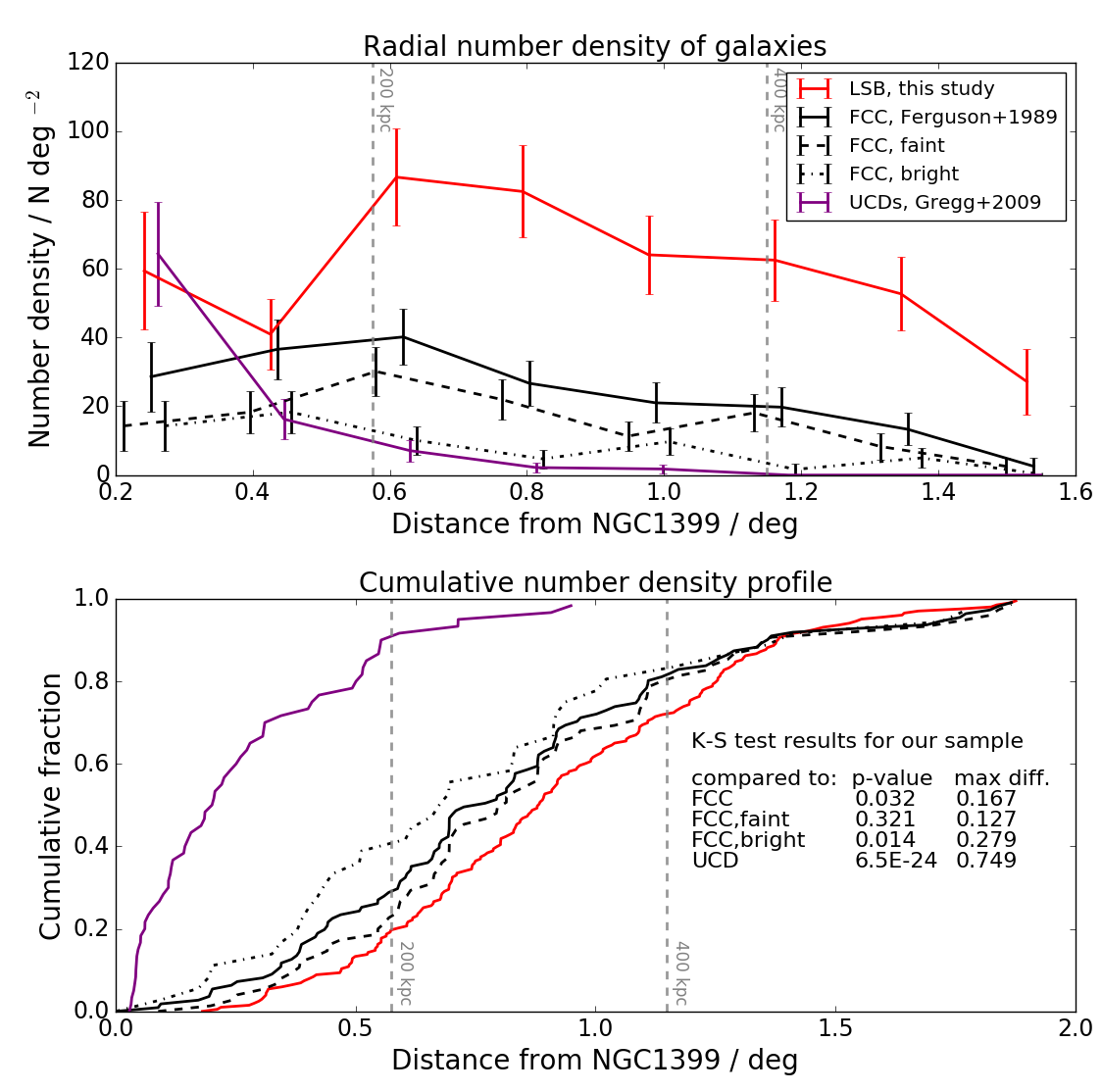}}
    \caption{Radial number density profile of the LSB dwarfs in our sample (red line) is compared with that of all galaxies in FCC (black line). We plot also the Ultra Compact Dwarfs from Gregg et al. (2009; purple line). Plotted separately are also the high surface brightness (with $\bar{\mu}_{e,r'}$ < 23 mag arcsec$^{-2}$) and low surface brightness ($\bar{\mu}_{e,r'}$ > 23 mag arcsec$^{-2}$) galaxies from FCC, indicated by semi-dotted and the dashed lines, respectively. The results of the K-S tests comparing our sample distribution with those of the other surveys are listed in the lower right corner of the lower panel. }
    	\label{fig:numberdensityprof}
\end{figure}

The locations of the galaxies identified by us in the four Fornax cluster fields are plotted over the combined i', r', and g'-band  image in Fig. \ref{fig:locations}. As explained in Section 4.1, we masked all the areas covered by stellar halos or bright extended galaxies. Then a cluster-centric radial surface density profile was made by counting the number of objects in radial bins and dividing those numbers by the non-masked  area within each bin. The number density profile and the corresponding cumulative profile are shown in Fig. \ref{fig:numberdensityprof}. We used NGC 1399 as the center of the cluster, since the hot intra-cluster x-ray gas is centered to it \citep{Paolillo2002}. Also, the smoothed FCC galaxy number density distribution peaks on top of NGC 1399 \citep{Drinkwater2001}. 

\indent We make a comparison between the radial distribution in the cluster of the LSB dwarf galaxies identified by us, and that of the FCC galaxies (see  Fig. 13). For the FCC galaxies we used those classified as "confirmed" or "probable cluster members" in \citet{Ferguson1989}. It appears that the LSB galaxies of this study are less centrally concentrated than the more luminous FCC galaxies. The Kolmogorov-Smirnov (K-S) test gives a p-value of 0.032 for the assumption that the two distributions are from the same underlying distribution, indicating that the difference is statistically significant (p<0.05). In principle there can be a bias in this comparison, because the FCC galaxies are identified also on top of the halos of bright stars and galaxies, which areas were excluded in our study. However, such bias would affect our result only if the  galaxies in the central parts of the cluster were more concentrated to the halos of bright galaxies than to the surrounding fields.

\indent We further divided the FCC into bright galaxies with mean effective surface brightness $\bar{\mu}_{e,r'}$ < 23 mag arcsec$^{-2}$, and to faint galaxies with $\bar{\mu}_{e,r'}$ > 23 mag arcsec$^{-2}$. By comparing the radial distributions of the galaxies in these two bins shows that the bright FCC galaxies are more centrally concentrated than the galaxies in our sample (K-S test p-value = 0.014). When comparing the radial distributions of the faint FCC galaxies with our sample galaxies, the K-S test gives a p-value of 0.32. This p-value means that these two distributions are not statistically different, which is expected as these two samples have several galaxies in common.

\subsection{Orientations}

\begin{figure}[!h]
    \centering
        \resizebox{\hsize}{!}{\includegraphics{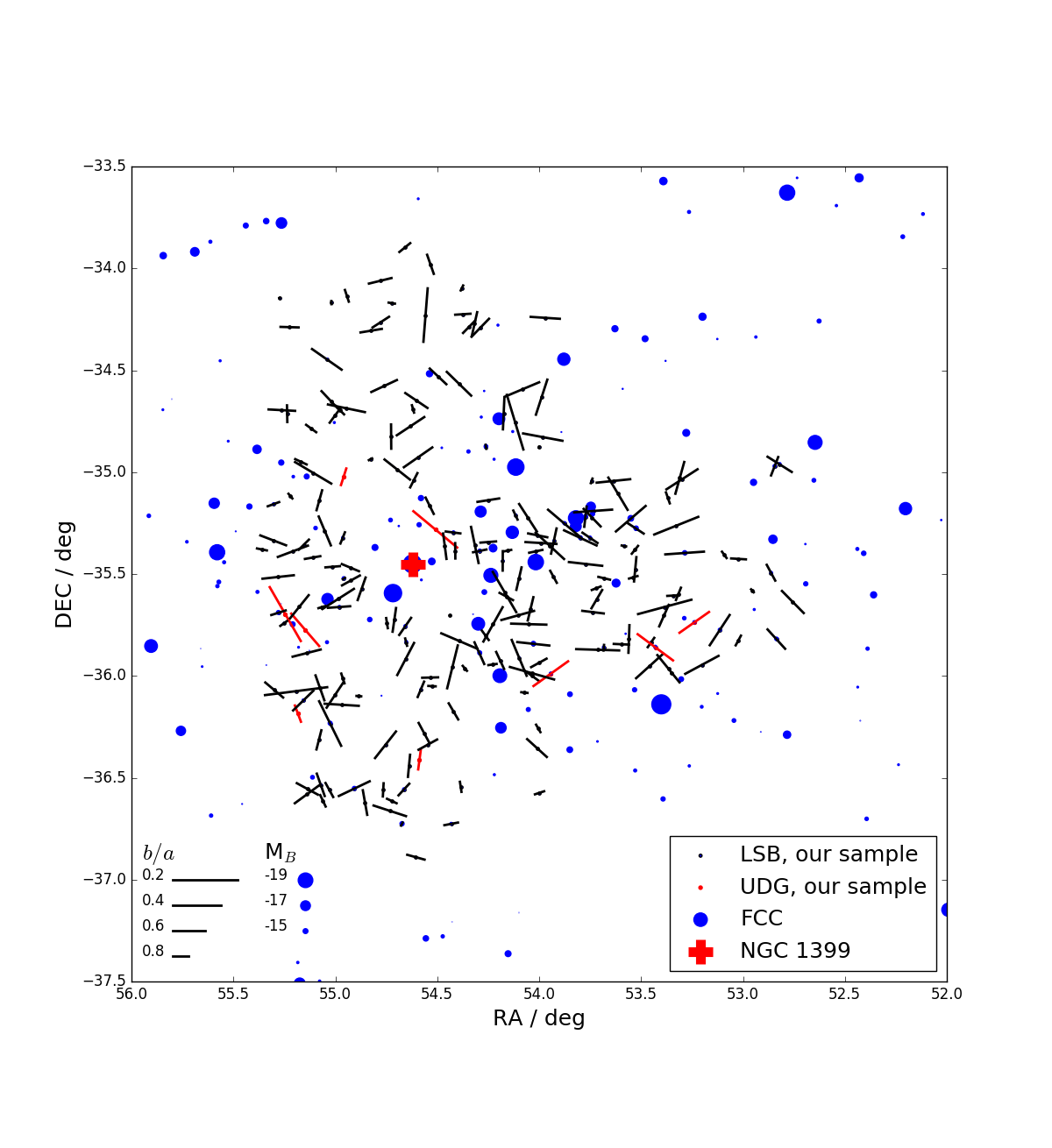}}
    \caption{The orientations of the LSB dwarf galaxies in our sample are shown with the black sticks, and the UDGs with the red sticks. The lengths of the sticks correspond to the actual ellipticities so that the ellipticity increases with increasing length of the stick. The galaxies appearing in FCC \citep{Ferguson1989} are plotted with the blue circles. The size of the circle corresponds to the brightness of the galaxies; the larger the circle is the brighter the galaxy is. The position of the NGC 1399 is marked with the orange cross.}
    	\label{fig:globalorientations}
\end{figure}

The orientations for the individual galaxies in our sample are shown in Fig. \ref{fig:globalorientations}, where over-plotted are also the locations of the FCC galaxies in the same field. The relative orientations with respect to the cluster center and the closest FCC galaxy with $\mathrm{M}_{\mathrm{r'}}$ < -18 mag\footnote{Transformed from B-band, see Appendix C for details.} are plotted in Fig. \ref{fig:orientations}.  However, as the galaxies with $b/a$ $\sim$ 1 may cause additional noise to the orientation plots, thus blurring possible underlying dependencies, only galaxies with $b/a$ < 0.9 are considered. It appears that when including all the galaxies up to $b/a$ = 0.9, there is no statistically significant preferred alignment, neither toward the closest bright galaxies (p-value = 0.657, for retaining the hypothesis that the alignments are random), nor toward NGC 1399 (p-value = 0.060). However, the galaxies with $b/a<$ 0.7 show a weak preferred alignment toward their bright nearby galaxies, for which a K-S test gives a p-value of 0.031. 

\begin{figure}[!h]
    \centering
        \resizebox{\hsize}{!}{\includegraphics{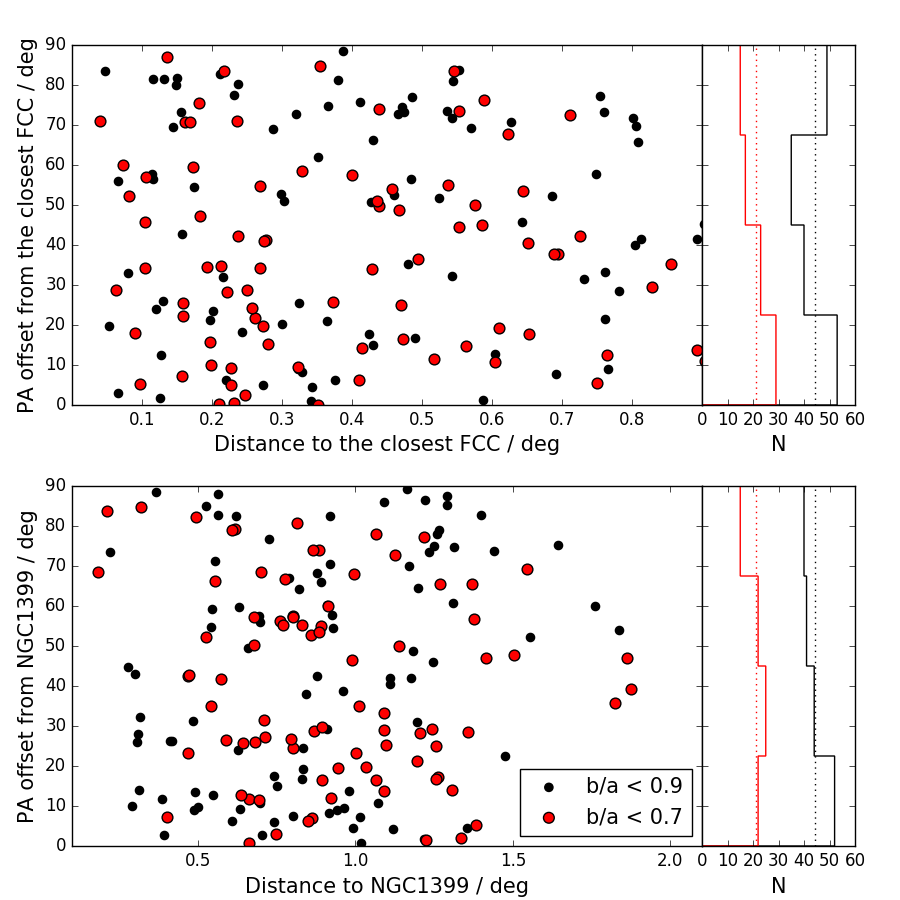}}
    \caption{In the upper plot we show the difference between the position angle of our sample galaxy, and the position angle of the line drawn from the galaxy to the closest FCC galaxy with  $\mathrm{M}_{\mathrm{r'}}<$-18 mag. In the bottom plot we show the difference between the position angle of our sample galaxies and the position angle of the line drawn from that galaxy to NGC 1399. The red and the black dots correspond to the galaxies with $b/a$ < 0.7 and < 0.9, respectively. The vertical histograms show the number distributions of the points in the scatter plots. The red and black histograms correspond to the red and black points, respectively.}
    	\label{fig:orientations}
\end{figure}

\section{Colors of the sample galaxies}

We measured the aperture magnitudes in the g', r' and i'-bands for the galaxies in our sample, using the background (GALFIT) subtracted and masked images. However, in this paper we only analyze the g'-r' colour, since in these bands the data are the deepest, and give accurate colours for most objects. We used elliptical apertures  defined by the parameters obtained from our r'-band GALFIT fits (center coordinates, and $\mathrm{R}_{e,r'}$ used as the major axis of the aperture). Instead of using total magnitudes
aperture colors were obtained. This is to minimize systematic errors from the sky background determination. As in \citet{Capaccioli2015}, we estimate the errors $\sigma_{g'-r'}$ for g'-r' colors as:

\begin{equation}
\begin{split}
\sigma_{g'-r'}^2 & = \sigma_{ZP,g'}^2+\sigma_{ZP,r'}^2 + \\
&\left(\frac{2.5}{I_{g'} \ln 10}\right)^2(\sigma_{I,g'}+\sigma_{sky,g'})^2+\left(\frac{2.5}{I_{r'} \ln 10}\right)^2(\sigma_{I,r'}+\sigma_{sky,r'})^2,
\end{split}
\end{equation} 

where $I_{g'}$ is g'-band mean intensity within the aperture, and $\sigma_{I,g'}$, $\sigma_{sky,g'}$ and $\sigma_{ZP,g'}$ are the errors for the surface brightness, the sky, and the photometric zero point in g'-band, respectively. $I_{r'}$, $\sigma_{I,r'}$, $\sigma_{sky,r'}$ and $\sigma_{ZP,r'}$, are the corresponding quantities in r'-band. For the mean intensity we assumed simple Poissonian behaviour, so that  $\sigma_{I,r'/g'}=\sqrt{I_{g'/r'} / (GAIN \times n)}\times GAIN$, where $n$ is the number of pixels within the aperture. $I$, $\sigma_{I}$ and $\sigma_{sky}$ are given in flux units, whereas $\sigma_{ZP}$ are in magnitudes.

\begin{figure}[!h]
    \centering
        \resizebox{\hsize}{!}{\includegraphics{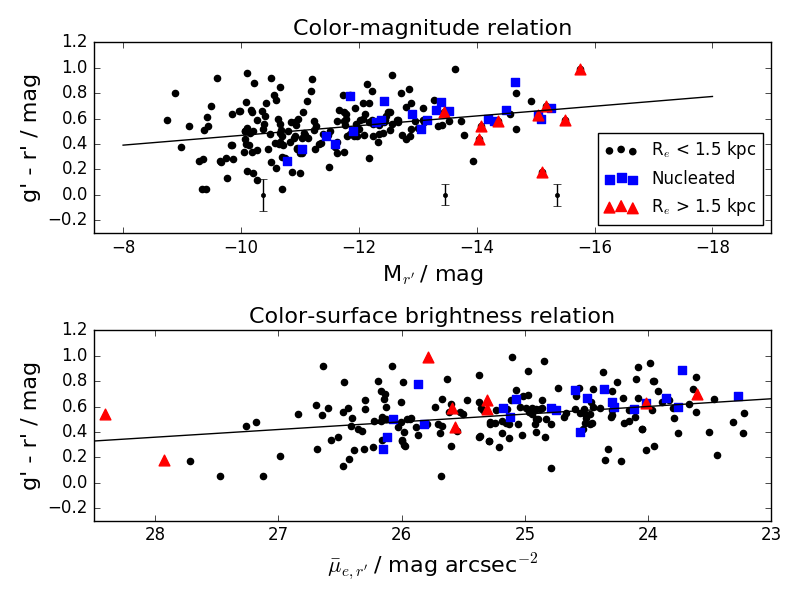}}
    \caption{Upper panel: The color-magnitude relation shown for the galaxies with effective radius R$_e$ < 1.5 kpc for nucleated (\textit{blue squares}) and non-nucleated galaxies ({\it black points}), and for the UDGs with R($_{e}$ > 1.5 kpc, (\textit{red triangles}) in our sample. The g'-r' colors are measured with an elliptical aperture with semi major axis of $\mathrm{R}_{e}$ taken from GALFIT. The errorbars below the points show the size of the errors (Eq. 7) along the color-magnitude relation. Lower panel: The color-surface brightness relation of the galaxies in our sample. The symbols are the same as in the upper plot.}
    \label{cmrelation}
\end{figure}

\indent The g'-r' colors of the galaxies as a function of their total absolute r'-band magnitudes M$_{r'}$ are shown in the upper panel of Fig. \ref{cmrelation}, with the typical errorbars of the colors shown below the points. Plotted separately are the nucleated and non-nucleated LSB dwarfs (with R$_e$ < 1.5 kpc), and the UDGs (with R$_e$ > 1.5 kpc). Pearson's correlation coefficient for the points shows a negative correlation $\rho$ = -0.33 $\pm$ 0.04, and a linear fit gives the relation $g'-r' = -0.04(\pm 0.01)\times (\mathrm{M}_{r'}+12)-0.48(\pm0.09)$. In the color-magnitude relation the UDGs are among the brightest galaxies, but follow the same relation with the dwarf LSB galaxies in our sample. The measured slope of the color-magnitude relation of our sample is the same  as that measured for the Virgo dwarf ellipticals by \citet{Kim2010} $d(g'-r')/d$M$_{r'}$=-0.04$\pm$ 0.01, and for the Virgo red sequence by \citet{Roediger2017} $d(g'-r')/d$M$_{r'}$=0.3--0.4.

\indent The g'-r' colors of the galaxies as a function of the mean effective surface brightness in the r'-band are shown in the lower panel of Fig. \ref{cmrelation}. We find that the colors correlate also with the surface brightness becoming redder with increasing surface brightness. Pearson's correlation coefficient for the points is $\rho$ = -0.26 $\pm$ 0.05, and a linear fit gives the relation $g'-r' = -0.06(\pm0.01)\times(\bar{\mu}_{e,r'}-24)+0.7(\pm0.3)$. A more thorough discussion of the colors of galaxies in the FDS catalog will follow in a future paper.

\section{Discussion}

The motivation of this work is to study the properties of the  LSB galaxies in the Fornax cluster, and how they compare with those observed in other clusters or galaxy groups. We are  particularly interested in the Ultra Diffuse Galaxies (UDGs), using a threshold surface brightness and size typical for the previously identified UDGs in clusters. 

\subsection{Concept of an UDG in the literature}

\indent To conduct a meaningful comparison between the UDGs in the Fornax cluster and in other galaxy environments, it is important to make sure that the objects we are comparing are selected similarly.  The definition of UDGs, adapted from \citet{VanDokkum2015}  for the Coma cluster, was that they are galaxies with R$_{e}$ > 1.5 kpc, and stellar mass of M$_{*,L} \approx 10^7 \mathrm{M}_{\odot}$, or -16.2 mag < $\mathrm{M}_{\mathrm{r'}}$ < -13.2 mag \footnote{Transformed (see Appendix C) from g'-band measurements of van Dokkum et al (2015).}. Works published earlier than that might contain a few similar galaxies, in which case they were simply called as LSB galaxies. Since the largest UDGs found so far have R$_{e} \approx$ 10 kpc \citep{Mihos2015}, and the smallest ones overlap with the typical dE galaxies, it is possible that some of the UDGs form  the low mass tail of the dEs with atypically large effective radii, and some of them form a genuinely distinct population. To study this, in the following we analyze separately the properties of the small UDGs with 1.5 kpc < R$_{e}$ < 3.0 kpc  (i.e., with typical sizes of UDGs in Coma), and large UDGs with R$_e$ > 3 kpc. 

\indent A comprehensive collection of UDG studies in the literature has been presented by \citet{Yagi2016}. However, not all of these works have sufficient image depth and the same measurements given as in this study. Also, most of these works contain very few UDGs. Here we discuss only those works to which we can make comparisons easily, without any auxiliary assumptions about the shapes or colors of these galaxies. The most complete available UDG samples have been made for the Coma cluster by \citet{Koda2015} (included in Yagi's collection), and for galaxy clusters at larger distances by \citet{VanDerBurg2016}. Both of these studies used SExtractor to generate object lists, and GALFIT to fit S\'ersic profiles to the galaxies.

\subsection{Comparison of UDGs in Fornax and in other environments}

\begin{figure*}[!ht]
	\centering
	\resizebox{\hsize}{!}{\includegraphics[width=17cm]{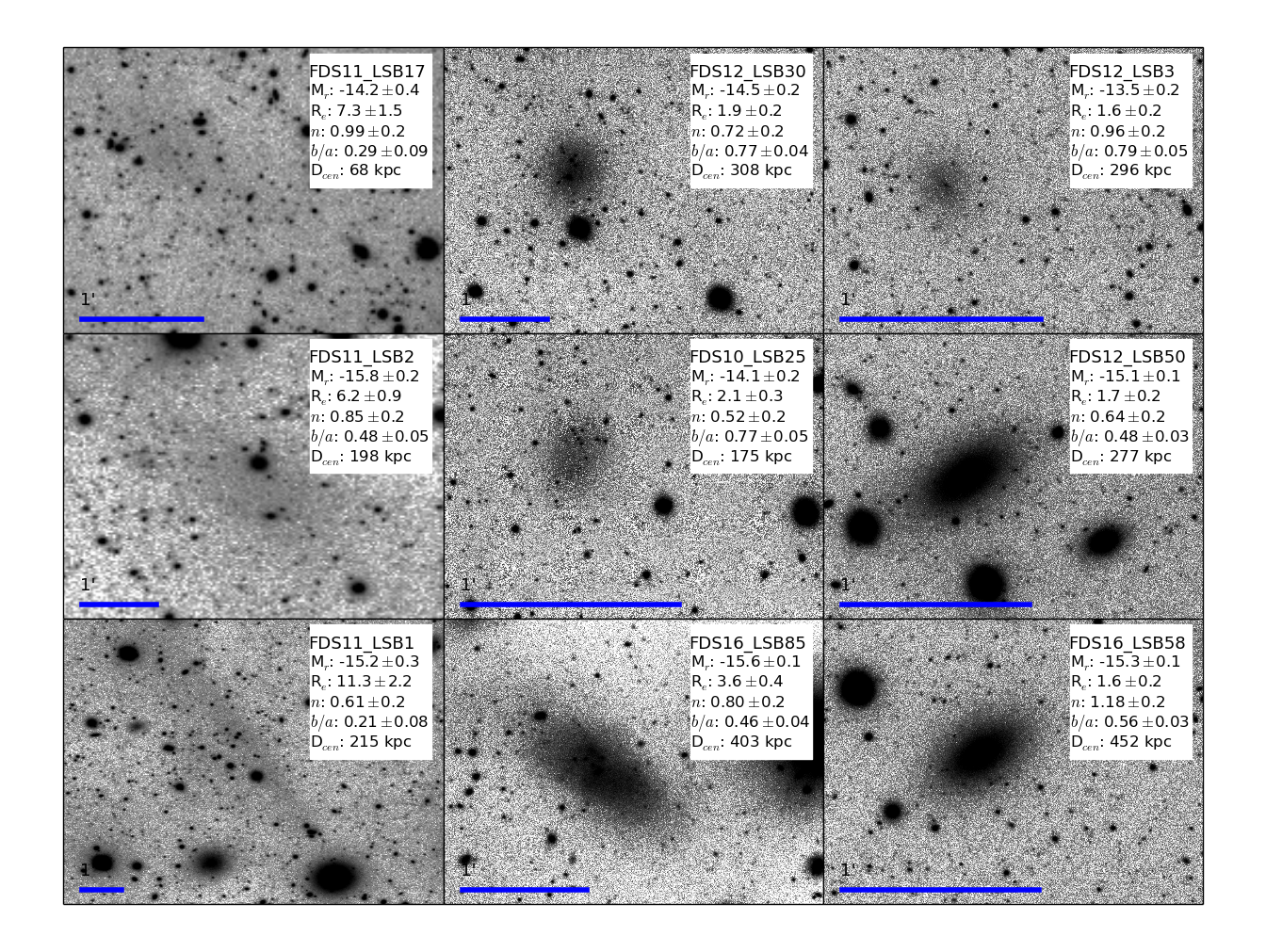}}
	\caption{Postage stamp images of the UDGs of our sample in r'-band, all shown in the same brightness scale (logarithmic scale between "-27.8" mag arcsec$^{-2}$ and 24 mag arcsec$^{-2}$ ). In the right corner shown are the name, the absolute r'-band magnitude M$_{r'}$, the effective radius R$_{e}$ in kilo parsecs, S\'ersic $n$, axis ratio $b/a$ and the projected cluster-centric distance D$_{cen}$. The postage stamps have different sizes on the sky, and therefore 1 arcmin scale bars are shown in the lower left corner of the images. FDS11\_LSB1, FDS11\_LSB2 and FDS12\_LSB3 do not appear in any previously published works. However, FDS11\_LSB1 and FDS11\_LSB2 have also been identified in Lisker et al. ({\it submitted}), although there only FDS11\_LSB2 is considered as a galaxy.}
	\label{fig:udg_cuts}
\end{figure*}

We found 9 UDG candidates in Fornax within the 4 deg$^2$ search area, of which 5 have R$_{e}$ < 3 kpc, and 4 have R$_{e}$ > 3 kpc. Three of these galaxies appear also in the sample of \citet{Munoz2015}, two of them are detected in \citet{Mieske2007}, five appear in the FCC, and four are detected by \citet{Bothun1991}. Additionally, two of the UDGs appear in Lisker et al. ({\it submitted}), but in that work FDS11\_LSB1 is not considered as a galaxy. We identified all the UDGs in \citet{Munoz2015} that were located within the area of our study. We identified also "FDS11\_LSB30", classified as UDG in \citet{Munoz2015}, but by its small size (R$_e$ < 1.5 kpc) it was not classified as UDG. Two of our UDGs are new detections. Fig. \ref{fig:udg_cuts} shows the r'-band stamp images of the detected UDGs, with their structural parameters and cluster centric distances listed in the upper right corner. The structural parameters of these UDGs are given in Table \ref{tab:properties}.

\begin{table}
\caption{Properties of the UDGs in the Fornax (N = 9, our sample) and Coma clusters (N = 288, the galaxies with R$_e$ > 1.5 kpc from the sample of \citealp{Yagi2016}). The columns show the median, standard deviation ($\sigma$), minimum (Min.), and maximum (Max.) values of a given quantity, repectively. }
\label{tab:properties}
\centering
\begin{tabular}{lcccc}
\hline\hline
Fornax         &  Median & $\sigma$ & Min. & Max. \\
\hline\hline
M$_{r'}$ / mag &  -15.1 &   0.7    & -15.8   & -13.5 \\
R$_{e}$ / kpc  &  2.09  & 3.22     & 1.59    & 11.25 \\
$b/a$          &  0.48  & 0.18     & 0.24    & 0.79  \\
g'-r' / mag    &  0.59  & 0.20     & 0.18    & 0.99  \\
S\'ersic $n$   &  0.80  & 0.22     & 0.40    & 1.18  \\
\hline\hline	
Coma \citep{Yagi2016}& Median & $\sigma$ & Min. & Max. \\
\hline\hline
M$_{r'}$ / mag &-14.8  & 0.9       & -16.8   & -11.8 \\
R$_{e}$ / kpc  & 1.86  & 0.57      & 1.51    & 6.12 \\
$b/a$          & 0.73  & 0.16      & 0.25    & 0.99 \\
g'-r' / mag    & 0.68  & 0.13      & 0.25    & 1.03 \\
S\'ersic $n$   & 0.89  & 0.33      & 0.17    & 2.71 \\
\end{tabular}
\end{table}

\subsubsection{Sizes}

\begin{figure*}[!ht]
	\centering
	\resizebox{\hsize}{!}{\includegraphics[width=17cm]{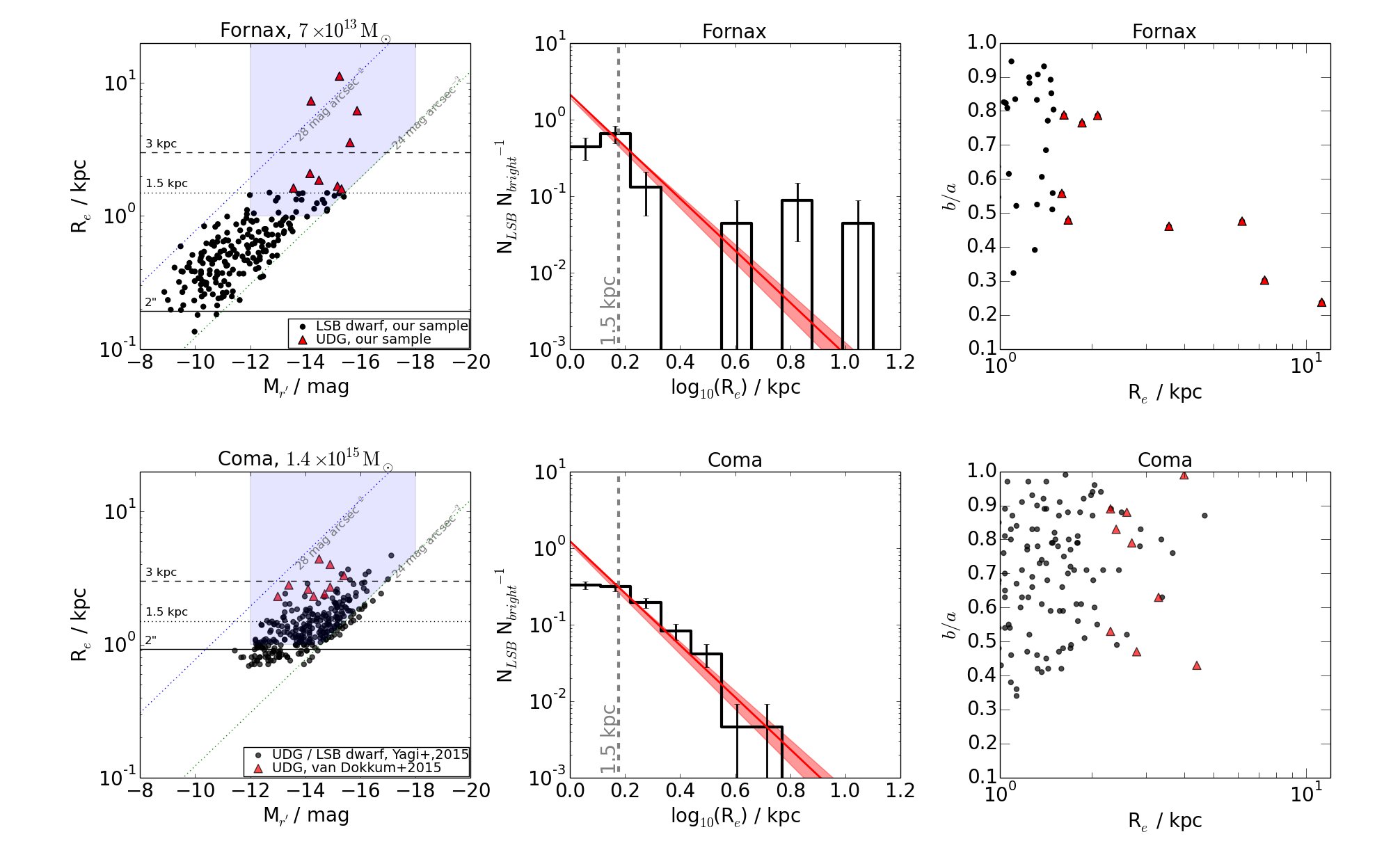}}
	\caption{{\it The left panels} show the luminosity-size relation of the known UDGs, and of the LSB dwarf galaxies within two cluster core radii in the Fornax (upper) and Coma (lower) clusters. The two horizontal lines represent the effective radii of 1.5 kpc and 3 kpc. The black points and the red triangles in the upper left panel show the LSB dwarfs and UDGs in our sample, respectively, and the black points, and the red triangles in the lower left plot correspond to the LSB dwarfs and UDGs from \citet{Yagi2016} and \citet{VanDokkum2015}, respectively. The blue areas show the selection limits of the middle and right panels. {\it The middle panels} show the number distribution of effective radii normalized by the number of bright galaxies in the studied area. The dashed lines show the scaled n[dex$^{-1}$] $\propto$ r$_{e}^{-3.4}$ relation found \citep{VanDerBurg2016} between the number of UDGs with a given effective radius in logarithmic bins. The solid vertical lines in the middle of the bins show the Poisson noise uncertainties in the bins. {\it The right panels} show axis ratios of the selected galaxies in Fornax (upper right panel) and Coma (lower right panel) against their effective radii.}
	\label{fig:udg_environments}
\end{figure*}

The sample of \citet{Yagi2016} includes 288 UDGs (the total number of galaxies is 854 galaxies) residing in the Coma cluster, of which 267 have R$_e$ < 3 kpc, and 21 are larger than 3 kpc. Most of these UDGs have R$_e$ $\sim$ 1.5 kpc, and the number of UDGs drops rapidly with increasing effective radius. The largest UDG in Coma has R$_{e}$ = 6.1 kpc, whereas in Fornax the largest one has R$_e$ $\sim$ 10 kpc, which is considerably larger. The Coma UDGs have r'-band magnitudes\footnote{transformed from Suprime-Cam R-band, see appendix C} between -16.8 mag < M$_{r'}$ < -11.8 mag, whereas in our sample for Fornax the magnitude range is -15.8 mag < M$_{r'}$ < -13.5 mag. The narrower magnitude range in Fornax is explained by the smaller sample size, as these two samples have similar medians and standard deviations (see Table \ref{tab:properties}).

\indent The size-magnitude relations of the UDGs in the Fornax and Coma clusters (\citealp{VanDokkum2015}, \citealp{Yagi2016}) are shown in the upper and lower left  panels of Fig. \ref{fig:udg_environments}, respectively. It appears that in the Coma cluster the division between UDGs and LSB dwarfs at 1.5 kpc is artificial, i.e., they form a continuous distribution in the size-magnitude parameter space. The same is not that apparent from the size-magnitude relation of the Fornax galaxies, where at least the two largest UDGs are clearly outliers.

\indent UDGs appear also in the Virgo cluster, although no systematic search of them has been done. \citet{Gavazzi2005} measured structural parameters of a sample of early-type galaxies in the Virgo cluster, and found 14 galaxies with 1.5 kpc < $\mathrm{R}_{e}$ < 3.0 kpc and $\bar{\mu}_{e,r'}$ > 24 mag arcsec$^{-2}$, which according to our criteria would be classified as UDGs. However, their study does not reach similar image depths as the deepest images obtained for the Coma and Fornax clusters, which probably explains why they do not find many UDGs larger than 3 kpc. Particularly the large UDGs in our study have low effective surface brightnesses, i.e. are fainter than $\bar{\mu}_{e,r'}$ > 26 mag arcsec$^{-2}$. However, \citet{Mihos2015} find three large UDGs (R$_e$ $\sim$ 3kpc -- 10 kpc) in the central parts of the Virgo cluster. Although these findings do not comprise a complete sample, they already demonstrate that UDGs in Virgo can be as large as the largest UDGs in the Fornax cluster.

\indent The observation that the largest UDGs in Coma are smaller than in Fornax or Virgo clusters, most likely is a detection bias related to the fact that the UDG identifications both in \citet{Yagi2016} and \citet{VanDokkum2015} are made using SExtractor. \citet{VanDerBurg2016} tested the detection efficiency of SExtractor using artificial LSB galaxies: they found that the detection efficiency is less than 0.5 for the UDGs with R$_e$ $\sim$ 3 kpc and $\bar{\mu}_{e,r'}$ $\sim$ 26 mag arcsec$^{-2}$, and further drops towards lower surface brightnesses and larger effective radii. Thus, if large UDGs like the ones detected in this study exist in Coma, most of them would not have been detected using automatic methods.

\indent In the middle panels of Fig. \ref{fig:udg_environments} we show the normalized number distributions of the effective radii (of a magnitude- and size-limited sample) of the LSB galaxies in Coma and Fornax clusters. The galaxies are chosen so that neither sample is limited by its selection criteria: we selected the galaxies within two cluster core radii from the cluster centers corresponding to 450 kpc and 700 kpc in the Fornax \citep{Ferguson1989b} and Coma clusters \citep{Kent1982}, respectively. Additionally, we required these galaxies to have R$_e$ > 1 kpc, $\bar{\mu}_{e,r'}$ > 24 mag arcsec$^{-2}$ and -18 mag < M$_{r'}$ -12 mag (shown with the blue area in the left panels). The histograms are normalized by the number of bright galaxies\footnote{We selected the galaxies in FCC that have M$_r'$ < -17 mag ( corresponding in M$_{B}$ $\sim$ -16 mag, see appendix C for details), and a membership status ``confirmed'' or ``likely member''. The galaxies in the Coma cluster were selected using the SDSS (DR10), at a redshift range of 0.0164 < z <  0.0232, and having r'-band magnitude M$_{r'}$ < -17 mag. This results to 23 galaxies in Fornax and 218 in Coma.} (N$_{bright}$) with M$_{r'}$ < -17 mag  in the selected area. These histograms highlight the difference between Fornax and Coma, the large UDGs in the former being clearly detached from the rest of the LSB population. We also plot the relation observed by  \citet{VanDerBurg2016}, which tells that the number of UDGs with given R$_e$ decreases as n[dex$^{-1}$] $\propto$ R$_{e}^{-3.4\pm0.2}$, where n[dex$^{-1}$] is the number of UDGs within a logarithmic bin. In the Fornax cluster, this relation clearly underestimates the number of large UDGs. This is not surprising as the sample of \citet{VanDerBurg2016} is known to miss many such galaxies, since they are using SExtractor. Using Monte Carlo modelling  and assuming the sizes of the UDGs in Fornax to follow the van der Burg (2016) size distribution, we find that the probability of 4 or more UDGs out of 9 having R$_e$ > 3 kpc is p = 0.01. However, the number statistics alone is not sufficient to tell if the distribution of our sample is different from the one found by Van der Burg et al. A sample that is not limited by the number statistics (like our sample is) nor missing the large UDGs (like the ones using SExtractor are) is clearly needed to understand the total contribution of large UDGs to the total galaxy populations  in clusters.

\subsubsection{Number of UDGs}

\begin{table*}
\caption{Fractions of UDGs in the Fornax and Coma clusters. The first column gives the UDG identifications in this study (within two core radii from the center $\sim$ 450 kpc), the second column gives the upper limits when those numbers are extrapolated to the virial radius of 0.7 Mpc \citep{Drinkwater2001}, being corrected also for the expected number of UDGs that are missed as parts of the area is excluded due to bright sources. The third column gives the number of UDGs in the Coma cluster within two core radii ($\sim$700 kpc) from the center. The fourth column gives the number of UDGs in the whole sample of \citet{Yagi2016}, which reaches up to 2.5 Mpc from the center of the Coma cluster. The Poisson uncertainties are based on the actual number of objects.}
\label{tab:frequency}
\centering
\begin{tabular}{ccccc}
\hline\hline
				& Fornax (r=450kpc) & Fornax (r=0.7Mpc) & Coma (r=700kpc)  & Coma (r=2.5Mpc) \\
\hline\hline
UDGs 			& 9$\pm$3 		& 42$\pm$12	&	98	& 288 \\
1.5 kpc < R$_e$ < 3 kpc & 5 		& 22$\pm$8  	&	91	& 267 \\
R$_e$ > 3 kpc & 4 				& 19$\pm$7 	&	7	& 21 \\
UDGs / Mpc$^{-2}$ &25$\pm$8		& 	-		&	64	& - \\
Normalized frequency, $\frac{\nu_{UDG}}{\nu_{bright}}$&  0.7$\pm$0.2   &  - & 0.45$\pm$0.05 &  - \\
\hline\hline
\end{tabular}
\end{table*}

Taking into account the homogeneity of the data and our tests made with the mock galaxies, we should be able to detect all the galaxies down to $\bar{\mu}_{e,g'}$ = 28.5 mag arcsec$^{-2}$. Given the fact that we are excluding 20 \% of the area to avoid possible source confusion, we are probably missing 1-3 UDGs, which would be bright enough to be detected in case that they were not overlapping with other sources. Adding these missed galaxies to the 9 identified UDGs we get surface number density of $\nu_{UDG}$ = 25$\pm$8 UDGs Mpc$^{-2}$, which compares to $\nu_{UDG}$ $\approx$ 64 UDGs Mpc$^{-2}$ in the Coma cluster (N(UDGs) = 98 within the innermost 700 kpc in \citealp{Yagi2016}). 
While normalizing these numbers with the surface densities of the bright galaxies\footnote{The same sample of bright galaxies was used as in section 8.2.1. Giving surface densities of $\nu_{bright}$ = 36 galaxies Mpc$^{-2}$ in Fornax and $\nu_{bright}$ = 140 galaxies Mpc$^{-2}$ in Coma.} in the studied areas, we get   $\frac{\nu_{UDG}}{\nu_{bright}}$ = 0.7$\pm$0.2 and 0.45$\pm$0.05 for Fornax and Coma clusters, respectively. This shows that the normalized surface number density of UDGs in the Fornax cluster is $\sim$ 2 times larger than that in the Coma cluster.

\begin{figure}[!h]
    \centering
        \resizebox{\hsize}{!}{\includegraphics{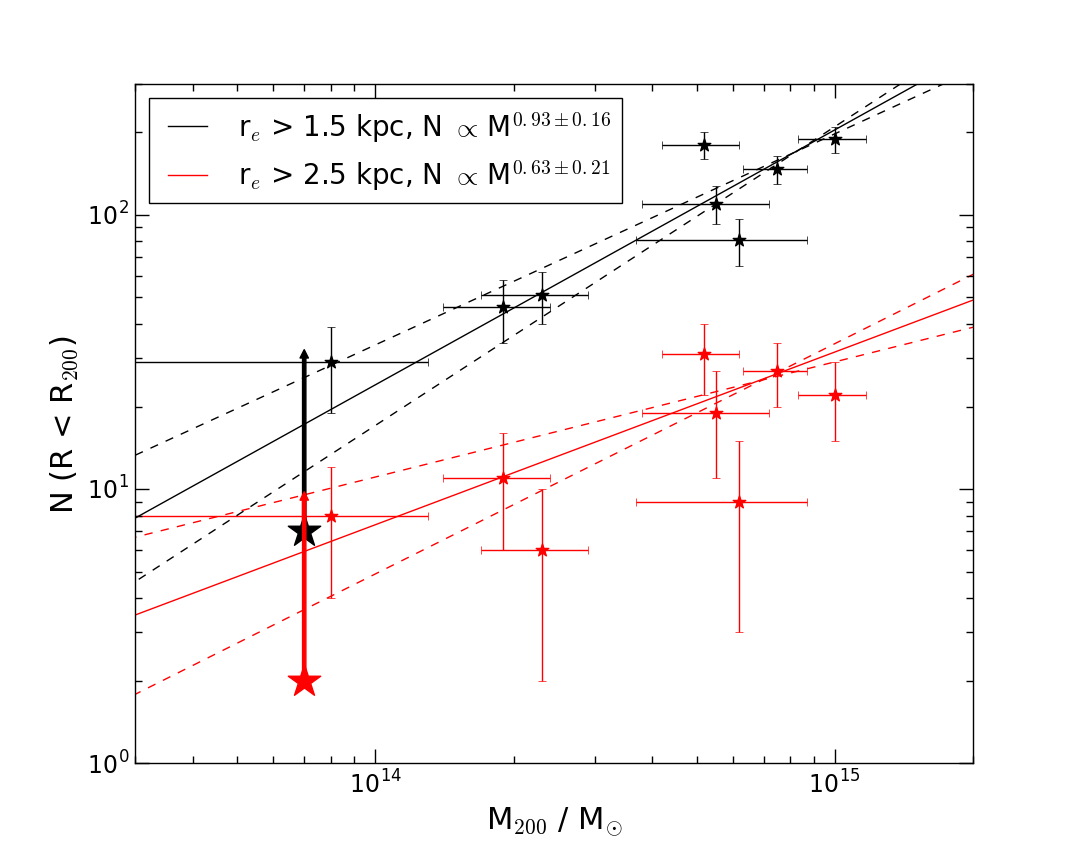}}
    \caption{The relation between the cluster mass (M$_{200}$) and the number of UDGs within the virial radius (R$_{200}$) in galaxy clusters at 0.044 < z < 0.063 is shown with black points \citep{VanDerBurg2016}.  {\it The red symbols} show the same relation for the UDGs with R$_{e}$ > 2.5 kpc. The observed values from our sample are marked with the large symbols, with the extrapolated upper limits shown with the arrows. }
    	\label{fig:vdburgrelation}
\end{figure}

\indent There exists a correlation between the virial mass (M$_{200}$) of the cluster and the number of UDGs within the virial radius (r$_{200}$) (\citealp{VanDerBurg2016}, see Fig. \ref{fig:vdburgrelation}). For a cluster halo mass of 7$\times10^{13}$ M$_{\odot}$, corresponding to that of the Fornax cluster \citep{Drinkwater2001}, the expected number of UDGs within r$_{200}$ is 10--20. If we assume uniform density of UDGs and extrapolate it up to the virial radius of Fornax r$_{200}$ (0.7 Mpc, or 2.2$^{\circ}$, \citealp{Drinkwater2001}), we get an upper limit of N = 42$\pm$12 UDGs. However, the sample of \citet{VanDerBurg2016} excludes galaxies that have $\bar{\mu}_{r',e}$ > 26.5 mag arcsec$^{-2}$,  or R$_e$ > 7 kpc. Therefore, in order to make a fair comparison to their study we need to drop 2 out of the 9 identified UDGs in our sample. Taking this into account leads to the upper limit of N = 33$\pm$9 UDGs inside r$_{200}$, which brings the Fornax cluster to the relation by van der Burg et al. (see Fig. \ref{fig:vdburgrelation}). For comparison, the Coma cluster has a mass of M = 1.4 $\times10^{15}$ M$_{\odot}$ \citep{Lokas2003}, and has 288 identified UDGs, which also matches well with the relation by \citet{VanDerBurg2016}.

\subsubsection{Shapes and orientations}

The stamp images in Fig. \ref{fig:udg_cuts} show that the large UDGs in our sample are less symmetric and more elongated than the smaller ones (see also the upper right panel of Fig. \ref{fig:udg_environments}). Their shapes can be caused by tidal interactions, but in principle they can also be inclined disks or prolate spheroidals as suggested by \citet{Burkert2016}. Some UDGs have earlier been identified as disrupted early-type dwarfs, like HCC-087 in the Hydra I cluster (with M$_{V}$ = -11.6 mag and R$_e$ = 3.1 kpc) studied by \citet{Koch2012}. Koch et al.  were able to reproduce the observed s-shaped morphology of this galaxy by modelling its gravitational interaction with the cluster center. Also several other studies have identified signs of tidal disruption in UDGs (\citealp{Mihos2015}, \citealp{Merrit2016}, \citealp{Toloba2016}, \citealp{Wittmann2017}). Obviously, at some level tidal interactions are shaping the morphology of the LSB galaxies in clusters.

\begin{figure*}[!ht]
	\centering
	\resizebox{\hsize}{!}{\includegraphics[width=17cm]{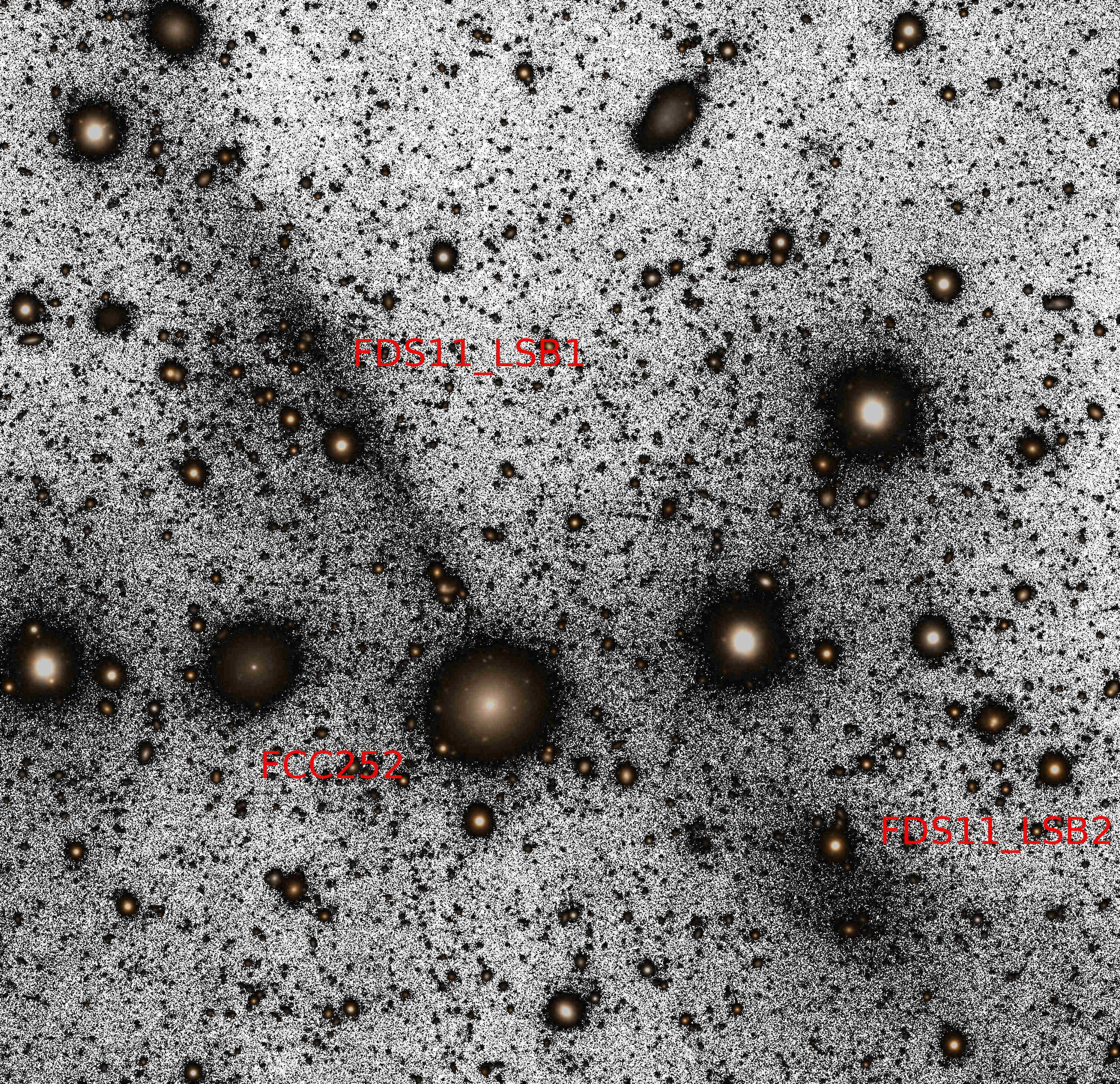}}
	\caption{A g'+r' color image showing  FDS11\_LSB1, FDS11\_LSB2 and FCC252. Colors are displayed  for the areas brighter than $\mu_{r'}$ > 26 mag arcsec$^{-2}$. Only r'-band image is shown for the levels fainter than that. FDS11\_LSB1 and FDS11\_LSB2 are separated by projected distance of 15 kpc from each other and $\sim$ 7 kpc from FCC252, which is a dwarf elliptical galaxy spectroscopically confirmed to be in Fornax (radial velocity = 1415 km s$^{-1}$, \citealp{Drinkwater2001b}). The morphology of these large UDGs is very elongated and may be due to tidal interactions.}
	\label{fig:UDGstreams}
\end{figure*}

\indent In the Coma cluster the UDGs are preferably elongated towards the cluster center  \citep{Yagi2016}, which is not the case for the 9 UDGs in the Fornax cluster (see Fig. 14). The number statistics are not good enough to conduct a conclusive analysis, but we found that some of the UDGs in Fornax are elongated toward their nearby dwarf galaxies (M$_{r'}>$-19 mag). In this sense an interesting pair is FDS11\_LSB1 and FDS11\_LSB2 (see Fig. \ref{fig:UDGstreams}), which galaxies are located at a projected distance of 15 kpc from each other. Both galaxies point toward a spectroscopically confirmed Fornax dE, FCC252 \citep{Drinkwater2001b}, which has a B-band total magnitude of M$_B$= -15.2 mag \citep{Ferguson1989}.  FCC252 itself does not show any signs of tidal interaction. Another UDG showing signs of galaxy interaction is FDS16\_LSB85 (see Fig. \ref{fig:lsbidentification}), which has a tidal tail pointing towards the nearby LSB dwarf, FCC 125 (\citealp{Ferguson1989}, or FDS 16\_LSB50 in our sample). FCC 125 has M$_{r'}$ = -13.3 mag assuming that it resides at the distance of Fornax, although it has no spectral confirmation of the cluster membership. On the other hand, the four smallest UDGs with round and symmetric appearance do not show any signs of tidal interactions. In order to test if the elongated shapes of the large UDGs are resulting from tidal interactions, a larger sample is needed.

\indent The properties of UDGs in our Fornax sample, and of those in
the Coma cluster given by \citet{Yagi2016}, are compared in Table
\ref{tab:properties}.  The mean axis ratio in Coma is $\langle b/a \rangle$ = 0.73$\pm$0.16, the small (R$_e$ < 3 kpc, $\langle b/a \rangle$ = 0.75), and large (R$_e$> 3 kpc, $\langle b/a\rangle$ = 0.69) UDGs having similar axis ratios. However, in our sample in Fornax the large UDGs are more elongated ($\langle b/a\rangle$ = 0.69 and $\langle b/a\rangle$ = 0.38, for small and large UDGs, respectively). 

\indent While comparing the b/a-values between the small UDGs and the dwarf LSB galaxies (R$_e$ < 1.5 kpc), we don't find any significant differences, neither in the Fornax nor in the Coma clusters. However, the large UDGs in Fornax have a significantly different $b/a$-distribution compared to the one of LSB dwarfs (K-S test p-value = 0.0005, for the $b/a$ distributions being the same). The same comparison between the two types of galaxies in Coma shows no significant difference (p-value = 0.24, see also the right panels in Fig. \ref{fig:udg_environments}). We conclude that in the Fornax cluster the large UDGs are significantly more elongated than the dwarf LSB galaxies, but the same is not obvious in Coma. We find no difference between the axis ratio distributions of the LSB dwarfs and small UDGs.

\indent As expected for the dwarf LSB galaxies (e.g. \citealp{Bothun1991}), the values of the S\'ersic index are below unity for the UDGs, both in the Coma and the Fornax clusters. In fact, the $n$-values are similar in the Fornax ( $\langle n\rangle$ = 0.8$\pm$0.2, $n$ = 0.4-1.2) and in the Coma ($\langle n\rangle$=0.9$\pm$0.2, $n$=0.4-2.0) clusters. The galaxies in both samples are fit using GALFIT, the only difference in the fitting approach being that Yagi et al. allowed the central {\it PSF} component of the galaxies to be off-centered from the S\'ersic component.

\subsubsection{Colors}

The colors of UDGs set constraints for their formation mechanisms. The UDGs in Coma \citep{Koda2015} have been shown to have similar colors as the normal red sequence dEs of the same luminosity. Also, the measured slope of the color-magnitude relation of our LSB sample matches well with the one of dE:s in Virgo cluster, which supports also them being mostly quiescent red-sequence galaxies. The mean color of the Coma UDGs is $\langle$g'--r'$\rangle$\footnote{Transformed from the Suprime-Cam $\langle$B--R$\rangle$ color. See Appendix C for details.} = 0.67 mag, which color is the same for the large and small UDGs. This is comparable to $\langle$g'--r'$\rangle$ = 0.59 mag (with the range of 0.2 -- 1.0) that we obtain for the UDGs in Fornax (see red triangles in Fig. 16). The only exception is FDS11\_LSB1, which has g'-r' = 0.18 mag, being a clear outlier towards blue colors. The observation that the UDGs do not clearly deviate from the colors of the LSB dwarf galaxies within the same luminosity range is consistent with them having similar origin. 

\subsection{How to explain the origin of UDGs?}

\subsubsection{Explanations in the literature}

Van Dokkum et al. (2015) first pointed out that UDGs form a population that is  continuous with the dwarf galaxies in the size-magnitude relation, and therefore can simply be a diffuse end tail in that relation. However, they do not think that UDGs originate from the same progenitors as the typical dwarf galaxies in clusters, since processes like harassment (\citealp{Moore1998}, \citealp{Smith2015}) and tidal stirring, rather make the galaxies more compact than diffuse. Their suggestion was that UDGs could be failed Milky Way mass (halo mass of $\sim$ 10$^{12}$ M$_{\odot}$) galaxies which lost their gas during their in-fall into the cluster environment, and therefore were not able to form a large amount of luminous matter, otherwise typical for their observed size. This means that compared to their sizes, these galaxies might have massive dark matter halos. However, the number of globular clusters (GC) around these galaxies (\citealp{Amorisco2016}, \citealp{Beasley2016}) rather suggest that they are Large Magellanic Could (LMC) type galaxies with halo mass of $\sim$ 10$^{10}$ M$_{\odot}$). Also, the similar cluster-centric distributions of dwarf galaxies and UDGs found by \citet{Roman2016} and by \citet{VanDerBurg2016} in the Coma cluster, is problematic in the interpretation of van Dokkum et al.. This is because due to the large dark matter halos dynamical friction should have made these originally Milky Way sized galaxies to sink deeper into the cluster potential well, which is not observed.

\indent \citet{Amorisco2016I} suggested that UDGs are dwarf galaxies with especially high original angular momentum. In this scenario, the high angular momentum makes the UDGs more flattened and extended than the typical dEs are. Otherwise the formed galaxies are expected to be similar, and to appear in similar environments as the other dwarf galaxies. Their model predicts that the sizes of the UDGs increase with increasing angular momentum, which makes the largest UDGs more disk-like (elongated when inclined) than the smaller ones. This prediction is well in agreement with our observation that the largest UDGs are more elongated. However, the analysis of \citet{Burkert2016} shows that the $b/a$-distribution of UDGs is more compatible with them being prolate spheroidals than disks. In \citet{Amorisco2016I} the baryonic physics is not modelled, but they were able to model the relation between the cluster halo mass and the total number of UDGs as seen by \citet{VanDerBurg2016}.

\indent On the other hand, \citet{DiCintio2016} suggested that UDGs are dwarf sized galaxies, which are extended due to their feedback driven gas outflows which give rise to an expanded stellar component. They modelled baryon physics of isolated galaxies, and for comparison with observations selected objects which fulfil the criteria of typical UDGs. Their models predict average B-R colors of 0.77 $\pm$ 0.12 mag (or g'-r' = 0.5 mag) for these galaxies. The predicted color is 0.13 mag bluer than the  mean observed value for the Coma UDGs, and 0.07 mag bluer than that the UDGs in the Fornax cluster in our sample. This discrepancy between the models and observations might be reduced if the galaxies were simulated in the cluster environment, where also external quenching takes place. An important prediction from their model is that the UDGs should mostly form outside the clusters, and also contain gas if it is not externally removed. This is well consistent with the observations of the ALFALFA survey \citep{Leisman2017}, where 115 HI rich isolated UDGs were studies. Additionally, \citet{Roman2017} find that the colors of the UDGs located in the periphery of group environments have bluer colors (g-i$\sim$0.45) than the UDGs in the clusters (g-i$\sim$0.76), which is also consistent with this formation scenario. One discrepancy between the models of \citet{DiCintio2016} and the observations is that the effective radii in the simulated galaxies vary between 1 kpc < $R_e$ < 3 kpc, whereas the observed $R_e$ can be as large as $\sim$ 10 kpc. This discrepancy can simply be a sampling issue of the simulations, or alternatively the largest UDGs have a different origin.

\indent \citet{Baushev2016} suggested that UDGs can form via straight-on collisions of two massive galaxies, which penetrate each other with large relative velocities. This collision has little effect on the stellar content, but will rip out the cold gas in these galaxies. The collision rate strongly depends on the number density of galaxies, and high relative velocities are needed to prevent merging of the colliding systems.  Therefore, this model is not good for explaining UDGs in galaxy groups nor in the field, where the relative velocities of galaxies are smaller. However, such collisions can hardly be prevented in galaxy clusters, where besides major mergers, also dwarf-sized galaxy collisions are expected. Since the strength of galaxy-galaxy interactions is stronger in low velocity encounters, the dwarf galaxies in groups are more likely to experience strong tidal forces than the ones in clusters, which could explain the large sizes of UDGs in the lower mass environments (e.g. \citealp{Merrit2016}). As these dwarf-dwarf interactions in clusters and groups are not well studied in simulations, further simulations of the importance of these tidal encounters are needed.

\indent It seems that both the high angular momentum origin, and feedback induced gas outflows, can reproduce most of the observed properties of UDGs at present. However, at this moment, we also should seriously consider the contribution of the tidal interactions, which could be the explanation of the most extended UDGs.

\subsubsection{Is the environment important for the formation of UDGs?}

The Coma and Fornax clusters are clearly different environments, which should affect the observed properties of UDGs, if their evolution is driven by their environment. The virial mass of Coma is 1.4$\times$10$^{15}$ M$_{\odot}$, and the virial radius is 2.9 Mpc \citep{Lokas2003}, while the corresponding values for the Fornax cluster are 7$\times10^{13}$ M$_{\odot}$ and 0.7 Mpc \citep{Drinkwater2001}. This means that although Coma is 20 times more massive than Fornax, it is less dense. The velocity dispersion of the Coma main group galaxies is $\sigma$ = 1082 km s$^{-1}$ \citep{Colless1996}, which is nearly three times larger than the one of the Fornax cluster galaxies with $\sigma$ = 374 km s$^{-1}$ \citep{Drinkwater2001}. Therefore galaxy-galaxy interactions are expected to be stronger in the Fornax than in the Coma cluster.

\indent As discussed above, the average magnitudes, g'-r' colors and S\'ersic $n$-values of the UDGs are similar in the Coma and Fornax clusters. The preferred alignment toward the cluster center that appears among the Coma UDGs, but not among those in the Fornax cluster, might suggest that the galaxy - cluster potential interaction might be more important in Coma.

\indent On the other hand, UDGs have been found also in lower density environments outside the clusters. Four tidally disturbed UDGs have been found in the NGC 5485 group \citep{Merrit2016}, one UDG in a filament of the Pisces-Perseus supercluster \citep{MartinezDelgado2016}, one UDG in the NGC 253 group \citep{Toloba2016}, and one as a satellite of Centaurus A \citep{Crnojevic2016}. All these galaxies are large having effective radii R$_e$ = 2.5 -- 5 kpc. The size distribution of the low density environment UDGs seems to be different from the one in clusters, but again this can be a bias of cluster samples missing the large UDGs. However, recently \citet{Wittmann2017} visually inspected the UDG population in the core of the Perseus cluster (M$_{vir}$ = 8.5$\times10^{14}$ M$_{\odot}$, \citealp{Mathews2006}), but did not find any UDGs larger than R$_e$ > 4.1 kpc. This observation might indicate that the difference in the effective radii distributions is not only an observational bias.  Otherwise, there seems to be no structural difference between the UDGs in the group environments and the ones in clusters.

\indent Recently, the observations of the isolated UDGs showed that they contain HI gas \citep{Leisman2017}, and posses bluer colors than the  ones in clusters \citep{Roman2017,Leisman2017}. Additionally, \citet{Leisman2017} observed that these HI rich UDGs may exist in high angular momentum halos. These observations are consistent with the scenario where UDGs are formed in dwarf-sized dark matter halos outside the clusters.

\indent To conclude, from the observational (photometric) point of view we do not find systematic differences in the structure of the bulk of the UDGs in different galaxy environments. However, in Fornax, Virgo and group environments there are elongated large UDGs, which have not been found in higher mass clusters like Coma or Perseus. The UDGs in the isolation and outskirts of groups possess bluer colors than the ones in groups and clusters, which might be an indication of environmental effects.

\subsection{LSB dwarfs in FDS}

Most of the galaxies in our sample are not UDGs but rather dwarf-sized LSB galaxies. The effective radii of these galaxies range from 120 pc to 1.5 kpc (an upper limit by definition), and the total r'-band magnitude between -8.9 mag > M$_{e,r'}$ > -15.2 mag, which means that we are reaching the sizes of the largest Milky Way satellite dSph galaxies. For comparison, Ursa Minor has R$_e$ = 181 pc and M$_V$ = -8.8 mag, and the Fornax dSph has R$_e$ = 710 pc  and M$_V$ = -13.4 mag \citep{McConnachie2012}. Having in mind that the dividing line between the LSB dwarfs and the other dwarf galaxies is physically not very clear, it is interesting to compare their radial distributions in the Fornax cluster. As the LSB dwarfs appear also in galaxy groups,  we can compare the Fornax LSB dwarfs with those appearing in the Centaurus group, where a sample comparable to ours exists.

\subsubsection{Radial distribution}

Dynamical friction \citep{Chandrasekhar1943} transfers kinetic energy from massive galaxies to smaller galaxies, making the massive galaxies to sink deeper into the cluster potential well. This mass segregation takes place in time scales shorter than Hubble time \citep{White1977}. Thus, assuming that all galaxies spend a similar amount of time in the cluster, the distribution of massive galaxies should be more centrally concentrated. Harassment should also make the galaxies themselves more centrally concentrated near the cluster center. For a dynamically relaxed cluster we would then expect that also among the dwarf galaxies the more massive and higher surface brightness galaxies preferentially appear in the central regions of the cluster. However, according to the cosmological simulations like the Millennium simulations by \citet{BoylanKolchin2009}, clusters grow hierarchically by accreting galaxy groups, which in reality makes the situation more complicated than that.

\indent To study if the above mentioned processes are relevant in the Fornax cluster we studied the radial number density profiles of the galaxies. We found that the bright FCC galaxies indeed are more centrally clustered than the fainter galaxies in our sample (see Fig. \ref{fig:numberdensityprof}). In the same figure plotted are also Ultra Compact Dwarf galaxies (UCDs) from the sample of \citet{Gregg2009}, which is complete up to 0.9$^{\circ}$ of the Fornax cluster. UCDs are compact dwarf galaxies, with sizes similar to GCs or nuclei of dwarf galaxies. It appears that the dwarf galaxy concentration in the central regions of the Fornax cluster increases with increasing compactness of the object, so that the most concentrated are the UCDs, and the least concentrated the LSB dwarfs.

\indent It seems that LSB dwarfs are the most numerous type of galaxies in Fornax, everywhere except in the area within 0.3$^{\circ}$ from NGC 1399 (note that only the area within 5 arcmin from NGC 1399 was masked). A similar phenomenon has been observed also in several other clusters (\citealp{VanDerBurg2016}, \citealp{Wittmann2017}). In principle this could be a bias due to difficulties in detecting the LSB galaxies in the central regions of the clusters, but it is also possible that the galaxies entering too close to the cluster center get tidally disturbed and  finally accreted to the central cD galaxy. Most probably we can exclude the first possibility, because bright stars, which are the main reason for possible overlapping with the LSB galaxies, are not clustered into the cluster center. We see a drop also in the number of bright FCC galaxies which should be easily visible throughout the Fornax cluster.

\subsubsection{Comparison with the Centaurus group}

In the following we compare the properties of the Fornax LSB dwarfs with those in the Centaurus group, located at the distance of 4 Mpc \citep{Jerjen2000}. The main body of the Centaurus group consists of four bright galaxies (brighter than M$_{r'}$ < -17 mag). The photometric measurements for the dwarf galaxies in this group in the g'- and r'-bands are provided by \citet{Muller2017}. The images reach the V-band surface brightness of 28 mag arcsec$^{-2}$. With that limiting surface brightness they should be able to find all similar LSB dwarfs as detected by us in the Fornax cluster. The Centaurus group is  located at  suitable distance so that the analysis can be done in a similar manner as in this paper. 

\begin{figure}[!h]
    \centering
        \resizebox{\hsize}{!}{\includegraphics{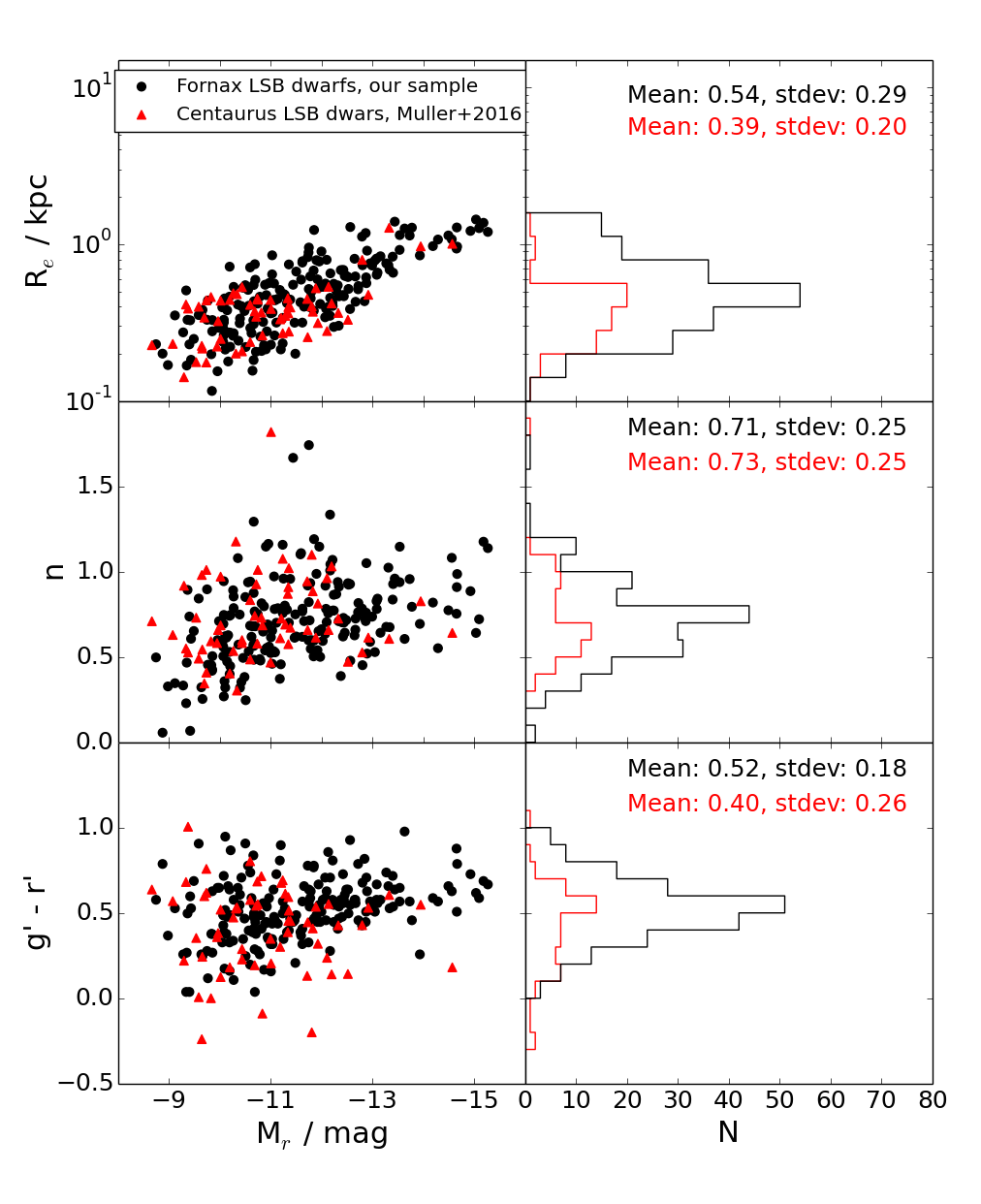}}
        \caption{The left panels show the effective radii (R$_e$), S\'ersic $n$-values and g'-r' colors as a function of absolute r'-band magnitude M$_{r'}$, for the LSB dwarfs of our sample (\textit{black points}), and for the Centaurus group dwarfs from \citet{Muller2017} (\textit{red triangles}). The histograms on the right show the distributions of these parameters for both samples with the corresponding colors.}
        \label{fig:sersic_n_vs_muller}
\end{figure}

\indent The size-magnitude, color-magnitude and S\'ersic $n$-magnitude relations of the LSB dwarf galaxies in the Fornax cluster, and in the Centaurus group are shown in Fig. \ref{fig:sersic_n_vs_muller}. The Centaurus group sample is complete for the sources which have diameters larger than 28 arcsec (0.6 kpc at 4 Mpc distance) at a surface brightness of 28 mag arcsec$^{-2}$. The effective radii of these galaxies range between  0.10 kpc < R$_{e}$ < 1.82 kpc, and the r'-band absolute magnitudes between -7.9 mag > M$_{r'}$ > -17.2 mag. To eliminate selection effects from the comparison, we removed from the Centaurus sample the galaxies with $\bar{\mu}_{e,r'}$ < 24 mag arcsec$^{-2}$, as these galaxies would not be included in our sample.

\indent The g'-r' colors and the values of S\'ersic $n$ of the Fornax and Centaurus LSB dwarf galaxies, are plotted as a function of the total r'-band magnitude in Fig. \ref{fig:sersic_n_vs_muller}. The number histograms of the two parameters are shown in the small right panels of both figures. It appears that the LSB dwarf galaxy colors in both environments are fairly similar. Only difference is that the average colors of galaxies in the Centaurus group become bluer and the effective radii smaller than in Fornax with increasing magnitude. Also, the scatter in the colors of the Centaurus group LSB dwarfs is clearly higher. Notable is that there is no difference in the colors , sizes or S\'ersic indices of the faintest LSB galaxies (M$_{r'}$ $\gtrsim$ -12 mag).

\indent The values of S\'eric $n$ parameters of these galaxies are also similar. In both environments the S\'ersic $n$ value drops from $n\sim$1 to 0.5 with decreasing galaxy magnitude. A similar trend is found also by \citet{Misgeld2009} for the Hydra I cluster, which has a virial mass of 4.4$^{+1.1}_{-1.0} \times 10^{14}$ M$_{\odot}$, thus being $\approx$6 times more massive than the Fornax cluster. 

\indent Indeed, there are no differences in the properties of the smallest LSB dwarf galaxies (M$_{r'}$ $\gtrsim$ -12 mag) in the Fornax cluster and the Centaurus group, regardless of the very different type of environments that they represent. This is consistent with the
simulations (\citealp{Zolotov2012}, \citealp{Onorbe2015}), which predict that strong background UV - radiation and star formation feedback in the early epoch of the universe suppresses the accretion of the cold gas into the smallest dwarf galaxies. Due to their low
mass, these galaxies are inefficient in accreting more cold gas later in their evolution, whereas the more massive ones can continue accreting gas (if available) and forming stars. 

\section{Summary and Conclusions}

New deep  g', r' and i'-band images have been obtained for the central region of the Fornax cluster using the OmegaCAM instrument attached to the VST telescope at ESO. Four 1$^{\circ}$ x 1$^{\circ}$ fields of the data, covering the innermost 4 deg$^2$ of the cluster, have been analysed in this work. The images, reaching a surface brightness of $\sim$28 mag arcsec$^{-2}$ (for {\it S/N} = 5 over 1 arcsec$^2$) in the r'-band, are used to compile a catalog of 205 LSB galaxies ($\mu_{0,r'}$ > 23  mag arcsec$^{-2}$), of which galaxies 59 are new detections . These detections complement the previous catalogs of dwarfs in the Fornax cluster, namely the Fornax Cluster Catalog by \citet{Ferguson1989}, the Next Generation Fornax Survey catalog by \citet{Munoz2015}, and the list of extended LSB structures in the Fornax cluster by Lisker et al. (\textit{submitted}). Galaxies are separated from tidal structures, and for these galaxies the photometric parameters are derived, by fitting S\'ersic functions to the azimuthally averaged light profiles, and by fitting the full 2D flux distributions of the galaxies using GALFIT. The reliability of the two methods is tested, and the completeness of the UDGs is evaluated by adding artificial mock galaxies to the images. Also, g'-r' aperture colors within the effective radii of the sample galaxies have been obtained.

The following conclusions are  obtained:

\begin{enumerate}

\item Based on visual inspection we identify nine galaxies fulfilling the definition for Ultra Diffuse Galaxies (UDGs) by \citet{VanDokkum2015}. Three of these galaxies are new detections.  The number density of UDGs normalized with the number of bright galaxies, is found to be higher in Fornax than in the Coma cluster (i.e 0.7$\pm$0.2 and 0.45$\pm$0.05, respectively). However, this can be a detection bias of the Coma sample in a sense that not all UDGs are detected in Coma.
\vskip 0.3cm

\item We made photometric measurements using GALFIT models and azimuthally averaged luminosity profiles. These two methods give consistent results in terms of effective radii (R$_{e}$), total r'-band magnitudes (m$_{r'}$), and S\'ersic $n$-values. We find that the GALFIT models are more reliable for obtaining the ellipticities ($b/a$), and are therefore used in the analysis of this paper.
\vskip 0.3cm

\item Assuming that the UDGs in our sample are located at the distance of Fornax they have -15.8 mag < M$_{r'}$ < -13.5 mag and 1.6 kpc < R$_e$ < 11 kpc, which are typical sizes of UDGs found in other environments. However, we find a larger fraction of large UDGs (R$_e$ > 3 kpc) than was found in the Coma cluster \citep{Yagi2016}, and in the distant clusters \citep{VanDerBurg2016}. As those studies, both using SExtractor, are not as effective as our visual inspection in detecting large UDGs, we can not exclude the possibility of the difference being a selection bias.   
\vskip 0.3cm

\item The five small UDGs with R$_e$ < 3 kpc detected in the Fornax cluster by us, are similar to the typical LSB dwarfs in terms of their magnitude, color and shape, being consistent with them having common origin. The four largest UDGs (with R$_e$ > 3 kpc) appear as outliers having 10 times larger effective radii than the typical LSB dwarfs of the same galaxy luminosities. Additionally, the large UDGs in our sample are significantly more elongated (0.2 < $b/a$ < 0.4) than the typical LSB dwarfs, in our sample and in Coma, and they also show signs of tidal interactions. 
\vskip 0.3cm

\item Contrary to the UDG orientations in Coma, we find that the LSB dwarfs in the Fornax cluster are randomly oriented, without any preferences neither toward the cluster center, nor toward their nearby bright galaxies. However, the most elongated galaxies show a preferred alignment towards their nearest bright galaxies.
\vskip 0.3cm

\item We find that the LSB galaxies in the center of the Fornax cluster follow a color-magnitude relation with a slope, which is same as the one of the red sequence in the Virgo cluster (\citealp{Kim2010}, \citealp{Roediger2017}
). The UDGs follow the same relation as the smaller LSB dwarfs, although there is one clear outlier towards blue.
\vskip 0.3cm

\item Our observations are in agreement with both the high-spin and the strong feedback models of UDG formation, where they are seen as an extension of the dwarf galaxy population. However, the large and elongated UDGs in our sample are likely indications that the tidal forces can also produce some of these UDGs, at least their largest, most elongated representatives.
\vskip 0.3cm 

\item  The LSB dwarf galaxies in our sample have magnitudes in the range of -15.4 mag < m$_{r'}$ < -8.9 mag (with the mean of -11.4 mag) and effective radii between 0.14 kpc < R$_{e}$ < 1.5 kpc (with the mean values of 0.5 kpc). Their colors and profile shapes were compared with the ones of the LSB dwarf galaxies in the Centaurus group \citep{Muller2017}: they were found to be similar for the faintest LSB galaxies (M$_{r'}$ $\gtrsim$ -12 mag). However, the LSB galaxies in the Centaurus become bluer and smaller with increasing magnitude, when compared to the ones in the Fornax cluster.
\vskip 0.3cm

\item We find that the distribution of LSB dwarf galaxies in the Fornax cluster is less centrally concentrated than the one of the bright galaxies with M$_{r'}$ < -17 mag. Also, the number density of LSB galaxies drops at a cluster-centric distance of r = 0.6$^{\circ}$ (180 kpc) inward, which can be an indication of tidal disruption of galaxies that enter close to the cluster center. The observed drop in the number of the LSB galaxies in the center of the Fornax cluster is consistent with the findings in other clusters, showing that it is a common phenomenon in galaxy clusters (\citet{Wittmann2017},\citet{VanDerBurg2016}).
\vskip 0.3cm

\end{enumerate}

\begin{acknowledgements} A. ~V. acknowledge the support from grant from the Vilho, Yrjö and Kalle Väisälä Foundation. J.~F-B. and R.~P. acknowledge support from grant AYA2016-77237-C3-1-P from the Spanish Ministry of Economy and Competitiveness (MINECO). A.~V., E.~L. and H.~S. acknowledge the support from the academy of Finland (grant n:o 297738). This paper has benefited strongly from discussions within the SUNDIAL ITN network, a EU Horizon 2020 research and innovation programme under the Marie Skłodowska-Curie grant agreement No 721463.  We would also like to thank the anonymous referee, whose detailed comments and suggestions helped to significantly improve this paper.  
\end{acknowledgements}

\bibliographystyle{aa}
\bibliography{LSB_paper}

\begin{appendix}

\section{Pixel value distribution moments.}

The center and initial shape parameters of a galaxy are measured by calculating the first and second moments of the galaxy's pixel value distribution. The pixels from which the moments are calculated are selected by manually placing an elliptical aperture over the galaxy that follows the apparent shape of the galaxy and covers the visible part of the galaxy. If we mark x- and y-coordinates and intensity of a pixel i as $x_i, y_i$ and $I_i$,respectively, we can calculate the first moments of pixel value distribution as follows
\begin{equation}
\begin{matrix}
\overline{x} = \frac{\sum_{i} I_i x_i }{\sum_i I_i}, \\ 
\overline{y} = \frac{\sum_{i} I_i y_i }{\sum_i I_i}.
\end{matrix} 
\end{equation}
The first moments correspond to the weighted center of the pixel value distribution. The second order moments 
\begin{equation}
\begin{matrix}
\overline{x^2} = \frac{\sum_{i} I_i x_i^2 }{\sum_i I_i} - \overline{x}^2, \\ 
\overline{y^2} = \frac{\sum_{i} I_i y_i^2 }{\sum_i I_i} - \overline{y}^2,  \\
\overline{xy} = \frac{\sum_{i} I_i x_i y_i }{\sum_i I_i} - \overline{x} \, \overline{y},
\end{matrix} 
\end{equation}
describe the spread of the distribution along the axes. They can be transformed to semi-minor axis $b$ and semi-major axis $a$ as
\begin{equation}
\begin{matrix}
a^2  =\frac{\overline{x^2} + \overline{y^2}}{2} + \sqrt{\left(\frac{\overline{x^2} - \overline{y^2}}{2}\right)^2+\overline{xy}^2}, \\ 
b^2  =\frac{\overline{x^2} + \overline{y^2}}{2} - \sqrt{\left(\frac{\overline{x^2} - \overline{y^2}}{2}\right)^2+\overline{xy}^2},
\end{matrix} 
\end{equation}
and the position angle $\theta$
\begin{equation}
\tan(2\theta) = 2 \frac{\overline{xy}}{\overline{x^2}-\overline{y^2}}.
\end{equation}

\section{Mock galaxy parameters}

The mock galaxies are generated by using 2D S\'ersic functions with the chosen structural parameters and random position angles. Also, the Poisson noise is added to the galaxies, and the models are then convolved with the OmegaCAM {\it PSF} (see section 5.2 for details) before adding them to the science mosaics. 

\indent The parameters of the UDG mock galaxies used for testing the detection efficiency are given in Table \ref{tab:mockgalaxies}. Also the indications of the detections in different fields are given in \ref{tab:mockgalaxies}. The additional $\sim$ 150 mock galaxies used for testing the photometric accuracy have m$_{r'}$= 15--24 mag, R$_e$ = 5--20 arcsec, S\'ersic $n$ = 0.3--2 and $b/a$ = 0.2--1. The differences between the input model and measured parameters for all the mock galaxies are shown in the Fig. \ref{fig:photomaccuracy}.

\begin{table}
\caption{Structural parameters of the mock UDGs adopted from \citet{VanDokkum2015} and \citet{Mihos2015}. The four last columns indicate if the mock galaxy was detected in the given field (Y = yes, N = no, O = not detected due to the selection criteria). }
\label{tab:mockgalaxies}
\centering
\begin{tabular}{lccccccccl}
\hline\hline
Name  & M$_g$ & R$_e$ [kpc] & $n$ & $b/a$ & D$_{10}$ & D$_{11}$ & D$_{12}$ & D$_{16}$ \\
\hline
VLSB-A\tablefootmark{(b)} & -15.0 & 9.7 & 1.2 & 0.5 & O & Y & Y & O \\

VLSB-B\tablefootmark{(b)} & -13.5 & 2.9 & 0.8 & 0.83 & Y & Y & Y & Y \\

VLSB-C\tablefootmark{(b)} & -14.9 & 5.5 & 0.7 & 0.88 & Y & N & Y & Y  \\

DF6\tablefootmark{(a)} & -14.3 & 4.4 & 1 & 0.47 & Y & Y & O & Y  \\

DF9\tablefootmark{(a)} & -14.0 & 2.8 & 1 & 0.92 & Y & Y & Y & Y \\

DF14\tablefootmark{(a)} & -14.4 & 3.8 & 1 & 0.51 & Y & Y & Y & Y  \\

DF1\tablefootmark{(a)} & -14.6 & 3.1 & 1 & 0.82 & Y & Y & Y & Y  \\

DF22\tablefootmark{(a)} & -13.8 & 2.1 & 1 & 0.84 & Y & Y & Y & Y \\

DF17\tablefootmark{(a)} & -15.2 & 4.4 & 1 & 0.71 & Y & Y & Y & Y \\

DF20\tablefootmark{(a)} & -13.0 & 2.3 & 1 & 0.53 & Y & Y & Y & Y  \\

DF24\tablefootmark{(a)} & -12.5 & 1.8 & 1 & 0.38 & Y & Y & Y & Y \\

DF30\tablefootmark{(a)} & -15.2 & 3.2 & 1 & 0.70 & Y & O & Y & Y  \\

DF34\tablefootmark{(a)} & -13.6 & 3.4 & 1 & 0.66 & Y & Y & Y & Y  \\

DF38\tablefootmark{(a)} & -14.3 & 1.8 & 1 & 0.84 & Y & Y & Y & Y \\

DF7\tablefootmark{(a)} & -16.0 & 4.3 & 1 & 0.76 & Y & Y & Y & Y  \\

\hline\hline
\end{tabular}
\tablefoot{
\tablefoottext{a}{\citet{VanDokkum2015}}
\tablefoottext{b}{\citet{Mihos2015}}
}
\label{t2}
\end{table}

\section{Transformations between photometric filters}

To transform the Subaru Suprime-Cam B and R magnitudes into the r'- and g'- band values we used the transformations from \citet{Yagi2010}. The r' band magnitude m$_{r'}$ is
\begin{equation}
\mathrm{m}_{r'} = \mathrm{m}_{R} - 0.0188 + 0.1492 \times\left( g' - r' \right) - 0.0128 \times \left( g'-r' \right)^2,
\end{equation}
where m$_{R}$ is the apparent magnitude in Suprime-Cam R-band. Also, the Suprime-Cam B-R color can be transformed into g'-r' color as follows:
\begin{equation}
B-R = 0.00440 + 1.32386\times\left( g' -r' \right) +0.0010734\times \left( g' -r' \right) ^2
\end{equation} 
For the transformations between the Johnson U,B,V,R and I and SDSS g',r' and i' colors we used the equations given in the SDSS website by Lupton (2005\footnote{http://www.sdss3.org/dr8/algorithms/sdssUBVRITransform.php\#Lupton2005}):
\begin{equation}
\mathrm{m}_{V} = \mathrm{m}_{r'} + 0.4216 \times\left( g' -r' \right) -0.0038,
\end{equation}
where m$_{V}$ and m$_{r'}$ are the apparent magnitudes in V- and r'- filters, respectively, and $g'-r'$ is g'-r' color. Similarly:
\begin{equation}
\mathrm{m}_{B} = \mathrm{m}_{r'} + 1.33130 \times\left( g' -r' \right) +0.2271,
\end{equation}
where m$_B$ is the apparent magnitude in B-band.

\section{List of galaxies}
We list the properties of the galaxies in our sample In Table  D.1. . 

\onecolumn 
\longtab{ 
\begin{landscape} 
\begin{longtable}{lccccccccl} 
\caption{The columns in the table correspond to the following properties: Name of the object (Name), right ascension (RA) and declination (DEC) of the object, axis ratio and error ($b/a$), position angle ($\theta$, east of North), apparent magnitude (m$_{r'}$) and error in r'-band , effective radius (R$_{e,r'}$) in r'-band and error in arc seconds , S\'ersic index ($n$) and error, g'-r' color within R$_e$ (g'-r') and error, and the possible FCC number and identifications in earlier works (References). The references in the table are  $^{(a)}$: \citet{Ferguson1989},  $^{(b)}$:\citet{Bothun1991}, $^{(c)}$: \citet{Munoz2015},  $^{(d)}$:\citet{Mieske2007}. The error estimations for each parameter are adopted from the tests made with the mock galaxies in section 5.4.2. The asterisks in the end of the object name indicates that it was fitted using a additional PSf component for the nucleus.} \\ 
\hline 
\hline 
Name & R.A. (J2000) & DEC (J2000) & $b/a$ & $\theta$ [deg] & m$_{r'}$ & R$_{e,r'}$ [arcsec]  & $n$ & g'-r'  & Ref \\ 
\hline 
\endfirsthead 
\caption{Continued.} \\ \hline 
Name & R.A. (J2000) & DEC (J2000) & $b/a$ & $\theta$ [deg] & m$_{r'}$ & R$_{e,r'}$ [arcsec] & $n$ & g'-r'  & Ref\\ 
\hline 
\endhead 
\hline 
\endfoot 
\hline 
\endlastfoot 
FDS10\_LSB2                              & 55.021 & -34.652 & 0.66$\pm$ 0.04 & 43.23 & 20.30 $\pm$ 0.16 & 4.4 $\pm$ 0.6 & 0.52 $\pm$ 0.17 & 0.33 $\pm$ 0.09 & \tablefootmark{(c)}  \\ 
FDS10\_LSB3                              & 54.945 & -34.687 & 0.54$\pm$ 0.09 & 78.32 & 21.96 $\pm$ 0.41 & 6.1 $\pm$ 1.3 & 0.23 $\pm$ 0.20 & 0.05 $\pm$ 0.11 &                      \\ 
FDS10\_LSB4                              & 54.977 & -34.696 & 0.86$\pm$ 0.03 & 44.55 & 19.68 $\pm$ 0.13 & 3.8 $\pm$ 0.4 & 0.62 $\pm$ 0.16 & 0.42 $\pm$ 0.09 & \tablefootmark{(c)}  \\ 
FDS10\_LSB5                              & 55.262 & -34.696 & 0.68$\pm$ 0.08 & 86.94 & 20.61 $\pm$ 0.37 & 9.0 $\pm$ 1.8 & 0.58 $\pm$ 0.20 & 0.05 $\pm$ 0.08 &                      \\ 
FDS10\_LSB6                              & 55.236 & -34.713 & 0.80$\pm$ 0.04 & 1.20 & 20.93 $\pm$ 0.16 & 2.9 $\pm$ 0.4 & 0.57 $\pm$ 0.17 & 0.52 $\pm$ 0.10 & \tablefootmark{(c)}  \\ 
FDS10\_LSB8                             $*$ & 55.002 & -34.723 & 0.79$\pm$ 0.05 & -36.59 & 19.86 $\pm$ 0.22 & 6.6 $\pm$ 1.0 & 1.67 $\pm$ 0.18 & 0.46 $\pm$ 0.08  & \tablefootmark{(c)}  \\ 
FDS10\_LSB9                              & 55.119 & -34.786 & 0.85$\pm$ 0.06 & 53.72 & 21.44 $\pm$ 0.27 & 4.0 $\pm$ 0.7 & 0.50 $\pm$ 0.19 & 0.28 $\pm$ 0.09 & \tablefootmark{(c)}  \\ 
FDS10\_LSB10                              & 55.109 & -35.003 & 0.48$\pm$ 0.04 & 59.29 & 20.05 $\pm$ 0.17 & 5.8 $\pm$ 0.8 & 0.96 $\pm$ 0.17 & 0.36 $\pm$ 0.11 & \tablefootmark{(c)}  \\ 
FDS10\_LSB11                              & 54.827 & -34.937 & 0.96$\pm$ 0.03 & -61.65 & 21.45 $\pm$ 0.11 & 1.4 $\pm$ 0.1 & 0.42 $\pm$ 0.16 & 0.39 $\pm$ 0.15 &                      \\ 
FDS10\_LSB12                              & 54.696 & -34.987 & 0.60$\pm$ 0.05 & 51.72 & 21.89 $\pm$ 0.20 & 2.8 $\pm$ 0.4 & 0.74 $\pm$ 0.18 & 0.05 $\pm$ 0.11 & \tablefootmark{(c)}  \\ 
FDS10\_LSB13                              & 54.594 & -34.927 & 0.58$\pm$ 0.05 & -55.11 & 21.22 $\pm$ 0.21 & 4.0 $\pm$ 0.6 & 0.27 $\pm$ 0.18 & 0.40 $\pm$ 0.09 & \tablefootmark{(c)}  \\ 
FDS10\_LSB14                              & 54.393 & -34.566 & 0.58$\pm$ 0.05 & 45.72 & 20.55 $\pm$ 0.20 & 5.0 $\pm$ 0.7 & 0.75 $\pm$ 0.18 & 0.29 $\pm$ 0.08 & \tablefootmark{(c)}  \\ 
FDS10\_LSB15                              & 54.496 & -34.530 & 0.72$\pm$ 0.05 & 46.66 & 20.21 $\pm$ 0.20 & 5.3 $\pm$ 0.8 & 0.53 $\pm$ 0.18 & 0.47 $\pm$ 0.08 &                      \\ 
FDS10\_LSB16                              & 54.602 & -34.649 & 0.67$\pm$ 0.04 & 56.08 & 20.63 $\pm$ 0.16 & 3.7 $\pm$ 0.5 & 1.29 $\pm$ 0.17 & 0.56 $\pm$ 0.09 &                      \\ 
FDS10\_LSB17                              & 54.631 & -34.774 & 0.60$\pm$ 0.04 & -56.43 & 21.03 $\pm$ 0.16 & 3.1 $\pm$ 0.4 & 0.89 $\pm$ 0.17 & 0.35 $\pm$ 0.10 & \tablefootmark{(c)}  \\ 
FDS10\_LSB20                              & 55.275 & -34.145 & 1.00$\pm$ 0.07 & -68.53 & 22.42 $\pm$ 0.28 & 2.4 $\pm$ 0.4 & 0.06 $\pm$ 0.19 & 0.80 $\pm$ 0.10 &                      \\ 
FDS10\_LSB21                              & 55.225 & -34.288 & 0.78$\pm$ 0.03 & 88.31 & 20.43 $\pm$ 0.12 & 2.7 $\pm$ 0.3 & 0.80 $\pm$ 0.16 & 0.18 $\pm$ 0.10 &                      \\ 
FDS10\_LSB23                              & 55.169 & -34.949 & 0.85$\pm$ 0.05 & 65.07 & 18.89 $\pm$ 0.18 & 8.2 $\pm$ 1.1 & 0.70 $\pm$ 0.17 & 0.49 $\pm$ 0.08 & \tablefootmark{(c)}  \\ 
FDS10\_LSB25                              & 54.959 & -35.023 & 0.69$\pm$ 0.05 & -28.43 & 17.26 $\pm$ 0.20 & 21.6 $\pm$ 3.1 & 0.40 $\pm$ 0.18 & 0.44 $\pm$ 0.09 & FCC226\tablefootmark{(a)}\tablefootmark{(b)}\tablefootmark{(c)} \\ 
FDS10\_LSB27                              & 55.042 & -34.447 & 0.56$\pm$ 0.03 & 55.06 & 16.64 $\pm$ 0.10 & 15.4 $\pm$ 1.6 & 0.91 $\pm$ 0.16 & 0.52 $\pm$ 0.08 & FCC234\tablefootmark{(a)} \\ 
FDS10\_LSB29                              & 55.019 & -34.168 & 0.96$\pm$ 0.03 & 2.73 & 17.89 $\pm$ 0.12 & 7.9 $\pm$ 0.9 & 0.93 $\pm$ 0.16 & 0.55 $\pm$ 0.08 & FCC231\tablefootmark{(a)} \\ 
FDS10\_LSB30                              & 54.943 & -34.135 & 0.85$\pm$ 0.03 & 18.23 & 20.48 $\pm$ 0.12 & 2.5 $\pm$ 0.3 & 0.81 $\pm$ 0.16 & 0.42 $\pm$ 0.10 &                      \\ 
FDS10\_LSB32                              & 54.659 & -33.898 & 0.83$\pm$ 0.03 & -50.83 & 20.29 $\pm$ 0.12 & 2.8 $\pm$ 0.3 & 0.62 $\pm$ 0.16 & 0.17 $\pm$ 0.10 &                      \\ 
FDS10\_LSB33                              & 54.780 & -34.061 & 0.72$\pm$ 0.04 & -76.23 & 21.19 $\pm$ 0.15 & 2.6 $\pm$ 0.3 & 0.32 $\pm$ 0.17 & 0.96 $\pm$ 0.11 &                      \\ 
FDS10\_LSB34                              & 54.723 & -34.171 & 0.92$\pm$ 0.04 & 80.84 & 21.34 $\pm$ 0.14 & 1.9 $\pm$ 0.2 & 0.60 $\pm$ 0.17 & 0.66 $\pm$ 0.12 &                      \\ 
FDS10\_LSB35                              & 54.777 & -34.264 & 0.76$\pm$ 0.02 & -57.05 & 18.96 $\pm$ 0.08 & 3.6 $\pm$ 0.3 & 0.71 $\pm$ 0.15 & 0.48 $\pm$ 0.09 &                      \\ 
FDS10\_LSB36                              & 54.824 & -34.306 & 0.74$\pm$ 0.03 & -80.59 & 20.63 $\pm$ 0.11 & 2.2 $\pm$ 0.2 & 0.87 $\pm$ 0.16 & 0.49 $\pm$ 0.12 &                      \\ 
FDS10\_LSB37                              & 54.761 & -34.577 & 0.65$\pm$ 0.03 & -64.92 & 20.10 $\pm$ 0.09 & 2.6 $\pm$ 0.2 & 0.46 $\pm$ 0.16 & 0.91 $\pm$ 0.11 &                      \\ 
FDS10\_LSB38                              & 54.726 & -34.826 & 0.70$\pm$ 0.06 & 0.40 & 19.34 $\pm$ 0.23 & 9.3 $\pm$ 1.4 & 1.15 $\pm$ 0.18 & 0.56 $\pm$ 0.07 & \tablefootmark{(c)}  \\ 
FDS10\_LSB39                              & 54.619 & -34.690 & 0.91$\pm$ 0.03 & 15.84 & 20.65 $\pm$ 0.10 & 1.9 $\pm$ 0.2 & 0.68 $\pm$ 0.16 & 0.40 $\pm$ 0.12 &                      \\ 
FDS10\_LSB40                              & 54.557 & -34.229 & 0.34$\pm$ 0.06 & -4.54 & 20.70 $\pm$ 0.27 & 8.6 $\pm$ 1.4 & 0.68 $\pm$ 0.19 & 0.21 $\pm$ 0.08 &                      \\ 
FDS10\_LSB41                              & 54.533 & -33.980 & 0.75$\pm$ 0.04 & 19.24 & 18.98 $\pm$ 0.15 & 6.6 $\pm$ 0.8 & 0.94 $\pm$ 0.17 & 0.42 $\pm$ 0.08 &                      \\ 
FDS10\_LSB43                              & 54.379 & -34.096 & 0.93$\pm$ 0.05 & -26.15 & 20.34 $\pm$ 0.21 & 4.6 $\pm$ 0.7 & 1.16 $\pm$ 0.18 & 0.36 $\pm$ 0.08 &                      \\ 
FDS10\_LSB44                              & 54.374 & -34.225 & 0.81$\pm$ 0.04 & -86.34 & 20.22 $\pm$ 0.13 & 3.2 $\pm$ 0.4 & 0.78 $\pm$ 0.17 & 0.46 $\pm$ 0.09 &                      \\ 
FDS10\_LSB45                              & 54.344 & -34.287 & 0.79$\pm$ 0.04 & -44.37 & 19.32 $\pm$ 0.13 & 5.0 $\pm$ 0.6 & 0.50 $\pm$ 0.17 & 0.46 $\pm$ 0.08 &                      \\ 
FDS10\_LSB46                              & 54.317 & -34.271 & 0.71$\pm$ 0.03 & -12.62 & 20.59 $\pm$ 0.11 & 2.5 $\pm$ 0.3 & 0.76 $\pm$ 0.16 & 0.39 $\pm$ 0.11 &                      \\ 
FDS10\_LSB47                              & 54.289 & -34.291 & 0.69$\pm$ 0.02 & -44.16 & 19.81 $\pm$ 0.08 & 2.4 $\pm$ 0.2 & 0.75 $\pm$ 0.15 & 0.22 $\pm$ 0.11 &                      \\ 
FDS10\_LSB49                              & 54.261 & -34.876 & 0.95$\pm$ 0.04 & 11.19 & 19.21 $\pm$ 0.13 & 4.7 $\pm$ 0.5 & 0.78 $\pm$ 0.17 & 0.47 $\pm$ 0.08 & FCC185\tablefootmark{(a)} \\ 
FDS10\_LSB51                              & 54.001 & -34.878 & 0.99$\pm$ 0.05 & -15.58 & 19.38 $\pm$ 0.21 & 7.2 $\pm$ 1.1 & 0.54 $\pm$ 0.18 & 0.46 $\pm$ 0.09 & \tablefootmark{(c)}  \\ 
FDS10\_LSB52                              & 53.982 & -34.828 & 0.51$\pm$ 0.04 & 79.41 & 17.52 $\pm$ 0.14 & 15.4 $\pm$ 1.9 & 0.79 $\pm$ 0.17 & 0.47 $\pm$ 0.08 & FCC159\tablefootmark{(a)}\tablefootmark{(c)} \\ 
FDS10\_LSB53                              & 54.174 & -34.711 & 0.61$\pm$ 0.04 & -2.59 & 19.98 $\pm$ 0.16 & 5.1 $\pm$ 0.6 & 0.78 $\pm$ 0.17 & 0.40 $\pm$ 0.08 & \tablefootmark{(c)}  \\ 
FDS10\_LSB54                              & 54.118 & -34.755 & 0.30$\pm$ 0.04 & 16.81 & 21.31 $\pm$ 0.14 & 3.5 $\pm$ 0.4 & 0.71 $\pm$ 0.17 & 0.66 $\pm$ 0.11 &                      \\ 
FDS10\_LSB55                              & 53.988 & -34.632 & 0.55$\pm$ 0.03 & -18.01 & 18.77 $\pm$ 0.11 & 6.2 $\pm$ 0.7 & 0.92 $\pm$ 0.16 & 0.51 $\pm$ 0.08 & \tablefootmark{(c)}  \\ 
FDS10\_LSB56                              & 54.083 & -34.594 & 0.55$\pm$ 0.07 & -67.20 & 21.64 $\pm$ 0.28 & 4.6 $\pm$ 0.8 & 0.25 $\pm$ 0.19 & 0.26 $\pm$ 0.09 & \tablefootmark{(c)}  \\ 
FDS10\_LSB57                              & 53.970 & -34.243 & 0.64$\pm$ 0.03 & 86.22 & 20.79 $\pm$ 0.13 & 2.7 $\pm$ 0.3 & 0.25 $\pm$ 0.16 & 0.26 $\pm$ 0.11 &                      \\ 
FDS11\_LSB1                              & 55.246 & -35.697 & 0.24$\pm$ 0.08 & 29.92 & 16.19 $\pm$ 0.37 & 116.4 $\pm$ 22.9 & 0.55 $\pm$ 0.20 & 0.18 $\pm$ 0.07 &                      \\ 
FDS11\_LSB2                              & 55.148 & -35.774 & 0.48$\pm$ 0.05 & 40.76 & 15.55 $\pm$ 0.22 & 63.9 $\pm$ 9.7 & 0.85 $\pm$ 0.18 & 0.99 $\pm$ 0.14 &                      \\ 
FDS11\_LSB4                              & 55.066 & -35.662 & 0.91$\pm$ 0.05 & -72.81 & 20.26 $\pm$ 0.19 & 4.4 $\pm$ 0.6 & 0.53 $\pm$ 0.18 & 0.33 $\pm$ 0.08 & \tablefootmark{(c)}  \\ 
FDS11\_LSB6                              & 54.711 & -35.725 & 0.71$\pm$ 0.07 & -8.23 & 21.19 $\pm$ 0.29 & 5.2 $\pm$ 0.9 & 0.62 $\pm$ 0.19 & 0.53 $\pm$ 0.08 &                      \\ 
FDS11\_LSB7                              & 54.035 & -36.000 & 0.77$\pm$ 0.07 & 54.15 & 20.74 $\pm$ 0.29 & 6.2 $\pm$ 1.1 & 0.94 $\pm$ 0.19 & 0.79 $\pm$ 0.08 &                      \\ 
FDS11\_LSB8                              & 54.001 & -35.936 & 0.78$\pm$ 0.07 & -61.15 & 21.23 $\pm$ 0.30 & 5.0 $\pm$ 0.9 & 0.56 $\pm$ 0.19 & 0.44 $\pm$ 0.08 & \tablefootmark{(c)}  \\ 
FDS11\_LSB10                              & 54.105 & -35.703 & 0.58$\pm$ 0.05 & -74.41 & 19.52 $\pm$ 0.18 & 7.5 $\pm$ 1.0 & 0.93 $\pm$ 0.18 & 0.60 $\pm$ 0.08 & \tablefootmark{(c)}  \\ 
FDS11\_LSB11                              & 54.161 & -35.610 & 0.81$\pm$ 0.06 & 61.68 & 20.63 $\pm$ 0.24 & 5.2 $\pm$ 0.8 & 0.48 $\pm$ 0.18 & 0.51 $\pm$ 0.08 & \tablefootmark{(c)}  \\ 
FDS11\_LSB12                              & 55.360 & -35.380 & 0.88$\pm$ 0.04 & 78.68 & 21.09 $\pm$ 0.17 & 2.7 $\pm$ 0.3 & 0.45 $\pm$ 0.17 & 0.50 $\pm$ 0.10 & \tablefootmark{(c)}  \\ 
FDS11\_LSB13                              & 55.112 & -35.419 & 0.81$\pm$ 0.05 & -79.83 & 20.36 $\pm$ 0.20 & 4.7 $\pm$ 0.7 & 0.69 $\pm$ 0.18 & 0.37 $\pm$ 0.08 & \tablefootmark{(c)}  \\ 
FDS11\_LSB14                              & 55.154 & -35.356 & 0.83$\pm$ 0.07 & -45.11 & 19.55 $\pm$ 0.30 & 10.7 $\pm$ 1.9 & 0.60 $\pm$ 0.19 & 0.34 $\pm$ 0.08 & \tablefootmark{(b)}\tablefootmark{(c)} \\ 
FDS11\_LSB15                              & 55.220 & -35.118 & 0.93$\pm$ 0.04 & 39.58 & 19.06 $\pm$ 0.17 & 6.7 $\pm$ 0.9 & 0.86 $\pm$ 0.17 & 0.46 $\pm$ 0.08 & \tablefootmark{(c)}  \\ 
FDS11\_LSB16                              & 54.615 & -35.040 & 0.80$\pm$ 0.07 & -26.87 & 18.73 $\pm$ 0.30 & 15.5 $\pm$ 2.7 & 0.48 $\pm$ 0.19 & 0.56 $\pm$ 0.09 &                      \\ 
FDS11\_LSB17                              & 54.510 & -35.281 & 0.30$\pm$ 0.09 & 50.31 & 17.22 $\pm$ 0.43 & 75.7 $\pm$ 16.1 & 1.00 $\pm$ 0.20 & 0.54 $\pm$ 0.09 &                      \\ 
FDS11\_LSB18                             $*$ & 54.249 & -35.138 & 0.73$\pm$ 0.06 & -81.05 & 20.51 $\pm$ 0.24 & 5.6 $\pm$ 0.9 & 0.49 $\pm$ 0.18 & 0.27 $\pm$ 0.11  & \tablefootmark{(c)}  \\ 
FDS11\_LSB30                             $*$ & 54.155 & -35.385 & 0.91$\pm$ 0.05 & -78.25 & 17.76 $\pm$ 0.19 & 13.8 $\pm$ 1.9 & 1.15 $\pm$ 0.18 & 0.66 $\pm$ 0.13  & FCC171\tablefootmark{(a)}\tablefootmark{(c)} \\ 
FDS11\_LSB32                              & 54.179 & -35.436 & 0.76$\pm$ 0.03 & -0.30 & 17.96 $\pm$ 0.12 & 8.3 $\pm$ 0.9 & 0.67 $\pm$ 0.16 & 0.54 $\pm$ 0.15 & FCC175\tablefootmark{(a)} \\ 
FDS11\_LSB35                              & 54.537 & -35.165 & 0.78$\pm$ 0.05 & 26.45 & 20.12 $\pm$ 0.22 & 6.2 $\pm$ 0.9 & 0.37 $\pm$ 0.18 & 0.82 $\pm$ 0.10 & \tablefootmark{(c)}  \\ 
FDS11\_LSB36                              & 55.078 & -35.138 & 0.75$\pm$ 0.07 & -15.66 & 20.84 $\pm$ 0.31 & 6.4 $\pm$ 1.1 & 0.49 $\pm$ 0.19 & 0.36 $\pm$ 0.08 & \tablefootmark{(b)}\tablefootmark{(c)}\tablefootmark{(d)} \\ 
FDS11\_LSB38                             $*$ & 55.304 & -35.157 & 0.85$\pm$ 0.03 & -69.87 & 16.20 $\pm$ 0.10 & 15.2 $\pm$ 1.6 & 0.72 $\pm$ 0.16 & 0.60 $\pm$ 0.08  & FCC260\tablefootmark{(a)}\tablefootmark{(b)}\tablefootmark{(c)} \\ 
FDS11\_LSB39                              & 55.305 & -35.335 & 0.66$\pm$ 0.07 & 68.99 & 21.88 $\pm$ 0.29 & 3.9 $\pm$ 0.7 & 0.07 $\pm$ 0.19 & 0.61 $\pm$ 0.09 &                      \\ 
FDS11\_LSB40                              & 55.208 & -35.389 & 0.61$\pm$ 0.06 & -69.42 & 22.02 $\pm$ 0.25 & 3.3 $\pm$ 0.5 & 0.33 $\pm$ 0.18 & 0.27 $\pm$ 0.10 & \tablefootmark{(c)}  \\ 
FDS11\_LSB41                              & 54.964 & -35.323 & 0.85$\pm$ 0.04 & -21.88 & 18.30 $\pm$ 0.15 & 8.8 $\pm$ 1.1 & 0.74 $\pm$ 0.17 & 0.53 $\pm$ 0.08 & \tablefootmark{(b)}\tablefootmark{(c)}\tablefootmark{(d)} \\ 
FDS11\_LSB42                              & 55.013 & -35.464 & 0.82$\pm$ 0.06 & -85.27 & 19.30 $\pm$ 0.27 & 10.8 $\pm$ 1.8 & 0.66 $\pm$ 0.19 & 0.52 $\pm$ 0.08 & \tablefootmark{(b)}\tablefootmark{(c)}\tablefootmark{(d)} \\ 
FDS11\_LSB43                              & 54.926 & -35.469 & 0.77$\pm$ 0.07 & 67.40 & 21.66 $\pm$ 0.31 & 4.3 $\pm$ 0.8 & 0.32 $\pm$ 0.19 & 0.27 $\pm$ 0.08 & \tablefootmark{(c)}\tablefootmark{(d)} \\ 
FDS11\_LSB44                              & 54.463 & -35.364 & 0.69$\pm$ 0.06 & 7.39 & 21.15 $\pm$ 0.25 & 4.6 $\pm$ 0.7 & 0.47 $\pm$ 0.18 & 0.49 $\pm$ 0.08 & \tablefootmark{(c)}\tablefootmark{(d)} \\ 
FDS11\_LSB45                              & 54.412 & -35.386 & 0.81$\pm$ 0.05 & 0.62 & 19.70 $\pm$ 0.22 & 7.2 $\pm$ 1.1 & 1.11 $\pm$ 0.18 & 0.38 $\pm$ 0.08 & \tablefootmark{(c)}\tablefootmark{(d)} \\ 
FDS11\_LSB46                              & 54.316 & -35.357 & 0.55$\pm$ 0.06 & 11.75 & 20.60 $\pm$ 0.27 & 7.1 $\pm$ 1.2 & 0.68 $\pm$ 0.19 & 0.46 $\pm$ 0.08 & \tablefootmark{(c)}\tablefootmark{(d)} \\ 
FDS11\_LSB47                             $*$ & 54.250 & -35.343 & 0.81$\pm$ 0.05 & -85.59 & 18.40 $\pm$ 0.19 & 11.0 $\pm$ 1.5 & 0.52 $\pm$ 0.18 & 0.64 $\pm$ 0.08  & \tablefootmark{(c)}\tablefootmark{(d)} \\ 
FDS11\_LSB48                              & 54.007 & -35.310 & 0.76$\pm$ 0.05 & 78.70 & 21.02 $\pm$ 0.18 & 3.3 $\pm$ 0.4 & 0.85 $\pm$ 0.18 & 0.12 $\pm$ 0.11 & \tablefootmark{(c)}\tablefootmark{(d)} \\ 
FDS11\_LSB49                              & 53.930 & -35.514 & 0.83$\pm$ 0.04 & 28.68 & 17.57 $\pm$ 0.16 & 13.7 $\pm$ 1.8 & 0.96 $\pm$ 0.17 & 0.58 $\pm$ 0.08 & FCC157\tablefootmark{(a)}\tablefootmark{(b)}\tablefootmark{(c)} \\ 
FDS11\_LSB50                              & 54.099 & -35.523 & 0.72$\pm$ 0.05 & -12.94 & 22.32 $\pm$ 0.20 & 2.0 $\pm$ 0.3 & 0.33 $\pm$ 0.18 & 0.38 $\pm$ 0.13 & \tablefootmark{(c)}  \\ 
FDS11\_LSB51                              & 54.170 & -35.589 & 0.43$\pm$ 0.08 & 30.21 & 21.10 $\pm$ 0.35 & 8.7 $\pm$ 1.7 & 0.40 $\pm$ 0.20 & 0.17 $\pm$ 0.08 & \tablefootmark{(d)}  \\ 
FDS11\_LSB53                              & 54.437 & -35.704 & 1.00$\pm$ 0.04 & 13.73 & 20.50 $\pm$ 0.16 & 3.1 $\pm$ 0.4 & 0.48 $\pm$ 0.17 & 0.50 $\pm$ 0.09 & \tablefootmark{(c)}\tablefootmark{(d)} \\ 
FDS11\_LSB55                              & 54.871 & -35.573 & 0.71$\pm$ 0.07 & -19.33 & 19.94 $\pm$ 0.28 & 8.9 $\pm$ 1.5 & 0.70 $\pm$ 0.19 & 0.41 $\pm$ 0.08 & \tablefootmark{(c)}\tablefootmark{(d)} \\ 
FDS11\_LSB56                              & 54.923 & -35.532 & 0.75$\pm$ 0.05 & -61.01 & 20.16 $\pm$ 0.20 & 5.4 $\pm$ 0.8 & 0.79 $\pm$ 0.18 & 0.45 $\pm$ 0.08 & \tablefootmark{(c)}\tablefootmark{(d)} \\ 
FDS11\_LSB57                             $*$ & 54.958 & -35.522 & 0.97$\pm$ 0.04 & -56.59 & 19.01 $\pm$ 0.15 & 6.1 $\pm$ 0.8 & 0.87 $\pm$ 0.17 & 0.57 $\pm$ 0.08  & FCC227\tablefootmark{(a)}\tablefootmark{(c)} \\ 
FDS11\_LSB58                             $*$ & 54.986 & -35.622 & 0.54$\pm$ 0.06 & -57.40 & 20.27 $\pm$ 0.25 & 7.7 $\pm$ 1.2 & 0.46 $\pm$ 0.18 & 0.36 $\pm$ 0.08  & \tablefootmark{(c)}\tablefootmark{(d)} \\ 
FDS11\_LSB59                              & 54.982 & -35.662 & 0.73$\pm$ 0.04 & -85.93 & 18.34 $\pm$ 0.15 & 9.1 $\pm$ 1.1 & 0.59 $\pm$ 0.17 & 0.58 $\pm$ 0.08 & FCC229\tablefootmark{(a)}\tablefootmark{(c)} \\ 
FDS11\_LSB60                              & 55.280 & -35.515 & 0.61$\pm$ 0.03 & -83.68 & 17.76 $\pm$ 0.13 & 11.1 $\pm$ 1.3 & 0.94 $\pm$ 0.17 & 0.58 $\pm$ 0.08 & FCC259\tablefootmark{(a)}\tablefootmark{(b)} \\ 
FDS11\_LSB61                              & 55.279 & -35.690 & 0.82$\pm$ 0.04 & -72.47 & 19.50 $\pm$ 0.17 & 6.0 $\pm$ 0.8 & 0.60 $\pm$ 0.17 & 0.46 $\pm$ 0.08 & FCC258\tablefootmark{(a)}\tablefootmark{(c)} \\ 
FDS11\_LSB62                             $*$ & 55.253 & -35.743 & 0.90$\pm$ 0.03 & -54.30 & 17.01 $\pm$ 0.13 & 12.9 $\pm$ 1.5 & 0.55 $\pm$ 0.16 & 0.58 $\pm$ 0.08  & FCC254\tablefootmark{(a)}\tablefootmark{(c)} \\ 
FDS11\_LSB63                              & 55.141 & -35.888 & 0.65$\pm$ 0.05 & -75.69 & 20.38 $\pm$ 0.21 & 5.7 $\pm$ 0.9 & 0.55 $\pm$ 0.18 & 0.46 $\pm$ 0.08 & \tablefootmark{(c)}  \\ 
FDS11\_LSB64                              & 54.747 & -35.741 & 0.88$\pm$ 0.03 & 8.90 & 20.39 $\pm$ 0.12 & 2.6 $\pm$ 0.3 & 1.15 $\pm$ 0.16 & 0.57 $\pm$ 0.10 & \tablefootmark{(c)}  \\ 
FDS11\_LSB65                             $*$ & 54.657 & -35.757 & 0.75$\pm$ 0.04 & -34.75 & 18.92 $\pm$ 0.17 & 8.2 $\pm$ 1.1 & 0.39 $\pm$ 0.17 & 0.59 $\pm$ 0.09  & FCC215\tablefootmark{(a)}\tablefootmark{(c)} \\ 
FDS11\_LSB66                              & 54.226 & -35.747 & 0.52$\pm$ 0.05 & -29.25 & 20.73 $\pm$ 0.19 & 4.9 $\pm$ 0.7 & 0.53 $\pm$ 0.18 & 0.41 $\pm$ 0.09 & \tablefootmark{(c)}\tablefootmark{(d)} \\ 
FDS11\_LSB67                             $*$ & 54.264 & -35.801 & 0.86$\pm$ 0.04 & 35.84 & 19.71 $\pm$ 0.17 & 5.1 $\pm$ 0.7 & 1.10 $\pm$ 0.17 & 0.40 $\pm$ 0.08  & \tablefootmark{(c)}\tablefootmark{(d)} \\ 
FDS11\_LSB68                              & 54.293 & -35.888 & 0.71$\pm$ 0.05 & 24.16 & 19.12 $\pm$ 0.18 & 7.8 $\pm$ 1.1 & 1.04 $\pm$ 0.17 & 0.57 $\pm$ 0.08 & FCC192\tablefootmark{(a)}\tablefootmark{(c)} \\ 
FDS11\_LSB69                              & 54.029 & -35.843 & 0.61$\pm$ 0.05 & 83.86 & 18.43 $\pm$ 0.21 & 14.2 $\pm$ 2.1 & 0.71 $\pm$ 0.18 & 0.46 $\pm$ 0.08 & FCC163\tablefootmark{(a)}\tablefootmark{(c)} \\ 
FDS11\_LSB70                              & 54.229 & -35.944 & 0.90$\pm$ 0.05 & -81.87 & 21.95 $\pm$ 0.19 & 2.0 $\pm$ 0.3 & 0.47 $\pm$ 0.18 & 0.28 $\pm$ 0.12 & \tablefootmark{(c)}\tablefootmark{(d)} \\ 
FDS11\_LSB71                              & 54.365 & -35.963 & 0.92$\pm$ 0.05 & 40.42 & 19.91 $\pm$ 0.20 & 5.3 $\pm$ 0.8 & 0.96 $\pm$ 0.18 & 0.48 $\pm$ 0.09 & \tablefootmark{(c)}\tablefootmark{(d)} \\ 
FDS11\_LSB72                              & 54.425 & -35.956 & 0.46$\pm$ 0.05 & -13.90 & 19.67 $\pm$ 0.21 & 9.2 $\pm$ 1.4 & 0.77 $\pm$ 0.18 & 0.33 $\pm$ 0.09 & \tablefootmark{(c)}\tablefootmark{(d)} \\ 
FDS11\_LSB73                              & 54.535 & -36.008 & 0.81$\pm$ 0.06 & -88.59 & 21.22 $\pm$ 0.23 & 3.8 $\pm$ 0.6 & 0.75 $\pm$ 0.18 & 0.50 $\pm$ 0.09 & \tablefootmark{(c)}\tablefootmark{(d)} \\ 
FDS11\_LSB74                              & 55.176 & -35.661 & 0.64$\pm$ 0.03 & -38.83 & 17.37 $\pm$ 0.10 & 10.2 $\pm$ 1.0 & 0.69 $\pm$ 0.16 & 0.27 $\pm$ 0.07 & FCC247\tablefootmark{(a)}\tablefootmark{(c)} \\ 
FDS11\_LSB76                              & 55.052 & -35.289 & 0.62$\pm$ 0.07 & 22.78 & 21.20 $\pm$ 0.28 & 5.4 $\pm$ 0.9 & 0.36 $\pm$ 0.19 & 0.19 $\pm$ 0.08 & \tablefootmark{(c)}  \\ 
FDS11\_LSB77                              & 54.653 & -35.834 & 0.91$\pm$ 0.03 & 14.21 & 18.63 $\pm$ 0.11 & 5.2 $\pm$ 0.5 & 0.73 $\pm$ 0.16 & 0.57 $\pm$ 0.08 & FCC214\tablefootmark{(a)}\tablefootmark{(c)} \\ 
FDS11\_LSB78                             $*$ & 54.392 & -35.829 & 0.52$\pm$ 0.03 & 66.83 & 17.11 $\pm$ 0.09 & 11.7 $\pm$ 1.1 & 0.82 $\pm$ 0.16 & 0.60 $\pm$ 0.07  & FCC196\tablefootmark{(a)}\tablefootmark{(c)} \\ 
FDS11\_LSB79                              & 54.292 & -35.387 & 0.83$\pm$ 0.03 & -78.41 & 18.86 $\pm$ 0.13 & 5.6 $\pm$ 0.6 & 0.70 $\pm$ 0.16 & 0.59 $\pm$ 0.08 & FCC191\tablefootmark{(a)}\tablefootmark{(c)} \\ 
FDS11\_LSB80                              & 54.422 & -35.296 & 0.82$\pm$ 0.03 & 82.65 & 18.57 $\pm$ 0.13 & 6.5 $\pm$ 0.7 & 0.82 $\pm$ 0.16 & 0.47 $\pm$ 0.08 & FCC197\tablefootmark{(a)}\tablefootmark{(c)} \\ 
FDS11\_LSB81                              & 54.116 & -35.211 & 0.88$\pm$ 0.04 & 25.14 & 18.20 $\pm$ 0.13 & 7.9 $\pm$ 0.9 & 0.84 $\pm$ 0.17 & 0.57 $\pm$ 0.07 & FCC168\tablefootmark{(a)}\tablefootmark{(c)} \\ 
FDS12\_LSB2                              & 55.300 & -36.069 & 0.71$\pm$ 0.04 & 49.87 & 21.23 $\pm$ 0.17 & 2.9 $\pm$ 0.4 & 0.50 $\pm$ 0.17 & 0.50 $\pm$ 0.10 & \tablefootmark{(c)}  \\ 
FDS12\_LSB3                              & 55.184 & -36.185 & 0.79$\pm$ 0.05 & 20.40 & 17.86 $\pm$ 0.22 & 16.8 $\pm$ 2.5 & 0.96 $\pm$ 0.18 & 0.65 $\pm$ 0.09 &                      \\ 
FDS12\_LSB4                              & 55.157 & -36.121 & 0.65$\pm$ 0.04 & -45.70 & 18.22 $\pm$ 0.14 & 9.8 $\pm$ 1.2 & 0.83 $\pm$ 0.17 & 0.59 $\pm$ 0.08 & FCC246\tablefootmark{(a)}\tablefootmark{(c)} \\ 
FDS12\_LSB5                              & 55.138 & -36.580 & 0.62$\pm$ 0.07 & -53.68 & 21.53 $\pm$ 0.32 & 5.3 $\pm$ 1.0 & 0.45 $\pm$ 0.19 & 0.13 $\pm$ 0.08 &                      \\ 
FDS12\_LSB6                              & 55.137 & -36.555 & 0.71$\pm$ 0.08 & 61.44 & 22.18 $\pm$ 0.36 & 4.2 $\pm$ 0.8 & 0.35 $\pm$ 0.20 & 0.54 $\pm$ 0.09 &                      \\ 
FDS12\_LSB8                              & 55.061 & -36.613 & 0.85$\pm$ 0.06 & 25.33 & 21.44 $\pm$ 0.23 & 3.4 $\pm$ 0.5 & 0.43 $\pm$ 0.18 & 0.64 $\pm$ 0.09 &                      \\ 
FDS12\_LSB9                             $*$ & 55.029 & -36.560 & 0.80$\pm$ 0.04 & 28.65 & 18.25 $\pm$ 0.15 & 9.5 $\pm$ 1.2 & 0.53 $\pm$ 0.17 & 0.52 $\pm$ 0.08  &                      \\ 
FDS12\_LSB10                             $*$ & 55.071 & -36.535 & 0.71$\pm$ 0.04 & 19.45 & 18.14 $\pm$ 0.15 & 10.1 $\pm$ 1.2 & 0.64 $\pm$ 0.17 & 0.59 $\pm$ 0.08  & FCC238\tablefootmark{(a)}\tablefootmark{(c)} \\ 
FDS12\_LSB11                              & 54.907 & -36.553 & 0.58$\pm$ 0.04 & -64.51 & 19.27 $\pm$ 0.14 & 6.4 $\pm$ 0.8 & 0.59 $\pm$ 0.17 & 0.59 $\pm$ 0.10 & FCC225\tablefootmark{(a)} \\ 
FDS12\_LSB12                             $*$ & 55.081 & -36.314 & 0.76$\pm$ 0.06 & -14.31 & 19.40 $\pm$ 0.24 & 9.5 $\pm$ 1.5 & 0.99 $\pm$ 0.18 & 0.50 $\pm$ 0.07  & \tablefootmark{(c)}  \\ 
FDS12\_LSB13                              & 55.025 & -36.234 & 0.39$\pm$ 0.04 & 26.28 & 18.50 $\pm$ 0.17 & 13.5 $\pm$ 1.8 & 0.78 $\pm$ 0.17 & 0.44 $\pm$ 0.08 & FCC233\tablefootmark{(a)}\tablefootmark{(c)} \\ 
FDS12\_LSB14                              & 54.969 & -36.142 & 0.58$\pm$ 0.04 & 86.51 & 21.11 $\pm$ 0.17 & 3.3 $\pm$ 0.4 & 0.53 $\pm$ 0.17 & 0.65 $\pm$ 0.10 & \tablefootmark{(c)}  \\ 
FDS12\_LSB16                              & 55.001 & -36.094 & 0.63$\pm$ 0.06 & -33.54 & 20.23 $\pm$ 0.23 & 6.8 $\pm$ 1.1 & 0.97 $\pm$ 0.18 & 0.49 $\pm$ 0.08 & \tablefootmark{(c)}  \\ 
FDS12\_LSB17                              & 54.885 & -36.099 & 0.94$\pm$ 0.05 & 85.27 & 20.02 $\pm$ 0.22 & 5.7 $\pm$ 0.9 & 0.80 $\pm$ 0.18 & 0.54 $\pm$ 0.08 & \tablefootmark{(c)}  \\ 
FDS12\_LSB18                              & 55.076 & -36.059 & 0.68$\pm$ 0.07 & 19.50 & 21.93 $\pm$ 0.30 & 4.0 $\pm$ 0.7 & 0.89 $\pm$ 0.19 & 0.51 $\pm$ 0.09 & \tablefootmark{(c)}  \\ 
FDS12\_LSB19                              & 54.963 & -36.013 & 0.87$\pm$ 0.05 & 18.17 & 21.23 $\pm$ 0.22 & 3.5 $\pm$ 0.5 & 0.69 $\pm$ 0.18 & 0.50 $\pm$ 0.09 & \tablefootmark{(c)}  \\ 
FDS12\_LSB20                              & 54.655 & -35.917 & 0.56$\pm$ 0.03 & -27.62 & 19.53 $\pm$ 0.12 & 4.8 $\pm$ 0.5 & 0.71 $\pm$ 0.16 & 0.55 $\pm$ 0.09 & \tablefootmark{(c)}\tablefootmark{(d)} \\ 
FDS12\_LSB21                              & 54.754 & -36.338 & 0.59$\pm$ 0.03 & -38.17 & 18.96 $\pm$ 0.11 & 5.6 $\pm$ 0.6 & 0.91 $\pm$ 0.16 & 0.55 $\pm$ 0.08 & \tablefootmark{(c)}  \\ 
FDS12\_LSB22                              & 54.664 & -36.558 & 0.81$\pm$ 0.04 & -42.76 & 18.63 $\pm$ 0.15 & 7.5 $\pm$ 0.9 & 0.66 $\pm$ 0.17 & 0.59 $\pm$ 0.08 & FCC216\tablefootmark{(a)} \\ 
FDS12\_LSB23                              & 54.723 & -36.614 & 0.87$\pm$ 0.05 & 64.90 & 20.24 $\pm$ 0.22 & 5.3 $\pm$ 0.8 & 0.53 $\pm$ 0.18 & 0.65 $\pm$ 0.08 &                      \\ 
FDS12\_LSB24                              & 54.854 & -36.622 & 0.69$\pm$ 0.07 & 10.07 & 21.11 $\pm$ 0.30 & 5.6 $\pm$ 1.0 & 0.74 $\pm$ 0.19 & 0.34 $\pm$ 0.08 &                      \\ 
FDS12\_LSB25                              & 54.765 & -36.560 & 0.84$\pm$ 0.07 & -3.23 & 21.55 $\pm$ 0.28 & 4.0 $\pm$ 0.7 & 0.90 $\pm$ 0.19 & 0.29 $\pm$ 0.09 &                      \\ 
FDS12\_LSB26                              & 54.673 & -36.727 & 0.96$\pm$ 0.04 & -27.40 & 19.08 $\pm$ 0.13 & 5.1 $\pm$ 0.6 & 1.07 $\pm$ 0.17 & 0.59 $\pm$ 0.08 & FCC217\tablefootmark{(a)} \\ 
FDS12\_LSB27                              & 54.733 & -36.662 & 0.58$\pm$ 0.04 & 71.91 & 21.46 $\pm$ 0.15 & 2.4 $\pm$ 0.3 & 0.45 $\pm$ 0.17 & 0.39 $\pm$ 0.12 &                      \\ 
FDS12\_LSB28                              & 54.605 & -36.890 & 0.78$\pm$ 0.03 & 74.44 & 19.13 $\pm$ 0.10 & 4.2 $\pm$ 0.4 & 1.33 $\pm$ 0.16 & 0.29 $\pm$ 0.09 &                      \\ 
FDS12\_LSB29                              & 54.432 & -36.726 & 0.83$\pm$ 0.04 & -79.16 & 18.42 $\pm$ 0.13 & 7.4 $\pm$ 0.9 & 0.76 $\pm$ 0.17 & 0.52 $\pm$ 0.09 & FCC199\tablefootmark{(a)} \\ 
FDS12\_LSB30                              & 54.587 & -36.412 & 0.76$\pm$ 0.04 & -7.53 & 16.94 $\pm$ 0.17 & 19.2 $\pm$ 2.5 & 0.72 $\pm$ 0.17 & 0.58 $\pm$ 0.08 & FCC212\tablefootmark{(a)} \\ 
FDS12\_LSB31                              & 54.638 & -36.441 & 0.73$\pm$ 0.07 & -5.50 & 21.71 $\pm$ 0.30 & 4.3 $\pm$ 0.8 & 0.84 $\pm$ 0.19 & 0.92 $\pm$ 0.09 &                      \\ 
FDS12\_LSB32                              & 54.546 & -36.337 & 0.74$\pm$ 0.05 & -61.14 & 20.81 $\pm$ 0.22 & 4.5 $\pm$ 0.7 & 0.38 $\pm$ 0.18 & 0.48 $\pm$ 0.08 & \tablefootmark{(c)}  \\ 
FDS12\_LSB33                              & 54.564 & -36.281 & 0.72$\pm$ 0.06 & 28.27 & 20.41 $\pm$ 0.26 & 6.6 $\pm$ 1.1 & 0.54 $\pm$ 0.19 & 0.45 $\pm$ 0.08 & \tablefootmark{(c)}  \\ 
FDS12\_LSB34                              & 54.420 & -36.174 & 0.77$\pm$ 0.08 & 30.46 & 19.44 $\pm$ 0.38 & 14.8 $\pm$ 3.0 & 1.19 $\pm$ 0.20 & 0.48 $\pm$ 0.08 & \tablefootmark{(c)}  \\ 
FDS12\_LSB35                              & 54.526 & -36.050 & 0.91$\pm$ 0.04 & 86.99 & 20.06 $\pm$ 0.16 & 4.1 $\pm$ 0.5 & 0.76 $\pm$ 0.17 & 0.58 $\pm$ 0.08 & \tablefootmark{(c)}\tablefootmark{(d)} \\ 
FDS12\_LSB42                              & 53.998 & -36.574 & 0.88$\pm$ 0.03 & -71.62 & 16.74 $\pm$ 0.11 & 12.9 $\pm$ 1.4 & 1.08 $\pm$ 0.16 & 0.64 $\pm$ 0.08 &                      \\ 
FDS12\_LSB44                              & 54.011 & -36.354 & 0.68$\pm$ 0.04 & 47.77 & 21.86 $\pm$ 0.17 & 2.2 $\pm$ 0.3 & 0.61 $\pm$ 0.17 & 0.54 $\pm$ 0.12 & \tablefootmark{(c)}  \\ 
FDS12\_LSB46                              & 54.004 & -36.258 & 0.89$\pm$ 0.04 & 30.21 & 20.37 $\pm$ 0.15 & 3.4 $\pm$ 0.4 & 0.55 $\pm$ 0.17 & 0.55 $\pm$ 0.09 & \tablefootmark{(c)}  \\ 
FDS12\_LSB47                              & 54.072 & -36.082 & 0.92$\pm$ 0.06 & 83.65 & 20.90 $\pm$ 0.23 & 4.0 $\pm$ 0.6 & 0.32 $\pm$ 0.18 & 0.62 $\pm$ 0.09 &                      \\ 
FDS12\_LSB50                             $*$ & 53.943 & -35.989 & 0.48$\pm$ 0.03 & -54.45 & 16.26 $\pm$ 0.09 & 17.3 $\pm$ 1.7 & 0.64 $\pm$ 0.16 & 0.63 $\pm$ 0.09  & FCC158\tablefootmark{(a)}\tablefootmark{(b)}\tablefootmark{(c)} \\ 
FDS12\_LSB52                              & 55.192 & -36.075 & 0.23$\pm$ 0.03 & -82.74 & 19.51 $\pm$ 0.09 & 5.4 $\pm$ 0.5 & 0.55 $\pm$ 0.16 & 0.64 $\pm$ 0.10 &                      \\ 
FDS12\_LSB53                              & 54.580 & -36.066 & 0.80$\pm$ 0.03 & -22.64 & 18.21 $\pm$ 0.12 & 7.7 $\pm$ 0.9 & 0.78 $\pm$ 0.16 & 0.68 $\pm$ 0.08 & FCC210\tablefootmark{(a)}\tablefootmark{(c)} \\ 
FDS12\_LSB54                              & 54.385 & -36.545 & 0.88$\pm$ 0.03 & 8.69 & 19.08 $\pm$ 0.12 & 4.9 $\pm$ 0.5 & 0.77 $\pm$ 0.16 & 0.82 $\pm$ 0.08 &                      \\ 
FDS12\_LSB55                              & 54.191 & -35.924 & 0.77$\pm$ 0.06 & 24.90 & 21.81 $\pm$ 0.24 & 3.0 $\pm$ 0.5 & 0.65 $\pm$ 0.18 & 0.70 $\pm$ 0.24 & \tablefootmark{(c)}\tablefootmark{(d)} \\ 
FDS16\_LSB6                              & 54.055 & -35.224 & 0.60$\pm$ 0.02 & 32.85 & 19.06 $\pm$ 0.08 & 3.6 $\pm$ 0.3 & 0.77 $\pm$ 0.15 & 0.56 $\pm$ 0.09 &                      \\ 
FDS16\_LSB7                              & 53.928 & -35.338 & 0.68$\pm$ 0.03 & -12.23 & 16.37 $\pm$ 0.10 & 14.6 $\pm$ 1.4 & 0.89 $\pm$ 0.16 & 0.74 $\pm$ 0.08 & FCC156\tablefootmark{(a)}\tablefootmark{(c)} \\ 
FDS16\_LSB10                              & 53.991 & -35.348 & 0.62$\pm$ 0.03 & 86.59 & 19.23 $\pm$ 0.10 & 4.1 $\pm$ 0.4 & 0.64 $\pm$ 0.16 & 0.72 $\pm$ 0.09 & \tablefootmark{(c)}\tablefootmark{(d)} \\ 
FDS16\_LSB11                              & 54.017 & -35.389 & 0.83$\pm$ 0.03 & -77.94 & 16.63 $\pm$ 0.10 & 11.6 $\pm$ 1.1 & 0.99 $\pm$ 0.16 & 0.80 $\pm$ 0.08 & FCC160\tablefootmark{(a)}\tablefootmark{(c)} \\ 
FDS16\_LSB12                              & 53.947 & -35.362 & 0.53$\pm$ 0.04 & 47.10 & 20.70 $\pm$ 0.14 & 3.5 $\pm$ 0.4 & 0.94 $\pm$ 0.17 & 0.75 $\pm$ 0.10 & \tablefootmark{(c)}\tablefootmark{(d)} \\ 
FDS16\_LSB14                              & 54.042 & -35.671 & 0.71$\pm$ 0.06 & -13.55 & 21.25 $\pm$ 0.27 & 4.8 $\pm$ 0.8 & 0.58 $\pm$ 0.19 & 0.34 $\pm$ 0.08 & \tablefootmark{(c)}  \\ 
FDS16\_LSB16                              & 54.051 & -35.746 & 0.57$\pm$ 0.04 & 88.31 & 21.17 $\pm$ 0.17 & 3.3 $\pm$ 0.4 & 0.61 $\pm$ 0.17 & 0.39 $\pm$ 0.10 & \tablefootmark{(c)}  \\ 
FDS16\_LSB20                             $*$ & 54.098 & -35.912 & 0.52$\pm$ 0.03 & 22.59 & 16.81 $\pm$ 0.10 & 13.7 $\pm$ 1.4 & 0.77 $\pm$ 0.16 & 0.67 $\pm$ 0.07  & FCC165\tablefootmark{(a)}\tablefootmark{(c)} \\ 
FDS16\_LSB23                              & 54.035 & -35.987 & 0.44$\pm$ 0.04 & 74.22 & 21.22 $\pm$ 0.15 & 3.2 $\pm$ 0.4 & 0.94 $\pm$ 0.17 & 0.73 $\pm$ 0.11 & \tablefootmark{(c)}  \\ 
FDS16\_LSB24                              & 53.713 & -35.872 & 0.47$\pm$ 0.03 & 88.30 & 19.83 $\pm$ 0.10 & 3.8 $\pm$ 0.4 & 0.61 $\pm$ 0.16 & 0.47 $\pm$ 0.10 &                      \\ 
FDS16\_LSB25                             $*$ & 53.683 & -35.862 & 0.93$\pm$ 0.03 & -5.67 & 16.03 $\pm$ 0.10 & 14.4 $\pm$ 1.4 & 1.14 $\pm$ 0.16 & 0.68 $\pm$ 0.08  & FCC137\tablefootmark{(a)}\tablefootmark{(b)}\tablefootmark{(c)} \\ 
FDS16\_LSB26                              & 53.736 & -35.687 & 0.74$\pm$ 0.05 & 83.04 & 21.08 $\pm$ 0.18 & 3.3 $\pm$ 0.5 & 0.75 $\pm$ 0.18 & 0.88 $\pm$ 0.09 & \tablefootmark{(c)}  \\ 
FDS16\_LSB28                              & 53.720 & -35.570 & 0.88$\pm$ 0.03 & -68.40 & 19.17 $\pm$ 0.13 & 4.8 $\pm$ 0.5 & 0.76 $\pm$ 0.16 & 0.87 $\pm$ 0.08 &                      \\ 
FDS16\_LSB29                              & 53.717 & -35.623 & 0.74$\pm$ 0.03 & -30.43 & 20.06 $\pm$ 0.10 & 2.8 $\pm$ 0.3 & 1.16 $\pm$ 0.16 & 0.51 $\pm$ 0.10 &                      \\ 
FDS16\_LSB30                              & 53.681 & -35.521 & 0.85$\pm$ 0.05 & 76.54 & 20.64 $\pm$ 0.19 & 3.9 $\pm$ 0.6 & 0.57 $\pm$ 0.18 & 0.85 $\pm$ 0.09 & \tablefootmark{(c)}  \\ 
FDS16\_LSB31                              & 53.772 & -35.451 & 0.59$\pm$ 0.06 & 83.56 & 19.57 $\pm$ 0.24 & 10.0 $\pm$ 1.6 & 0.92 $\pm$ 0.18 & 0.79 $\pm$ 0.08 & \tablefootmark{(c)}\tablefootmark{(d)} \\ 
FDS16\_LSB32                              & 53.751 & -35.322 & 0.78$\pm$ 0.03 & 55.68 & 18.58 $\pm$ 0.12 & 6.2 $\pm$ 0.7 & 0.76 $\pm$ 0.16 & 0.80 $\pm$ 0.08 & FCC144\tablefootmark{(a)}\tablefootmark{(c)} \\ 
FDS16\_LSB33                             $*$ & 53.798 & -35.323 & 0.56$\pm$ 0.03 & 65.01 & 18.88 $\pm$ 0.11 & 5.7 $\pm$ 0.6 & 0.62 $\pm$ 0.16 & 0.74 $\pm$ 0.08  & FCC146\tablefootmark{(a)}\tablefootmark{(c)} \\ 
FDS16\_LSB34                              & 53.877 & -35.252 & 0.48$\pm$ 0.04 & 49.69 & 18.65 $\pm$ 0.15 & 9.7 $\pm$ 1.2 & 0.63 $\pm$ 0.17 & 0.59 $\pm$ 0.08 & FCC154\tablefootmark{(a)}\tablefootmark{(c)} \\ 
FDS16\_LSB35                              & 53.740 & -35.224 & 0.70$\pm$ 0.03 & 44.00 & 19.45 $\pm$ 0.12 & 4.5 $\pm$ 0.5 & 0.60 $\pm$ 0.16 & 0.79 $\pm$ 0.08 & \tablefootmark{(c)}\tablefootmark{(d)} \\ 
FDS16\_LSB36                             $*$ & 53.735 & -35.191 & 0.54$\pm$ 0.03 & -85.89 & 17.90 $\pm$ 0.12 & 10.0 $\pm$ 1.1 & 0.84 $\pm$ 0.16 & 0.73 $\pm$ 0.07  & FCC140\tablefootmark{(a)}\tablefootmark{(c)} \\ 
FDS16\_LSB37                              & 53.773 & -35.219 & 0.75$\pm$ 0.03 & -6.53 & 18.45 $\pm$ 0.09 & 5.2 $\pm$ 0.5 & 0.73 $\pm$ 0.16 & 0.83 $\pm$ 0.08 & FCC145\tablefootmark{(a)}\tablefootmark{(c)} \\ 
FDS16\_LSB38                             $*$ & 53.742 & -35.043 & 0.94$\pm$ 0.04 & -20.06 & 17.99 $\pm$ 0.14 & 8.9 $\pm$ 1.1 & 1.02 $\pm$ 0.17 & 0.67 $\pm$ 0.07  & FCC142\tablefootmark{(a)}\tablefootmark{(c)} \\ 
FDS16\_LSB39                              & 53.636 & -35.043 & 0.59$\pm$ 0.06 & -84.10 & 20.63 $\pm$ 0.26 & 6.6 $\pm$ 1.1 & 0.56 $\pm$ 0.19 & 0.42 $\pm$ 0.08 & \tablefootmark{(c)}  \\ 
FDS16\_LSB40                              & 53.613 & -35.106 & 0.54$\pm$ 0.05 & 30.80 & 20.87 $\pm$ 0.21 & 4.9 $\pm$ 0.7 & 0.35 $\pm$ 0.18 & 0.72 $\pm$ 0.09 & \tablefootmark{(c)}  \\ 
FDS16\_LSB41                              & 53.550 & -35.229 & 0.52$\pm$ 0.03 & -49.41 & 19.64 $\pm$ 0.12 & 4.8 $\pm$ 0.5 & 0.71 $\pm$ 0.16 & 0.67 $\pm$ 0.09 & FCC131\tablefootmark{(a)}\tablefootmark{(c)} \\ 
FDS16\_LSB42                              & 53.525 & -35.277 & 0.77$\pm$ 0.04 & 56.82 & 18.78 $\pm$ 0.15 & 7.4 $\pm$ 0.9 & 0.93 $\pm$ 0.17 & 0.65 $\pm$ 0.08 & FCC127\tablefootmark{(a)}\tablefootmark{(c)} \\ 
FDS16\_LSB43                             $*$ & 53.584 & -35.362 & 0.95$\pm$ 0.03 & 77.94 & 16.65 $\pm$ 0.10 & 11.3 $\pm$ 1.1 & 0.75 $\pm$ 0.16 & 0.89 $\pm$ 0.08  & FCC133\tablefootmark{(a)}\tablefootmark{(c)} \\ 
FDS16\_LSB44                              & 53.529 & -35.477 & 0.70$\pm$ 0.07 & -5.49 & 20.79 $\pm$ 0.30 & 6.6 $\pm$ 1.2 & 0.85 $\pm$ 0.19 & 0.92 $\pm$ 0.10 & \tablefootmark{(c)}  \\ 
FDS16\_LSB45                              & 53.538 & -35.516 & 0.89$\pm$ 0.05 & -73.46 & 17.67 $\pm$ 0.19 & 15.2 $\pm$ 2.1 & 0.61 $\pm$ 0.18 & 0.99 $\pm$ 0.13 & FCC130\tablefootmark{(a)}\tablefootmark{(b)}\tablefootmark{(c)} \\ 
FDS16\_LSB47                              & 53.559 & -35.819 & 0.66$\pm$ 0.06 & -2.03 & 19.67 $\pm$ 0.25 & 9.4 $\pm$ 1.5 & 0.78 $\pm$ 0.19 & 0.58 $\pm$ 0.07 & \tablefootmark{(c)}  \\ 
FDS16\_LSB49                              & 53.597 & -35.845 & 0.81$\pm$ 0.06 & 87.44 & 19.79 $\pm$ 0.26 & 8.1 $\pm$ 1.3 & 0.62 $\pm$ 0.19 & 0.50 $\pm$ 0.09 & \tablefootmark{(c)}  \\ 
FDS16\_LSB50                              & 53.451 & -35.836 & 0.86$\pm$ 0.04 & -34.91 & 18.03 $\pm$ 0.14 & 9.1 $\pm$ 1.1 & 0.68 $\pm$ 0.17 & 0.75 $\pm$ 0.07 & FCC125\tablefootmark{(a)}\tablefootmark{(b)}\tablefootmark{(c)} \\ 
FDS16\_LSB52                              & 53.458 & -35.951 & 0.55$\pm$ 0.07 & -47.87 & 21.02 $\pm$ 0.29 & 6.4 $\pm$ 1.1 & 0.79 $\pm$ 0.19 & 0.59 $\pm$ 0.08 & \tablefootmark{(c)}  \\ 
FDS16\_LSB54                              & 53.350 & -35.987 & 1.00$\pm$ 0.07 & -14.49 & 20.33 $\pm$ 0.30 & 6.8 $\pm$ 1.2 & 0.56 $\pm$ 0.19 & 0.60 $\pm$ 0.08 &                      \\ 
FDS16\_LSB55                              & 53.202 & -35.946 & 0.54$\pm$ 0.04 & -61.66 & 18.81 $\pm$ 0.13 & 7.6 $\pm$ 0.9 & 0.72 $\pm$ 0.17 & 0.65 $\pm$ 0.08 &                      \\ 
FDS16\_LSB56                              & 53.366 & -35.965 & 0.57$\pm$ 0.04 & 39.07 & 20.67 $\pm$ 0.14 & 3.3 $\pm$ 0.4 & 0.62 $\pm$ 0.17 & 0.60 $\pm$ 0.10 &                      \\ 
FDS16\_LSB58                              & 53.239 & -35.737 & 0.56$\pm$ 0.03 & -54.81 & 16.12 $\pm$ 0.09 & 16.5 $\pm$ 1.6 & 1.18 $\pm$ 0.16 & 0.70 $\pm$ 0.09 & FCC110\tablefootmark{(a)}\tablefootmark{(b)} \\ 
FDS16\_LSB59                              & 53.388 & -35.704 & 0.71$\pm$ 0.04 & -23.84 & 20.19 $\pm$ 0.14 & 3.8 $\pm$ 0.5 & 0.70 $\pm$ 0.17 & 0.74 $\pm$ 0.09 & \tablefootmark{(c)}  \\ 
FDS16\_LSB60                              & 53.384 & -35.661 & 0.32$\pm$ 0.05 & -74.92 & 19.55 $\pm$ 0.21 & 11.5 $\pm$ 1.7 & 1.74 $\pm$ 0.18 & 0.65 $\pm$ 0.08 & \tablefootmark{(c)}  \\ 
FDS16\_LSB62                              & 53.317 & -35.597 & 0.84$\pm$ 0.04 & -22.02 & 21.13 $\pm$ 0.14 & 2.2 $\pm$ 0.3 & 0.63 $\pm$ 0.17 & 0.67 $\pm$ 0.11 & \tablefootmark{(c)}  \\ 
FDS16\_LSB63                              & 53.286 & -35.397 & 0.53$\pm$ 0.04 & -85.96 & 19.36 $\pm$ 0.14 & 6.3 $\pm$ 0.8 & 0.65 $\pm$ 0.17 & 0.68 $\pm$ 0.08 & FCC114\tablefootmark{(a)}\tablefootmark{(c)} \\ 
FDS16\_LSB64                              & 53.327 & -35.263 & 0.42$\pm$ 0.04 & -68.11 & 19.20 $\pm$ 0.17 & 9.6 $\pm$ 1.3 & 0.70 $\pm$ 0.17 & 0.64 $\pm$ 0.08 &                      \\ 
FDS16\_LSB65                              & 53.368 & -35.126 & 0.86$\pm$ 0.06 & 19.07 & 20.94 $\pm$ 0.27 & 5.0 $\pm$ 0.8 & 1.08 $\pm$ 0.19 & 0.66 $\pm$ 0.08 &                      \\ 
FDS16\_LSB66                              & 53.299 & -35.035 & 0.55$\pm$ 0.07 & -57.14 & 20.27 $\pm$ 0.32 & 10.2 $\pm$ 1.9 & 0.65 $\pm$ 0.19 & 0.45 $\pm$ 0.10 &                      \\ 
FDS16\_LSB67                              & 53.310 & -35.028 & 0.60$\pm$ 0.05 & -15.56 & 20.91 $\pm$ 0.18 & 3.8 $\pm$ 0.5 & 0.75 $\pm$ 0.17 & 0.56 $\pm$ 0.09 &                      \\ 
FDS16\_LSB70                             $*$ & 53.092 & -35.406 & 0.90$\pm$ 0.06 & 35.90 & 19.45 $\pm$ 0.24 & 8.2 $\pm$ 1.3 & 0.50 $\pm$ 0.18 & 0.78 $\pm$ 0.07  &                      \\ 
FDS16\_LSB71                              & 53.022 & -35.426 & 0.82$\pm$ 0.03 & 86.48 & 18.42 $\pm$ 0.12 & 6.8 $\pm$ 0.8 & 1.05 $\pm$ 0.16 & 0.72 $\pm$ 0.08 &                      \\ 
FDS16\_LSB72                              & 52.954 & -35.583 & 0.95$\pm$ 0.04 & 41.89 & 19.13 $\pm$ 0.15 & 5.5 $\pm$ 0.7 & 1.01 $\pm$ 0.17 & 0.72 $\pm$ 0.08 &                      \\ 
FDS16\_LSB74                              & 53.114 & -35.775 & 0.56$\pm$ 0.04 & -33.29 & 18.87 $\pm$ 0.15 & 8.3 $\pm$ 1.0 & 0.73 $\pm$ 0.17 & 0.57 $\pm$ 0.09 & FCC103\tablefootmark{(a)}\tablefootmark{(b)} \\ 
FDS16\_LSB75                              & 53.023 & -35.826 & 0.87$\pm$ 0.06 & -28.43 & 20.61 $\pm$ 0.24 & 5.0 $\pm$ 0.8 & 0.77 $\pm$ 0.18 & 0.30 $\pm$ 0.08 &                      \\ 
FDS16\_LSB77                              & 52.836 & -35.819 & 0.68$\pm$ 0.04 & 42.24 & 19.22 $\pm$ 0.16 & 6.8 $\pm$ 0.9 & 0.92 $\pm$ 0.17 & 0.60 $\pm$ 0.08 & FCC93\tablefootmark{(a)} \\ 
FDS16\_LSB78                              & 52.755 & -35.638 & 0.62$\pm$ 0.04 & 44.54 & 20.45 $\pm$ 0.15 & 3.8 $\pm$ 0.5 & 0.65 $\pm$ 0.17 & 0.57 $\pm$ 0.09 &                      \\ 
FDS16\_LSB79                              & 52.863 & -35.494 & 0.78$\pm$ 0.04 & 28.26 & 18.49 $\pm$ 0.16 & 8.8 $\pm$ 1.1 & 0.45 $\pm$ 0.17 & 0.69 $\pm$ 0.08 & FCC97\tablefootmark{(a)} \\ 
FDS16\_LSB83                              & 52.845 & -34.971 & 0.76$\pm$ 0.04 & -21.21 & 18.84 $\pm$ 0.16 & 7.6 $\pm$ 1.0 & 0.64 $\pm$ 0.17 & 0.62 $\pm$ 0.08 & FCC94\tablefootmark{(a)} \\ 
FDS16\_LSB84                              & 52.820 & -34.962 & 0.65$\pm$ 0.04 & 57.86 & 19.55 $\pm$ 0.16 & 6.2 $\pm$ 0.8 & 0.82 $\pm$ 0.17 & 0.58 $\pm$ 0.08 &                      \\ 
FDS16\_LSB85                              & 53.430 & -35.860 & 0.46$\pm$ 0.04 & 53.43 & 15.81 $\pm$ 0.15 & 36.9 $\pm$ 4.5 & 0.80 $\pm$ 0.17 & 0.59 $\pm$ 0.07 & \tablefootmark{(b)}\tablefootmark{(c)} \\ 
FDS16\_LSB86                              & 53.530 & -35.381 & 0.88$\pm$ 0.07 & -38.63 & 22.55 $\pm$ 0.32 & 2.8 $\pm$ 0.5 & 0.50 $\pm$ 0.19 & 0.59 $\pm$ 0.10 & \tablefootmark{(c)}  \\ 
FDS16\_LSB87                              & 53.343 & -35.606 & 0.86$\pm$ 0.03 & -86.91 & 18.74 $\pm$ 0.10 & 4.6 $\pm$ 0.5 & 0.93 $\pm$ 0.16 & 0.94 $\pm$ 0.08 & \tablefootmark{(c)}  \\ 
\end{longtable} 
\end{landscape} 
}

\end{appendix}

\end{document}